%% file: main.tex
\documentclass[twocolumn,10pt]{asme2ej}
\usepackage{todonotes}

\usepackage{graphics} 
\usepackage{epsfig} 
\usepackage{times} 
\usepackage{amsmath} 
\usepackage{amssymb}  
\usepackage{hyperref} 
\usepackage{color}
\usepackage[T1]{fontenc}
\usepackage{algorithm}
\usepackage{algpseudocode}
\usepackage{multirow}
\usepackage{ctable} 
\usepackage{subcaption}
\usepackage{amsfonts}
\usepackage{footnote}

\title{Safe Learning Reference Governor: Theory and Application to Fuel Truck Rollover Avoidance}
\author{Kaiwen Liu
    \affiliation{
	Department of Aerospace Engineering\\
	University of Michigan\\
	Ann Arbor, Michigan 48109\\
    Email: kwliu@umich.edu
    }	
}

\author{Nan Li
    \affiliation{
	Department of Aerospace Engineering\\
	University of Michigan\\
	Ann Arbor, Michigan 48109\\
    Email: nanli@umich.edu
    }	
}

\author{Ilya Kolmanovsky
    \affiliation{
	Department of Aerospace Engineering\\
	University of Michigan\\
	Ann Arbor, Michigan 48109\\
    Email: ilya@umich.edu
    }	
}

\author{Denise Rizzo
    \affiliation{
	U.S. Army DEVCOM GVSC\\
	Warren, Michigan 48397\\
    Email: denise.m.rizzo2.civ@mail.mil
    }	
}

\author{Anouck Girard
    \affiliation{
	Department of Aerospace Engineering\\
	University of Michigan\\
	Ann Arbor, Michigan 48109\\
    Email: anouck@umich.edu
    }	
}

\input{notation.tex}

\begin{document}

\maketitle
\thispagestyle{empty}
\pagestyle{empty}

\input{sections/00_abstract.tex}

\input{sections/01_introduction.tex}

\input{sections/02_problem_formulation.tex}

\input{sections/03_learning_reference_governor.tex}

\input{sections/04_fuel_truck_dynamics.tex}

\input{sections/05_results.tex}

\input{sections/06_conclusion.tex}

\input{sections/07_appendix.tex}

\bibliographystyle{ieeetr}
\bibliography{references}

\addtolength{\textheight}{-12cm}  

\end{document}

%% file: notation.tex
\usepackage{xcolor}

\newcommand{\state}{\ensuremath{x}}

\newcommand{\inputref}{\ensuremath{\nu}}
\newcommand{\outputsys}{\ensuremath{y}}
\newcommand{\command}{\ensuremath{r}}



%% file: sections/00_abstract.tex
\begin{abstract}
This paper proposes a learning reference governor (LRG) approach to enforce state and control constraints in systems for which an accurate model is unavailable; and this approach enables the reference governor to gradually improve command tracking performance through learning while enforcing the constraints during learning and after learning is completed. The learning can be performed either on a black-box type model of the system or directly on the hardware. After introducing the LRG algorithm and outlining its theoretical properties, this paper investigates LRG application to fuel truck (tank truck) rollover avoidance. Through simulations based on a fuel truck model that accounts for liquid fuel sloshing effects, we show that the proposed LRG can effectively protect fuel trucks from rollover accidents under various operating conditions. 

\end{abstract}

%% file: sections/01_introduction.tex
\section{Introduction} \label{sec:intro}

Safety-critical systems are common in various areas, including transportation control, aerospace applications, nuclear plants, medical applications, etc. The safety conditions can often be expressed as pointwise-in-time state and control constraints, such as actuator range and rate limits, thermal and power limits, safety and comfort limits, as well as obstacle avoidance requirements. In these applications, the controller needs to enforce constraints during the system operation, as constraint violation can lead to catastrophic consequences.

Reference Governor (RG) is an add-on scheme to a nominal closed-loop system, acting as a pre-filter to modify set-point commands to guard the system against potential constraint violations \cite{Garone_Survey}. Such an add-on scheme is suitable for integration with the legacy systems as it does not need to change the existing/legacy controllers. Typical RG designs are model-based. For the case when the model is unknown, a model-free learning-based approach to enforcing state/output constraints based on RG scheme has been proposed \cite{Kaiwen_Model_Free_Learning} for non-safety-critical control systems, where constraint violations are undesirable but do not lead to catastrophic consequences. The learning algorithm proposed in \cite{Kaiwen_Model_Free_Learning} evolves RG based on observed constraint violations during the learning phase and gradually eliminates these violations. 

In this paper, we focus on the learning reference governor (LRG) design for safety-critical systems, where constraints need to be enforced both during and after learning. As an application of LRG, we focus on the steering control of tank trucks that are used in transporting chemical and petroleum products. Accidents associated with these tank trucks may lead to severe people injury and property damage. According to \cite{treichel2006safety}, rollover is the most frequent accident type associated with tank trucks. Since these trucks are partially filled most of the time, liquid sloshing is the main cause of the rollover accidents \cite{salem1999review}, \cite{kang2002cargo} as the free space in the partially-filled tank allows liquid sloshing to happen when the vehicle state changes, and the truck's driving stability is severely affected due to the sloshing force. Although many techniques have been pursued to reduce the effect of sloshing, such as placing baffles inside the tank, these methods cannot cancel the sloshing effects. Consequently, we focus on solutions that minimally modify vehicle steering for vehicles with active front steering (including automated vehicles) if it becomes necessary to avoid vehicle rollover. Model-based reference governor solutions for rollover protection have been proposed in \cite{Bencatel_rollover}, \cite{tian2017coordinated}. As accurate and suitable for online use models of fuel sloshing effects are presently unavailable, we consider the application of LRG which does not need an accurate model.

In spacecraft applications, a pendulum-mass analogy \cite{abramson1961sloshing} or a spring-mass analogy \cite{sidi1997spacecraft} have been considered to capture sloshing modes. A trammel pendulum model is proposed in \cite{slosh_model} to study the driving stability of tank trucks under sloshing effects. We choose the latter approach to establish a model for simulations of LRG, which itself does not require an accurate model.

Model predictive control (MPC) is a common approach to enforce constraints \cite{camacho2013model}, \cite{borrelli2017predictive}, \cite{rawlings2017model}. For systems which do not have an accurate model, learning-based model predictive control (LMPC) has been proposed, which integrates model-based MPC with learning. Safety guarantees for LMPC have been developed in \cite{Aswani_Gonzalez_LearningMPC}, however, a nominal linear model and a bound on deviations of the system dynamics is required. LMPC algorithms frequently learn/estimate a model that can represent the system dynamics, and then use this learned/estimated model to compute control signals \cite{Robust_cons_LNMPC}. Gaussian processes have been employed for LMPC in \cite{Gaussian_MPC}, \cite{LMPC_safe_explor}, however, a theoretical guarantee of safety is missing from these approaches. Adaptive MPC approaches exist which can handle model uncertainty through online parameter estimation \cite{guay2015robust}, \cite{nguyen2020output}. However, such approaches which assume that the model is known except for the parameters require more knowledge of the system compared to the proposed approach.

In addition to LMPC, adaptive and learning control barrier function methods have been proposed to handle systems with parametric uncertainty \cite{taylor2020adaptive}, \cite{taylor2019episodic}, but they rely on a known nominal system model and only ensure safety for a certain class of model uncertainty. Safe RL approaches have been proposed, aiming at ensuring constraint satisfaction during learning. These approaches are typically either model-based  \cite{berkenkamp2017safe} or require a robust controller for the nominal dynamics of the system \cite{cheng2019end}. Iterative learning control (ILC) is another technique that learns from prior experience to improve the controller performance \cite{bristow2006survey}. However, the focus of ILC is typically on improving the tracking performance over a repeated operation \cite{norrlof2002adaptive}. Extensions of ILC that do not require repetitions such as in \cite{kawamura2002analysis} still require that trajectories executed by the system are related by a time-scale transformation.

{\color{black}
The contributions of this paper are as follows:
\begin{enumerate}
    \item We present a novel safe LRG algorithm that integrates safe learning with the RG framework to achieve constraint management of safety-critical systems. This safe LRG algorithm is different from the non-safety-critical algorithm proposed in our previous work \cite{Kaiwen_Model_Free_Learning}. The proposed safe LRG algorithm relies on minimal prior knowledge of the system, where the required prior knowledge can be obtained by running a series of experiments before learning. Then, such LRG can be used to perform learning on systems' hardware or their black box models.
    \item We relax the assumption relied upon in our previous conference paper \cite{liu2020model}, which enables us to treat a broader class of systems. At the same time, we provide theoretical guarantees of constraint enforcement during learning and after learning is completed. We also provide theoretical guarantees of convergence of the proposed learning algorithm, and of finite time convergence of the modified reference (output of LRG) to the original constant commands. 
    \item We demonstrate an application of LRG to fuel truck rollover avoidance under sloshing effects. An equivalent mechanical model for lateral fuel sloshing based on \cite{equiv_model} is first described, and a vehicle dynamics model is integrated with the lateral fuel sloshing model for the simulation study. Simulation results are reported which illustrate the learning process and vehicle responses after learning for step commands, sine-and-dwell tests and when driving conditions change.
\end{enumerate} 

This paper is distinguished from our previous conference paper \cite{liu2020model} by relaxing a Lipschitz continuity assumption to a H\"older continuity assumption, which enables us to treat a broader class of systems, providing more details of the LRG algorithms, extending the theoretical analysis of LRG, including detailed proofs, and illustrating a practical LRG application to fuel truck rollover avoidance. This paper is also different from \cite{liu2021tanktruck}: Although this paper and \cite{liu2021tanktruck} both consider LRG applications to the problem of fuel truck rollover avoidance, \cite{liu2021tanktruck} focuses on practical implementation of LRG to such an application and does not contain the theoretical results, Propositions 1-4, of this paper. For instance, \cite{liu2021tanktruck} introduces the use of neural network as a function approximator to store the LRG learning results; however, with integration with a neural network, the strict constraint enforcement guarantee of LRG may no longer hold due to neural network approximation error. 

}

The paper is organized as follows. In Section~\ref{sec:problem_formulation}, we formulate the problem considered in this paper. In Section~\ref{sec:learning_rg}, we introduce LRG and establish its theoretical properties. Section~\ref{sec:truck_dyn} describes the equivalent mechanical model of fuel sloshing as well as the tank truck dynamics model. Section~\ref{sec:results} illustrates the simulation results obtained from the fuel truck model and LRG. Finally, concluding remarks are made in Section~\ref{sec:conclusion}.

The notations used in this paper are standard. In particular, for a right-continuous signal $\theta:[0,\infty) \to \Theta$, we use $\theta(t^-)$ to denote the left-sided limit $\lim_{t' \nearrow t} \theta(t')$, and for a better distinction, we also use $\theta(t^+)$ to denote $\theta(t)$.

%% file: sections/02_problem_formulation.tex
\section{Problem Formulation} \label{sec:problem_formulation}

\begin{figure}[h]
\begin{center}
\begin{picture}(240.0, 60.0)
\put(  0,  -4){\epsfig{file=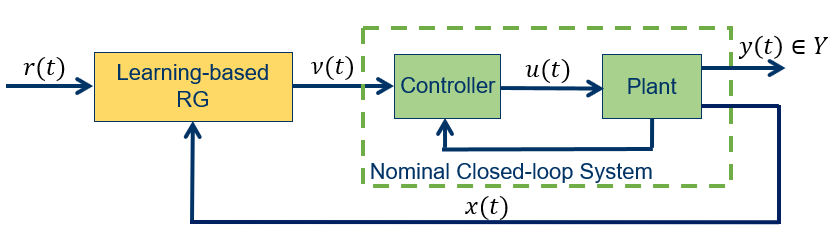,width=0.975\linewidth}}  
\end{picture}
\end{center}
      \caption{\small{Diagram of a nominal closed-loop system augmented with a learning-based reference governor for handling constraints.}}
      \label{fig:control_diagram}
\end{figure}

In this paper, we consider a stable (or a pre-stabilized) system the dynamics of which can be represented by the following equations: 
\begin{subequations}\label{eq:pf_sys}
	\begin{align}
    \dot{\state}(t) &= f\big(\state(t), \inputref(t)\big) \label{eq:pf_sys_1}, \\
    \outputsys(t) &= g\big(\state(t), \inputref(t)\big) \label{eq:pf_sys_2},
    \end{align}
\end{subequations}
where $\state(t) \in \mathbb{R}^n$ denotes the system state at time $t \in [0,\infty)$, $\outputsys(t) \in \mathbb{R}^m$ denotes the system output, $\inputref(t)$ denotes the reference input, taking values in a compact and convex set $V\subset \mathbb{R}^{n_\inputref}$ and determining the set-point of the system, and $f:\mathbb{R}^n \times \mathbb{R}^{n_\inputref} \rightarrow \mathbb{R}^n$ and $g:\mathbb{R}^n\times\mathbb{R}^{n_\inputref} \rightarrow \mathbb{R}^m$ are nonlinear functions. We note that such a pre-stabilized system typically consists of a plant to be controlled and a nominal controller, illustrated as in Fig.~\ref{fig:control_diagram}.

We assume that the system must operate without violating a prescribed set of specifications, which are represented as pointwise-in-time constraints on the outputs of the form:
\begin{equation}\label{eq:constraint}
\outputsys(t) \in Y, \quad \forall t \in [0, \infty),
\end{equation}
where $Y \subset \mathbb{R}^m$ is a closed set with a nonempty interior. 

We make the following assumptions:

{\it (A1)} For any initial condition $\state(0) = \state_0 \in \mathbb{R}^n$ and piecewise continuous reference input signal $\inputref: [0, \infty) \rightarrow V$, the solution to the differential equation \eqref{eq:pf_sys_1}, $\state:[0, \infty) \rightarrow \mathbb{R}^n$, exists and is unique.

{\it (A2)} For any constant reference input $\inputref \in V$, the autonomous system,
\begin{equation}\label{eq:pf_sys_init}
    \dot{\state}(t) = f\big(\state(t), \inputref\big),
\end{equation}
has a unique equilibrium point, $\state_\inputref = \state_\inputref(\inputref)$, which is globally asymptotically stable (GAS).

We note that since \eqref{eq:pf_sys} represents a system that has been stabilized by a nominal controller (see Fig.~\ref{fig:control_diagram}), assumptions {\it (A1)} and {\it (A2)} are reasonable.

{\it (A3)} The steady-state mapping in {\it (A2)}, $\state_\inputref: V \rightarrow \mathbb{R}^n$, and the output function in \eqref{eq:pf_sys_2}, $g:\mathbb{R}^n \times \mathbb{R}^{n_\inputref} \rightarrow \mathbb{R}^m$, are continuous functions.

The above assumptions {\it (A1)}-{\it (A3)} characterize the class of systems to be treated. Although the system \eqref{eq:pf_sys} is stable (or pre-stabilized), the nominal controller may not have the ability to enforce the imposed constraints \eqref{eq:constraint}. In addition, for many practical systems, the functions $f$ and $g$ in their corresponding models \eqref{eq:pf_sys} can be highly complex or not given explicitly (e.g., when the model is given as a black-box simulation code). Moreover, in some circumstances, the system may not even have an accurate model as is the case for a vehicle which has undergone in-field modifications. Therefore, the proposed LRG scheme relies on learning rather than explicit knowledge of $f$ and $g$ to achieve constraint enforcement. For situations where the system does not have a model, the learning process may be performed directly on the hardware. In such a case, it can be important to ensure constraint satisfaction even during the learning process, especially for safety-critical systems, in which constraint violations may cause catastrophic consequences such as damage to the hardware or to the human operator. The proposed control scheme achieves such a safe learning based on the assumptions {\it (A4)}-{\it (A6)} stated below.




Firstly, let $\psi(\cdot, \state_0, \inputref): [0, \infty) \rightarrow \mathbb{R}^n$ denote the solution to \eqref{eq:pf_sys_init} corresponding to the initial condition $\state(0) = \state_0$ and constant reference input $\inputref \in V$, and let $\phi(\cdot, \state_0, \inputref) = g\big(\psi(\cdot, \state_0, \inputref),\inputref\big)$ denote the corresponding output trajectory $\outputsys$. Then, let $D: \mathbb{R}^{n_\inputref} \times \mathbb{R}^{n_\inputref} \times \mathbb{R}^n \to \mathbb{R}$ be defined as
\begin{equation} \label{eq:def_D}
D(\inputref, \delta\inputref, \delta\state) := \sup_{t \in [0,\infty)} \big\|\phi\big(t, \state_\inputref(\inputref)+\delta\state, \inputref+\delta\inputref \big) - \outputsys_\inputref(\inputref) \big\|,
\end{equation}
where $\outputsys_\inputref(\inputref) = g\big(\state_\inputref(\inputref),\inputref\big)$ denotes the steady-state output corresponding to the reference $\inputref$, and $\|\cdot\| = \|\cdot\|_{\mathbb{R}^{m}}$ denotes an arbitrary vector norm on $\mathbb{R}^{m}$. 

The above function $D$ represents the maximum deviation of the output trajectory $\outputsys$ from the steady-state output $\outputsys_\inputref(\inputref)$ when the initial condition $x(0)$ is deviated from the equilibrium $\state_\inputref(\inputref)$ by $\delta\state$ and the reference input $\inputref(t)$ is deviated from $\inputref$ by a constant $\delta\inputref$. Note that according to its definition, $D(\inputref, 0, 0) = 0$ for all $\inputref \in \mathbb{R}^{n_\inputref}$. 

This function $D$ will be exploited in the proposed control scheme for constraint enforcement. However, in general, there does not exist an explicit formula for $D$, even when the functions $f$ and $g$ in the model \eqref{eq:pf_sys} are explicitly known. Therefore, we employ a data-driven approach to learn/estimate $D$. In particular, as discussed above, constraint satisfaction should be ensured during the process of data collection and learning. For this, we rely on the following assumptions on the function $D$:

{\color{black}
{\it (A4)} There is a known pair of constants $L>0, \beta \geq 1$ such that for any $z_1, z_2 \in \mathbb{R}^{n_\inputref} \times \mathbb{R}^{n_\inputref} \times \mathbb{R}^n$, it holds that
\begin{equation} \label{eq:lipschitz}
\big| D(z_1) - D(z_2) \big| \leq L\, \| z_1 - z_2 \|^{\frac{1}{\beta}},
\end{equation}
where $\|\cdot\| = \|\cdot\|_{\mathbb{R}^{n_\inputref} \times \mathbb{R}^{n_\inputref} \times \mathbb{R}^n}$ denotes an arbitrary vector norm on $\mathbb{R}^{n_\inputref} \times \mathbb{R}^{n_\inputref} \times \mathbb{R}^n$. Note that this H\"older continuity assumption is a weaker and more general assumption than the Lipschitz continuity assumption relied upon in \cite{liu2020model} and \cite{liu2021tanktruck}, because when $\beta = 1$, \eqref{eq:lipschitz} reduces to Lipschitz continuity condition.

The above H\"older continuity assumption for $D$ is reasonable. For instance, in Lemma~2 and 3 in the Appendix we show that this assumption holds true for all asymptotically stable linear time-invariant (LTI) systems, where we also provide formulas to compute estimates of the constants $L$ and $\beta$ in \eqref{eq:lipschitz} when the LTI system has an explicit model. For nonlinear systems with explicit models, the  technique in \cite{8880489}, which derives an explicit bound on the system responses to deviations in the initial condition and reference input using the logarithmic norms, can be used to check {\it (A4)} and estimate $L$ and $\beta$. In practice, the constants $L$ and $\beta$ may also be estimated based on engineering insight or by a data-driven approach using sampled trajectories, in both cases avoiding the need for explicit knowledge of the functions $f$ and $g$ in \eqref{eq:pf_sys}. On the one hand, estimating constants $L$ and $\beta$ can be much easier than identifying $f$ and $g$ globally from data. On the other hand, in principle, for a fixed $\beta$, the estimate of $L$ can be arbitrarily conservative as long as it is a finite number. Note, however, that a more conservative estimate of $L$ may result in a slower learning rate; this will be further shown in our simulation case study in Section~\ref{sec:results}.
}

We next make the following assumption:

{\it (A5)} At each time instant $t \in [0,\infty)$, the state $\state(t)$, the output $\outputsys(t)$, and the distance from the steady-state output, $\outputsys_\inputref\big(\inputref(t)\big)$, associated with the current reference input, $\inputref(t)$, to the constraint boundary, 
\begin{equation}\label{eq:distance}
d\big(\inputref(t)\big) := \text{dist}\big( \outputsys_\inputref\big(\inputref(t)\big), Y^C \big) = \inf_{\outputsys \in Y^C} \big\|\outputsys_\inputref\big(\inputref(t)\big)-\outputsys\big\|,
\end{equation}
can all be measured. In \eqref{eq:distance}, $Y^C$ denotes the complement of the constraint set $Y$, i.e., $Y^C = \mathbb{R}^m \backslash Y$.

We note that measuring $d\big(\inputref(t)\big)$ requires knowledge of the steady-state output mapping $\outputsys_\inputref(\cdot) = g\big(\state_\inputref(\cdot),\cdot\big): V \rightarrow \mathbb{R}^m$. When $f$ and $g$ are not explicitly known, the mapping $\outputsys_\inputref$, or the distance mapping $d(\cdot) = \text{dist}\big( \outputsys_\inputref(\cdot), Y^C \big): V \rightarrow \mathbb{R}$, can be estimated using data of preliminary steady-state experiments with the system, and be function-fitted or stored as a look-up table for online use, as done in \cite{vahidi2006constraint}.

{\it (A6)} Finally, we assume there is a known pair $(T, \varepsilon)$, with $\varepsilon>0$ being sufficiently small, such that for any $(\inputref, \delta\inputref, \delta\state) \in \mathbb{R}^{n_\inputref} \times \mathbb{R}^{n_\inputref} \times \mathbb{R}^n$ satisfying $\inputref, \inputref + \delta\inputref \in V$, it holds that
\begin{align}\label{eq:A6}
& \tilde{D}(\inputref, \delta\inputref, \delta\state) := \nonumber \\[2pt]
& \max_{t\in[0, T]} \big\|\phi\big(t, \state_\inputref(\inputref)+\delta\state, \inputref+\delta\inputref \big) - \outputsys_\inputref(\inputref) \big\| + \varepsilon \nonumber \\[2pt]
& \geq D(\inputref, \delta\inputref, \delta\state).
\end{align}

The last assumption {\it (A6)} is reasonable and ensures that the value of $D(\inputref, \delta\inputref, \delta\state)$, which is defined in \eqref{eq:def_D} as a supremum over an infinite interval, can be estimated using trajectory data of finite length (of length $T$) with high accuracy (with error bounded by $\varepsilon$). Note that according to their definitions in \eqref{eq:def_D} and \eqref{eq:A6}, $D(\inputref, \delta\inputref, \delta\state)$ and $\tilde{D}(\inputref, \delta\inputref, \delta\state)$ also satisfy $D(\inputref, \delta\inputref, \delta\state) + \varepsilon \ge \tilde{D}(\inputref, \delta\inputref, \delta\state)$. Therefore, we have $D \le \tilde{D} \le D + \varepsilon$ pointwise on $\mathbb{R}^{n_\inputref} \times \mathbb{R}^{n_\inputref} \times \mathbb{R}^n$.




%% file: sections/03_learning_reference_governor.tex
\section{Learning Reference Governor} \label{sec:learning_rg}

We adopt a reference governor approach to enforce the constraints. The RG is an add-on scheme to the nominal closed-loop system, illustrated by Fig.~\ref{fig:control_diagram}. It acts as a pre-filter, which monitors the commanded reference input, $\command(t)$, and adjusts it to a modified reference input, $\inputref(t)$, to ensure constraint satisfaction. Differently from conventional RG schemes, the design of which requires explicit knowledge of the model \eqref{eq:pf_sys} \cite{Garone_Survey}, the proposed RG relies on learning and is thereby referred to as {\it learning reference governor (LRG)}.

\subsection{Reference Governor} \label{sec:learning_rg_1}


The LRG updates the reference input at sample time instants $\{t_k\}_{k = 0}^{\infty} \subset [0, \infty)$ (with $t_k \rightarrow \infty$ as $k \rightarrow \infty$) based on the following reference update law,
\begin{equation}\label{equ:adjustment}
\inputref(t^+) = 
	\inputref(t^-) + \kappa\big(\state(t),\command(t),\inputref(t^-)\big) \big( \command(t) - \inputref(t^-) \big),
\end{equation}
where $\inputref(t^-)$ and $\inputref(t^+)$ denote the reference input values before and after the update, respectively, and $\kappa: \mathbb{R}^n \times \mathbb{R}^{n_\inputref} \times \mathbb{R}^{n_\inputref} \rightarrow [0, 1]$ is a scalar function. The LRG maintains $\inputref(t)$ as constant over each interval $[t_k, t_{k+1})$. This way, the resulting reference input signal $\inputref: [0, \infty) \rightarrow \mathbb{R}^{n_\inputref}$ is piecewise constant and right continuous, and by {\it (A1)}, the system \eqref{eq:pf_sys} has a unique solution. 

The following lemma is exploited by the LRG to enforce the constraints \eqref{eq:constraint}.

{\bf Lemma 1:} Given a pair $(\state_0, \inputref) \in \mathbb{R}^n \times V$, suppose that
$\phi(t, \state_0, \inputref) \in Y$ for all $t \in [0,\infty)$. Then, any adjustment $\delta \inputref$ satisfying $\inputref + \delta \inputref \in V$ and
\begin{equation}\label{equ:lemma_1}
D\big(\inputref, \delta\inputref, \state_0 - x_\inputref(\inputref)\big) \le d(\inputref),
\end{equation} 
guarantees $\phi(t, \state_0, \inputref + \delta \inputref) \in Y$ for all $t \in [0,\infty)$.

{\bf Proof:} Firstly, $\phi(t, \state_0, \inputref) \in Y$ for all $t \in [0,\infty)$ implies $\outputsys_\inputref(\inputref) \in Y$, since $\state_\inputref(\inputref)$ is GAS by {\it (A2)}, $g$ is continuous by {\it (A3)}, and $Y$ is closed. 

Then, we consider two cases separately: 1) $d(\inputref) = 0$, and 2) $d(\inputref) > 0$. For the former case, i.e., if $d(\inputref) = 0$, we have
\begin{align}
0 &\le D\big(\inputref, \delta\inputref, \state_0 - \state_\inputref(\inputref)\big) \nonumber \\
&= \!\! \sup_{t \in [0,\infty)} \!\! \big\|\phi(t, \state_0, \inputref+\delta\inputref) - \outputsys_\inputref(\inputref) \big\| \le d(\inputref) = 0, 
\end{align}
which implies $\phi(t, \state_0, \inputref+\delta\inputref) \equiv \outputsys_\inputref(\inputref) \in Y$. 

For the latter case $d(\inputref) > 0$, according to the definition of $d(\inputref)$ in \eqref{eq:distance}, it holds that the closed ball centered at $\outputsys_\inputref(\inputref)$ with radius $d(\inputref)$ is contained entirely in the constraint set $Y$, i.e., $\overline{B}\big(\outputsys_\inputref(\inputref),d(\inputref)\big):= \big\{\outputsys \in \mathbb{R}^m \big| \|\outputsys_\inputref(\inputref) - \outputsys \| \le d(\inputref)\big\} \subseteq Y$. This is because for any $\outputsys' \in Y^C$, we have $\|\outputsys_\inputref(\inputref) - \outputsys'\| \ge \inf_{\outputsys \in Y^C} \|\outputsys_\inputref(\inputref) - \outputsys \big\| = d(\inputref)$, which implies the open ball $B\big(\outputsys_\inputref(\inputref),d(\inputref)\big):= \big\{\outputsys \in \mathbb{R}^m \big| \|\outputsys_\inputref(\inputref) - \outputsys\| < d(\inputref)\big\} \subseteq Y$. Since $\overline{B}\big(\outputsys_\inputref(\inputref),d(\inputref)\big)$ is the closure of $B\big(\outputsys_\inputref(\inputref),d(\inputref)\big)$ and $Y$ is closed, it holds that $\overline{B}\big(\outputsys_\inputref(\inputref),d(\inputref)\big) \subseteq Y$.

Then, according to the definition of $D$ in \eqref{eq:def_D}, $D\big(\inputref, \delta\inputref, \state_0 - x_\inputref(\inputref)\big) \le d(\inputref)$ implies $\phi(t,\state_0, \inputref+\delta\inputref) \in \overline{B}\big(\outputsys_\inputref(\inputref),d(\inputref)\big) \subseteq Y$ for all $t \in [0,\infty)$. $\blacksquare$

In principle, while guaranteeing the satisfaction of \eqref{eq:constraint}, minimizing the deviation of the modified reference $\inputref(t^+)$ from the commanded value $\command(t)$ is desired, which, according to the reference update law \eqref{equ:adjustment}, corresponds to maximizing the value of $\kappa\big(\state(t),\command(t),\inputref(t^-)\big)$ in $[0, 1]$. Note also that when $\kappa\big(\state(t),\command(t),\inputref(t^-)\big) = 1$, we have $\inputref(t^+) = \command(t)$, i.e., the commanded reference input is reached. 

On the basis of Lemma~1, if the function $D$, which is defined in \eqref{eq:def_D}, is known, then one can design the function $\kappa$ in the reference update law \eqref{equ:adjustment} (or equivalently, determine the reference adjustment $\delta\inputref$ at each sample time instant) such that the condition \eqref{equ:lemma_1} is satisfied, which in turn guarantees the satisfaction of \eqref{eq:constraint} after the reference update. However, as discussed in Section~\ref{sec:problem_formulation}, the function $D$ is typically not known a priori. Therefore, in what follows we introduce a safe learning algorithm to learn/estimate $D$ from data, and at the same time evolve the design of $\kappa$. Here, ``safe'' means that the algorithm guarantees the constraints \eqref{eq:constraint} to be satisfied over the entire process of data collection and learning.

\subsection{Safe Learning Algorithm} \label{sec:learning_rg_2}

The proposed learning algorithm is formally presented as Algorithm~\ref{alg:SafeLearning_RG_1} with the \texttt{Kappa} function in line~5 determined according to Algorithm~\ref{alg:ComputeKappa}. This learning algorithm continually improves the estimate of $D$ and at the same time evolves $\kappa$ as more data are collected.

\begin{algorithm}
\caption{Safe learning algorithm}
\label{alg:SafeLearning_RG_1}
\begin{algorithmic}[1]
\State Initialize the system \eqref{eq:pf_sys} with a strictly constraint-admissible steady state $\state_\inputref\big(\inputref(0^-)\big)$ as the initial condition, i.e., $\state(0) = \state_\inputref\big(\inputref(0^-)\big)$ with $\inputref(0^-) \in V$ and satisfying $y(0^-) = g\big(\state(0), \inputref(0^-)\big) \in \text{int}(Y)$, and initialize the dataset $\mathcal{D} \leftarrow \emptyset$; 
\For{$n = 0:n_{\max}-1$}
\State Generate $\command_n \in V \setminus \big\{\inputref\big(n k_{\max} T^-\big)\big\}$ either randomly (e.g., based on a uniform distribution) or according to a training profile $\{\command_i\}_{i=0}^{\infty}$ that covers the operating range of the system;
\For{$k = 0:k_{\max}-1$}
\State At the sample time instant $t = (n k_{\max} + k)T$, compute 
\begin{equation} \label{eq:alg_kappa}
\kappa(t) = \texttt{Kappa}\big(\state(t), \command_n, \inputref(t^-), d\big(\inputref(t^-)\big), \mathcal{D}\big);
\end{equation}
\State Adjust the reference input according to 
\begin{equation} \label{eq:alg_nu}
	\inputref(t^+) = 
	\inputref(t^-) + \kappa(t) \big( \command_n - \inputref(t^-) \big);
\end{equation}
\State At the next sample time instant $t' = t + T$, measure $\tilde{D}(t) = \tilde{D}\big(\inputref(t^-), \delta \inputref(t), \delta \state(t)\big)$, where $\delta \inputref(t) = \kappa(t) \big( \command_n - \inputref(t^-) \big)$ and $\delta \state(t) = \state(t)-\state_\inputref\big(\inputref(t^-)\big)$;
\State Add the new data point $\big(\inputref(t^-),\delta \inputref(t),\delta \state(t),\tilde{D}(t)\big)$ to the dataset $\mathcal{D}$, i.e.,
\begin{equation}
\mathcal{D} \leftarrow \mathcal{D} \cup \big(\inputref(t^-),\delta \inputref(t),\delta \state(t),\tilde{D}(t)\big).
\end{equation}
\EndFor
\EndFor
\end{algorithmic}
\end{algorithm}

\setlength{\textfloatsep}{1pt}
\begin{algorithm}[t]
\caption{\texttt{Kappa}($\state, \command, \inputref, d, \mathcal{D}$)}
\label{alg:ComputeKappa}
\color{black}
\begin{algorithmic}[1]
\For{$(\inputref_i, \delta\inputref_i, \delta\state_i, \tilde{D_i}) \in \mathcal{D}$}
\State Compute $\kappa_i$ as the solution to the following optimization problem,
{\small
\begin{align} \label{eq:alg_alpha}
& \max\,\, \kappa \in [0,1], \quad \text{subject to} \\
&\!\!\!\! \big\|\kappa (r-\inputref) - \delta\inputref_i \big\| \le \bigg(\frac{d-\tilde{D_i}}{L}\bigg)^\beta - \bigg\|\begin{bmatrix}
\inputref \\ \state - \state_\inputref(\inputref)
\end{bmatrix} - \begin{bmatrix}
\inputref_i \\
\delta\state_i
\end{bmatrix} \bigg\|, \nonumber
\end{align}
}
if a solution exists, and $\kappa_i = 0$ otherwise;
\EndFor
\State \begin{equation}\label{eq:alg_alpha_2}
\kappa' = \mathop{\text{sat}}_{[0,1]} \bigg(\frac{d}{L}\bigg)^\beta \frac{1}{\| r-\inputref\|} - \frac{\|\state - \state_\inputref(\inputref)\|}{\| r-\inputref\|},
\end{equation}
\State \Return $\kappa = \max\, (\max_i \kappa_i, \kappa')$.
\end{algorithmic}
\end{algorithm}

{\color{black}
In Algorithm~\ref{alg:SafeLearning_RG_1}, the sampling period $T$ in lines~3, 5 and 7 is the $T$ in {\it (A6)}. The $\|\cdot\|$'s in \eqref{eq:alg_alpha} and \eqref{eq:alg_alpha_2} are the norm $\|\cdot\| = \|\cdot\|_{\mathbb{R}^{n_\inputref} \times \mathbb{R}^{n_\inputref} \times \mathbb{R}^n}$ used in \eqref{eq:lipschitz} restricted to the subspaces $\{0\} \times \mathbb{R}^{n_\inputref} \times \{0\}$, $\mathbb{R}^{n_\inputref} \times \{0\} \times \mathbb{R}^n$, and $\{0\} \times \{0\} \times \mathbb{R}^n$, respectively, i.e., $\|\delta \inputref\| = \|(0,\delta \inputref,0)\|$, $\|(\inputref,\delta x)\| = \|(\inputref,0,\delta x)\|$, and $\|\delta x\| = \|(0,0,\delta x)\|$. We note also that \eqref{eq:alg_alpha_2} is the solution to the following optimization problem,
\begin{align}
& \max\,\, \kappa \in [0,1], \quad \text{subject to} \\
&\!\!\!\! \big\|\kappa (r-\inputref) - 0 \big\| \le \bigg(\frac{d - 0}{L}\bigg)^\beta - \bigg\|\begin{bmatrix}
\inputref \\ \state - \state_\inputref(\inputref)
\end{bmatrix} - \begin{bmatrix}
\inputref \\ 0
\end{bmatrix} \bigg\|, \nonumber
\end{align}
which has the same form as \eqref{eq:alg_alpha}, but with $(\inputref_i, \delta\inputref_i, \delta\state_i, \tilde{D_i})$ replaced by $(\inputref, 0, 0, 0)$.

We note that Algorithm~\ref{alg:SafeLearning_RG_1} continually improves the following estimate of $D$,
\begin{align}\label{equ:upper_D}
\overline{D}(\inputref, \delta\inputref, \delta\state) &= \min \Bigg( \min_{i \in \mathcal{D}} \Bigg( \tilde{D}_i + L\, \Bigg\|\begin{bmatrix}
\inputref \\ \delta\inputref \\ \delta\state
\end{bmatrix} - \begin{bmatrix} 
\inputref_i \\
\delta\inputref_i \\ \delta\state_i
\end{bmatrix} \Bigg\|^\frac{1}{\beta}\, \Bigg), \nonumber \\
&\quad\quad\quad\quad L\, \bigg\|\begin{bmatrix}
\delta\inputref \\ \delta\state
\end{bmatrix} \bigg\|^\frac{1}{\beta}\, \Bigg).
\end{align}
In particular, under assumptions {\it (A4)} and {\it (A6)}, \eqref{equ:upper_D} is an upper bound for $D$. This is because for any $(\inputref, \delta\inputref, \delta\state) \in \mathbb{R}^{n_\inputref} \times \mathbb{R}^{n_\inputref} \times \mathbb{R}^n$, on the one hand, {\it (A4)} and {\it (A6)} imply
\begin{align}\label{equ:upper_D_2}
D(\inputref, \delta\inputref, \delta\state) &\le D(\inputref', \delta\inputref', \delta\state') + L\, \Bigg\|\begin{bmatrix}
\inputref \\ \delta\inputref \\ \delta\state
\end{bmatrix} - \begin{bmatrix} 
\inputref' \\
\delta\inputref' \\ \delta\state'
\end{bmatrix} \Bigg\|^\frac{1}{\beta} \nonumber \\
&\le \tilde{D}(\inputref', \delta\inputref', \delta\state') + L\, \Bigg\|\begin{bmatrix}
\inputref \\ \delta\inputref \\ \delta\state
\end{bmatrix} - \begin{bmatrix} 
\inputref' \\
\delta\inputref' \\ \delta\state'
\end{bmatrix} \Bigg\|^\frac{1}{\beta},
\end{align}
for all measured data $\big(\inputref', \delta\inputref', \delta\state', \tilde{D}(\inputref', \delta\inputref', \delta\state')\big)$; and, on the other hand, {\it (A4)} and the fact that $D(\inputref, 0, 0) = 0$ for all $\inputref \in \mathbb{R}^{n_\inputref}$ imply
\begin{align}\label{equ:upper_D_3}
D(\inputref, \delta\inputref, \delta\state) &\le D(\inputref, 0, 0) + L\, \Bigg\|\begin{bmatrix}
\inputref \\ \delta\inputref \\ \delta\state
\end{bmatrix} - \begin{bmatrix} 
\inputref \\
0 \\ 0
\end{bmatrix} \Bigg\|^\frac{1}{\beta} \nonumber \\
&= L\, \bigg\|\begin{bmatrix} \delta\inputref \\ \delta\state
\end{bmatrix} \bigg\|^\frac{1}{\beta}.
\end{align}
Then, \eqref{eq:alg_alpha} and \eqref{eq:alg_alpha_2} maximize $\kappa \in [0,1]$ such that $\overline{D}(\inputref, \delta\inputref, \delta\state) \le d(\inputref)$.

The optimization problem \eqref{eq:alg_alpha} is a convex program with a scalar decision variable and can generally be solved with a bisection method. For the following two special cases, which are commonly encountered in practical situations, closed-form solutions to \eqref{eq:alg_alpha} exist:

Firstly, if the norm $\|\cdot\| = \|\cdot\|_{\mathbb{R}^{n_\inputref} \times \mathbb{R}^{n_\inputref} \times \mathbb{R}^n}$ in \eqref{eq:lipschitz} restricted to the subspace $\{0\} \times \mathbb{R}^{n_\inputref} \times \{0\}$ is a quadratic norm, i.e., $\big\|\kappa (r-\inputref) - \delta\inputref_i \big\| = \big\|\kappa (r-\inputref) - \delta\inputref_i \big\|_Q = \sqrt{(\kappa (r-\inputref) - \delta\inputref_i)^\top Q (\kappa (r-\inputref) - \delta\inputref_i)}$ for some positive-definite matrix $Q \in \mathbb{R}^{n_\inputref \times n_\inputref}$, then \eqref{eq:alg_alpha} admits the following closed-form solution, 
\begin{align}\label{equ:kappa_closed_1}
& \kappa_i = \mathop{\text{sat}}_{[0,1]}\, \frac{1}{(r-\inputref)^{\top} Q(r-\inputref)} \bigg( (r-\inputref)^{\top} Q\, \delta\inputref_i\, +  \\ 
& \sqrt{\big((r-\inputref)^{\top} Q\, \delta\inputref_i\big)^2 - (r-\inputref)^{\top} Q(r-\inputref) \big(\delta\inputref_i^{\top} Q\, \delta\inputref_i -\Theta_i\big)}\, \bigg), \nonumber
\end{align}
where
\begin{equation}
\Theta_i = \bigg( \Big(\frac{d-\tilde{D_i}}{L} \Big)^\beta - \bigg\|\begin{bmatrix}
\inputref \\ \state - \state_\inputref(\inputref)
\end{bmatrix} - \begin{bmatrix}
\inputref_i \\
\delta\state_i
\end{bmatrix} \bigg\|\, \bigg)^2.    
\end{equation}

Secondly, if the reference input is a scalar, i.e., $n_\inputref = 1$, and $\big\|\kappa (r-\inputref) - \delta\inputref_i \big\| = \big|\kappa (r-\inputref) - \delta\inputref_i \big|$, then the closed-form solution \eqref{equ:kappa_closed_1} further reduces to
\begin{align}\label{equ:kappa_closed_2}
&\kappa_i = \mathop{\text{sat}}_{[0,1]} \max \\ 
&\frac{1}{\command-\inputref}\bigg( \delta\inputref_i \pm \bigg( \Big(\frac{d-\tilde{D_i}}{L} \Big)^\beta - \bigg\|\begin{bmatrix}
\inputref \\ \state - \state_\inputref(\inputref)
\end{bmatrix} - \begin{bmatrix}
\inputref_i \\
\delta\state_i
\end{bmatrix} \bigg\|\, \bigg) \bigg). \nonumber
\end{align}
}



The learning algorithm terminates either after the maximum number of training commands, $n_{\max}$, has been reached or when the moving average of the tracking error between the commanded reference input $\command(t)$ and the modified reference input $\inputref(t)$, defined as $\frac{1}{T_{\text{win}}} \int_{t-T_{\text{win}}}^t \|\command(s)-\inputref(s)\|\, \text{d}s$ with $T_{\text{win}}>0$ denoting the window size, converges lower than a specified threshold value. After termination of the learning phase, the system is ready for operation. In principle, learning can be stopped at any time and the resulting reference governor will enforce the constraints, however, its performance in terms of moving average of the tracking error can be more conservative than after training is completed.

During the operating phase, the LRG uses \eqref{equ:adjustment} to adjust the reference input $\inputref(t)$ at each sample time instant $t \in \{t_k\}_{k = 0}^{\infty}$, where $\kappa\big(\state(t),\command(t),\inputref(t^-)\big)$ is determined according to \eqref{eq:alg_kappa} and Algorithm~2 with the dataset $\mathcal{D}$ obtained from the learning phase. We note that the LRG can use a different sampling period for the operating phase than for the learning phase and smaller values in the operating phase can improve performance.

We are now ready to discuss theoretical properties of the proposed LRG algorithm.

\subsection{Theoretical Properties} \label{sec:learning_rg_3}

The LRG algorithm introduced in Sections~\ref{sec:learning_rg_1} and \ref{sec:learning_rg_2} guarantees constraint satisfaction during both the learning and the operating phases, pointwise convergence of the estimate of $D$, and finite-time convergence of the modified reference input $\inputref(t)$ to constant, strictly steady-state constraint-admissible reference command $\command(t)$ during the operating phase. These properties are presented as the following propositions.

{\color{black}
{\bf Proposition 1:} During the learning phase, constraints $\outputsys(t) \in Y$ are guaranteed to be satisfied for all $t \in [0, \infty)$.

{\bf Proof:} Let $\tau = \sigma T$, $\sigma = 0,1,2,\dots$, 
be an arbitrary sample time instant during the learning phase. Suppose $\phi(t', \state(\tau), \inputref(\tau^-)) \in Y$ for all $t' \in [0, \infty)$, where $\phi(\cdot, \state(\tau), \inputref(\tau^-)) = g\big(\psi(\cdot, \state(\tau), \inputref(\tau^-)),\inputref(\tau^-)\big)$ denotes the output trajectory of \eqref{eq:pf_sys_init} with the initial condition $\state_0 = \state(\tau)$ and constant reference input $\inputref = \inputref(\tau^-)$. Then, for any $\delta \inputref$ satisfying 
{\small 
\begin{equation}\label{eq:P1_1}
\big\|\delta \inputref - \delta\inputref_i \big\| \le \bigg(\frac{d\big(\inputref(\tau^-)\big) \!-\! \tilde{D_i}}{L}\bigg)^\beta - \bigg\|\!\begin{bmatrix}
\inputref(\tau^-) \\ \state(\tau) \!-\! \state_\inputref\big(\inputref(\tau^-)\big)
\end{bmatrix} - \begin{bmatrix}
\inputref_i \\
\delta\state_i
\end{bmatrix} \!\bigg\|_{}
\end{equation} }
for some $i \in \mathcal{D}$, it holds that
\begin{align}\label{eq:P1_2}
& D\big(\inputref(\tau^-), \delta\inputref, x(\tau)-x_\inputref(\inputref(\tau^-))\big) \le \nonumber \\
& \tilde{D}_i + L\, \Bigg\|\begin{bmatrix}
\inputref(\tau^-) \\ \delta\inputref \\ x(\tau)-x_\inputref(\inputref(\tau^-))
\end{bmatrix} - \begin{bmatrix} 
\inputref_i \\
\delta\inputref_i \\ \delta\state_i
\end{bmatrix} \Bigg\|^\frac{1}{\beta} \le \nonumber \\
& \tilde{D}_i + L\, \bigg(\bigg\|\begin{bmatrix}
\inputref(\tau^-) \\ x(\tau)-x_\inputref(\inputref(\tau^-))
\end{bmatrix} - \begin{bmatrix} 
\inputref_i \\
\delta\state_i
\end{bmatrix} \bigg\| + \|
\delta\inputref - \delta\inputref_i
\| \bigg)^\frac{1}{\beta} \nonumber \\
& \le d\big(\inputref(\tau^-)\big),
\end{align}
where we have used the triangle inequality to derive the second inequality. Then, according to Lemma~1, \eqref{eq:P1_2} implies $\phi(t', x(\tau), \inputref(\tau^-)+ \delta\inputref) \in Y$ for all $t' \in [0,\infty)$. Similarly, for any $\delta \inputref$ satisfying 
\begin{equation}\label{eq:P1_3}
\|\delta \inputref \| \le \bigg(\frac{d\big(\inputref(\tau^-)\big)}{L}\bigg)^\beta - \big\|\state(\tau) - \state_\inputref\big(\inputref(\tau^-)\big)
\big\|,
\end{equation}
it holds that
\begin{align}\label{eq:P1_4}
& D\big(\inputref(\tau^-), \delta\inputref, x(\tau) \!-\! x_\inputref(\inputref(\tau^-))\big) \!\le\! L\, \bigg\|\!\begin{bmatrix} \delta\inputref \\ x(\tau) \!-\! x_\inputref(\inputref(\tau^-))
\end{bmatrix} \!\bigg\|^\frac{1}{\beta} \nonumber \\
& \le L\, \Big(\big\| x(\tau)-x_\inputref\big(\inputref(\tau^-)\big)
\big \| + \|
\delta\inputref\| \Big)^\frac{1}{\beta}
\le d\big(\inputref(\tau^-)\big),
\end{align}
which implies $\phi(t', x(\tau), \inputref(\tau^-)+ \delta\inputref) \in Y$ for all $t' \in [0,\infty)$. Therefore, for $\inputref(\tau^+)$ determined according to \eqref{eq:alg_kappa}, \eqref{eq:alg_nu}, and Algorithm~2, it must hold that $\phi(t', x(\tau), \inputref(\tau^+)) \in Y$ for all $t' \in [0,\infty)$. Note that if there exists no $\delta \inputref$ satisfying \eqref{eq:P1_1} or \eqref{eq:P1_3}, then Algorithm~2 returns $\kappa(t) = 0$. In this case, we have $\inputref(\tau^+) = \inputref(\tau^-)$, and thus, $\phi(t', x(\tau), \inputref(\tau^+)) = \phi(t', x(\tau), \inputref(\tau^-)) \in Y$.

Since the learning algorithm is initialized with $\state(0) = \state_\inputref\big(\inputref(0^-)\big)$ such that $\phi(t', \state(0), \inputref(0^-)) \equiv g\big(\state(0), \inputref(0^-)\big) \in Y$ for all $t' \in [0, \infty)$, the above result says $\phi(t', x(0), \inputref(0^+)) \in Y$ for all $t' \in [0,\infty)$, which also implies $\phi(t', x(T), \inputref(T^-)) = \phi(t'+T, x(0), \inputref(0^+)) \in Y$ for all $t' \in [0,\infty)$. Then, by induction on $\tau = 0, T, 2T, \dots$, we obtain that $\phi(t', x(\tau), \inputref(\tau^+)) \in Y$ for all $t' \in [0,\infty)$ and all sample time instants $\tau$. Now let $t \in [0, \infty)$ be arbitrary and consider the sample time instant $\tau$ immediately before $t$ (i.e., $\tau \le t < \tau+T$). It holds that $\outputsys(t) = \phi(t-\tau, x(\tau), \inputref(\tau^+)) \in Y$. This proves $\outputsys(t) \in Y$ for all $t \in [0, \infty)$. $\blacksquare$




Proposition~1 certifies that our algorithm is a safe learning algorithm, i.e., ensures constraint satisfaction over the entire learning process.
}


{\bf Corollary 1:} During the operating phase, constraints $\outputsys(t) \in Y$ are guaranteed to be satisfied for all $t \in [0, \infty)$. In particular, this guarantee does not depend on the length of the learning phase.

{\bf Proof:} The constraint satisfaction result established in the proof of Proposition~1 depends only on the reference update law used by LRG, neither on the sampling period $T$ nor on the number of data points in the dataset $\mathcal{D}$. Therefore, the same proof can be used to prove Corollary~1, with $T$ now corresponding to the sampling period for the operating phase. $\blacksquare$



Corollary~1 establishes the constraint enforcement property of our LRG after training. Although the guaranteed constraint satisfaction does not depend on the length of learning, the performance, in terms of reference tracking, will be improved as learning proceeds. The following proposition formalizes such a result.

{\color{black}
{\bf Proposition 2:} Let $\overline{D}^{\sigma}$ denote the estimate of $D$, \eqref{equ:upper_D}, after the sample time instant $\tau = \sigma T$ during the learning phase. The following properties hold: (i) $D \le \overline{D}^{\sigma+1} \le \overline{D}^{\sigma}$ pointwise on $\mathbb{R}^{n_\inputref} \times \mathbb{R}^{n_\inputref} \times \mathbb{R}^n$ for all $\sigma = 0, 1, \dots$; (ii) $\overline{D}^{\sigma}$ converges pointwise on $\mathbb{R}^{n_\inputref} \times \mathbb{R}^{n_\inputref} \times \mathbb{R}^n$ as $\sigma \to \infty$; and (iii) the pointwise limit $\overline{D}^{\infty} := \lim_{\sigma \to \infty} \overline{D}^{\sigma}$ satisfies $\overline{D}^{\infty} \ge D$ and is H\"older continuous with constants $L$ and $\beta$ on $\mathbb{R}^{n_\inputref} \times \mathbb{R}^{n_\inputref} \times \mathbb{R}^n$.

{\bf Proof:} Firstly, $D \le \overline{D}^{\sigma}$ for all $\sigma = 0, 1, \dots$ follows from the expression \eqref{equ:upper_D} and the inequalities \eqref{equ:upper_D_2} and \eqref{equ:upper_D_3}.
Then, $\overline{D}^{\sigma+1} \le \overline{D}^{\sigma}$ follows from the fact that the dataset corresponding to $\overline{D}^{\sigma+1}$, denoted as $\mathcal{D}^{\sigma+1}$, over which the second minimum in the expression \eqref{equ:upper_D} is taken, is a superset of that corresponding to $\overline{D}^{\sigma}$, $\mathcal{D}^{\sigma}$. We have $\mathcal{D}^{\sigma+1} \supseteq \mathcal{D}^{\sigma}$ because as learning proceeds, more data points are collected into $\mathcal{D}$. This proves (i).

For each $(\inputref, \delta\inputref, \delta\state) \in \mathbb{R}^{n_\inputref} \times \mathbb{R}^{n_\inputref} \times \mathbb{R}^n$, since $0 \le D(\inputref, \delta\inputref, \delta\state) \le \overline{D}^{\sigma+1}(\inputref, \delta\inputref, \delta\state) \le \overline{D}^{\sigma}(\inputref, \delta\inputref, \delta\state)$ for all $\sigma = 0, 1, \dots$, by the monotone convergence theorem, the sequence $\big\{\overline{D}^{\sigma}(\inputref, \delta\inputref, \delta\state)\big\}_{\sigma = 0}^{\infty}$ must converge. This proves the pointwise convergence of $\overline{D}^{\sigma}$ on $\mathbb{R}^{n_\inputref} \times \mathbb{R}^{n_\inputref} \times \mathbb{R}^n$ as $\sigma \to \infty$.

For (iii), $\overline{D}^{\infty} \ge D$ follows from the fact that $\overline{D}^{\sigma} \ge D$ for all $\sigma = 0, 1, \dots$ It remains to show that $\overline{D}^{\infty}$ is H\"older continuous with constants $L$ and $\beta$.

Let $\sigma = 0, 1, \dots$ be arbitrary and denote the corresponding dataset as $\mathcal{D}^{\sigma}$. It is easy to see that for each $i \in \mathcal{D}^{\sigma}$, the following function,
\begin{equation}
\overline{D}^{\sigma,i}(\inputref, \delta\inputref, \delta\state) := \tilde{D}_i + L\, \Bigg\|\begin{bmatrix}
\inputref \\ \delta\inputref \\ \delta\state
\end{bmatrix} - \begin{bmatrix} 
\inputref_i \\
\delta\inputref_i \\ \delta\state_i
\end{bmatrix} \Bigg\|^\frac{1}{\beta},
\end{equation}
is H\"older continuous with constants $L$ and $\beta$. Because $\overline{D}^{\sigma,i}$ for all $i \in \mathcal{D}^{\sigma}$ and $\overline{D}'(\inputref, \delta\inputref, \delta\state) := L\, \bigg\|\begin{bmatrix} \delta\inputref \\ \delta\state
\end{bmatrix} \bigg\|^\frac{1}{\beta}$ share the same H\"older constants $L$ and $\beta$, we have that $\overline{D}^{\sigma} = \min\, \big(\min_{i \in \mathcal{D}^{\sigma}} \overline{D}^{\sigma,i}, \overline{D}' \big)$ is also H\"older continuous with constants $L$ and $\beta$. Note that this result holds for all $\sigma = 0, 1, \dots$

Since $\overline{D}^{\infty}$ is the pointwise limit of the non-increasing sequence of functions $\{\overline{D}^{\sigma}\}_{\sigma = 0}^{\infty}$, it can also be expressed as
$\overline{D}^{\infty} = \inf_{\sigma = 0,1,\dots} \overline{D}^{\sigma}$. In this case, because $\overline{D}^{\sigma}$ for all $\sigma = 0,1,\dots$ share the same H\"older continuous with constants $L$ and $\beta$, we can conclude that $\overline{D}^{\infty} = \inf_{\sigma = 0,1,\dots} \overline{D}^{\sigma}$ is H\"older continuous with constants $L$ and $\beta$. This completes the proof of (iii). $\blacksquare$

Recall that Algorithm~2 maximizes $\kappa \in [0,1]$ such that $\overline{D}(\inputref, \delta\inputref, \delta\state) \le d(\inputref)$. According to Proposition~2(i), as learning proceeds, the estimate $\overline{D}(\inputref, \delta\inputref, \delta\state)$ becomes less conservative (i.e., smaller), and correspondingly, the feasible region for $\kappa$ is enlarged. In turn, $\kappa$ can take more aggressive values for $\inputref(t)$ to track $\command(t)$, which transfers to improved reference tracking performance.

After the learning process, the dataset $\mathcal{D}$ may contain a large number of data points. For online implementation of our LRG, Algorithm~2 needs to run fast enough to produce $\kappa(t)$ for updating the reference input $\inputref(t)$ in real time. This is possible because, on the one hand, the optimization problem \eqref{eq:alg_alpha} may admit closed-form solutions, such as \eqref{equ:kappa_closed_1} and \eqref{equ:kappa_closed_2}; and on the other hand, the computational tasks of \eqref{eq:alg_alpha} for different data points $i \in \mathcal{D}$ are ready to be performed in parallel. Furthermore, one may partition the space $\mathbb{R}^{n_\inputref} \times \mathbb{R}^{n_\inputref} \times \mathbb{R}^n$ by bounded subsets $\{U_i\}_{i = 1}^{\infty}$, such as cubes, with diameter $\text{diam}(U_i) = \sup\{\|z_1 - z_2\| \,|\, z_1, z_2 \in U_i\} \le m$ for all $i$, and post-process the dataset $\mathcal{D}$ such that if there are multiple data points in the same $U_i$, then only one of them is kept and the others are dropped. This way, the number of data points in $\mathcal{D}$ will be reduced, and in turn, the computational cost of Algorithm~2 will become lower.

Denote the estimate $\overline{D}$ corresponding to the dataset before post-processing as $\overline{D}^{\text{pre}}$ and that corresponding to the dataset after post-processing as $\overline{D}^{\text{post}}$. The following result can be used to guide the design of the parameter $m$ to balance the tradeoff between estimation performance and dataset complexity.

{\bf Proposition 3:} After post-processing, the estimate, $\overline{D}^{\text{post}}$, satisfies $\overline{D}^{\text{pre}} \le  \overline{D}^{\text{post}} \le \overline{D}^{\text{pre}} + 2 L m^\frac{1}{\beta} + \varepsilon$ pointwise on $\mathbb{R}^{n_\inputref} \times \mathbb{R}^{n_\inputref} \times \mathbb{R}^n$.

{\bf Proof:} Firstly, the dataset before post-processing, $\mathcal{D}^{\text{pre}}$, is a superset of that after post-processing, $\mathcal{D}^{\text{post}}$. Thus, according to the expression \eqref{equ:upper_D}, we have $\overline{D}^{\text{pre}} \le  \overline{D}^{\text{post}}$.

Secondly, for each $(\inputref, \delta\inputref, \delta\state) \in \mathbb{R}^{n_\inputref} \times \mathbb{R}^{n_\inputref} \times \mathbb{R}^n$, according to \eqref{equ:upper_D}, we have
\begin{align}
\overline{D}^{\text{pre}}(\inputref, \delta\inputref, \delta\state) \!&=\! \min \Bigg(\tilde{D}_j + L\, \Bigg\|\!\begin{bmatrix}
\inputref \\ \delta\inputref \\ \delta\state
\end{bmatrix} - \begin{bmatrix} 
\inputref_j \\
\delta\inputref_j \\ \delta\state_j
\end{bmatrix}\!\Bigg\|^\frac{1}{\beta}, \nonumber \\
&\quad\quad\quad\quad L\, \bigg\|\!\begin{bmatrix}
\delta\inputref \\ \delta\state
\end{bmatrix}\!\bigg\|^\frac{1}{\beta}\, \Bigg),   
\end{align}
for some data point $j \in \mathcal{D}^{\text{pre}}$. The post-processing procedure introduced above ensures that there exists some data point $k \in \mathcal{D}^{\text{post}}$ such that $\|(\inputref_k,\delta\inputref_k,\delta\state_k) - (\inputref_j,\delta\inputref_j,\delta\state_j)\| \le m$. Therefore, we have
\begin{align}
& \overline{D}^{\text{post}}(\inputref, \delta\inputref, \delta\state) = \nonumber \\
& \min \Bigg( \min_{i \in \mathcal{D}^{\text{post}}} \Bigg( \tilde{D}_i + L\, \Bigg\|\begin{bmatrix}
\inputref \\ \delta\inputref \\ \delta\state
\end{bmatrix} - \begin{bmatrix} 
\inputref_i \\
\delta\inputref_i \\ \delta\state_i
\end{bmatrix} \Bigg\|^\frac{1}{\beta}\, \Bigg), L\, \bigg\|\begin{bmatrix}
\delta\inputref \\ \delta\state
\end{bmatrix} \bigg\|^\frac{1}{\beta}\, \Bigg) \nonumber \\
&\le \min \Bigg(\tilde{D}_k + L\, \Bigg\|\!\begin{bmatrix}
\inputref \\ \delta\inputref \\ \delta\state
\end{bmatrix} - \begin{bmatrix} 
\inputref_k \\
\delta\inputref_k \\ \delta\state_k
\end{bmatrix}\!\Bigg\|^\frac{1}{\beta}, L\, \bigg\|\!\begin{bmatrix}
\delta\inputref \\ \delta\state
\end{bmatrix}\!\bigg\|^\frac{1}{\beta}\, \Bigg) \nonumber \\
&\le \min \Bigg(\tilde{D}_k + (\tilde{D}_j - D_j) + L\, \Bigg(\Bigg\|\!\begin{bmatrix}
\inputref \\ \delta\inputref \\ \delta\state
\end{bmatrix} - \begin{bmatrix} 
\inputref_j \\
\delta\inputref_j \\ \delta\state_j
\end{bmatrix}\!\Bigg\| \nonumber \\
&+ \Bigg\|\!\begin{bmatrix} 
\inputref_k \\
\delta\inputref_k \\ \delta\state_k
\end{bmatrix} - \begin{bmatrix} 
\inputref_j \\
\delta\inputref_j \\ \delta\state_j
\end{bmatrix}\!\Bigg\| \Bigg)^\frac{1}{\beta}, L\, \bigg\|\!\begin{bmatrix}
\delta\inputref \\ \delta\state
\end{bmatrix}\!\bigg\|^\frac{1}{\beta}\, \Bigg) \nonumber \\
&\le \min \Bigg(\tilde{D}_k + (\tilde{D}_j - D_j) + L\,\Bigg\|\!\begin{bmatrix}
\inputref \\ \delta\inputref \\ \delta\state
\end{bmatrix} - \begin{bmatrix} 
\inputref_j \\
\delta\inputref_j \\ \delta\state_j
\end{bmatrix}\!\Bigg\|^\frac{1}{\beta} \nonumber \\
&+ L\,\Bigg\|\!\begin{bmatrix} 
\inputref_k \\
\delta\inputref_k \\ \delta\state_k
\end{bmatrix} - \begin{bmatrix} 
\inputref_j \\
\delta\inputref_j \\ \delta\state_j
\end{bmatrix}\!\Bigg\|^\frac{1}{\beta}, L\, \bigg\|\!\begin{bmatrix}
\delta\inputref \\ \delta\state
\end{bmatrix}\!\bigg\|^\frac{1}{\beta}\, \Bigg) \nonumber \\
&\le \min \Bigg(\tilde{D}_j + L\, \Bigg\|\!\begin{bmatrix}
\inputref \\ \delta\inputref \\ \delta\state
\end{bmatrix} - \begin{bmatrix} 
\inputref_j \\
\delta\inputref_j \\ \delta\state_j
\end{bmatrix}\!\Bigg\|^\frac{1}{\beta}, L\, \bigg\|\!\begin{bmatrix}
\delta\inputref \\ \delta\state
\end{bmatrix}\!\bigg\|^\frac{1}{\beta}\, \Bigg) \nonumber \\
&\quad\quad + (\tilde{D}_k - D_j) + L\, \|(\inputref_k,\delta\inputref_k,\delta\state_k) - (\inputref_j,\delta\inputref_j,\delta\state_j)\|^\frac{1}{\beta} \nonumber \\
&\le \overline{D}^{\text{pre}}(\inputref, \delta\inputref, \delta\state) + 2Lm^\frac{1}{\beta} + \varepsilon,
\end{align}
where we have used $\tilde{D}_j - D_j \ge 0$, by {\it (A6)}, and the triangle inequality to derive the second inequality, and the triangle inequality variant with exponents $1/\beta \in (0, 1]$ to derive the third inequality, and have used the following result to derive the last inequality,
\begin{align}
\tilde{D}_k - D_j &\le D_k + \varepsilon - D_j \nonumber \\
&\le \big|D(\inputref_k, \delta\inputref_k, \delta\state_k) - D(\inputref_j, \delta\inputref_j, \delta\state_j)\big| + \varepsilon \nonumber \\
&\le L\, \|(\inputref_k,\delta\inputref_k,\delta\state_k) - (\inputref_j,\delta\inputref_j,\delta\state_j)\|^\frac{1}{\beta} + \varepsilon \nonumber \\
&\le Lm^\frac{1}{\beta} + \varepsilon,
\end{align}
in which the third inequality is due to {\it (A4)}. This proves $\overline{D}^{\text{post}} \le \overline{D}^{\text{pre}} + 2 L m^\frac{1}{\beta} + \varepsilon$. $\blacksquare$

We next study the convergence of the modified reference $\inputref(t)$ to the commanded value $\command(t)$. To achieve an enhanced convergence property, following the approach in \cite{8880489}, we exploit the following two sets, 
\begin{align}
V_1(r) &:= \big\{ \inputref \in V \,\big|\, d(\inputref) \geq \min\,(d(r),\delta) \big\}, \\
V_2(\overline{\inputref}, \command) &:= \big\{\inputref\in V \,\big|\, \|\inputref - \command \| \leq \max\, (\|\overline{\inputref} - \command \| - \lambda, 0 ) \big\}, \nonumber
\end{align}
where $\delta > 0$ is arbitrarily small and $\lambda = \lambda(r) \in \Big(0, \big(\frac{\min\,(d(r),\delta)}{L}\big)^\beta\Big)$, and we slightly modify the reference update law \eqref{equ:adjustment} during the operating phase to 
\begin{align}\label{equ:adjustment_2}
&\inputref(t^+) = \tilde{\inputref}(t^+)\, \chi(t) + \inputref(t^-) \big(1-\chi(t)\big), \\
&\tilde{\inputref}(t^+)	= \inputref(t^-) + \kappa\big(\state(t),\command(t),\inputref(t^-)\big) \big( \command(t) - \inputref(t^-) \big), \nonumber
\end{align}
where $\chi(t) = 1$ if $\tilde{\inputref}(t^+) \in V_1\big(r(t)\big) \cap V_2\big(\inputref(t^-), r(t)\big)$, and $\chi(t) = 0$ otherwise.


{\bf Proposition 4:} Consider the operation of the reference governor based on the update law \eqref{equ:adjustment_2}. Suppose that there exists $t_s \in [0,\infty)$ such that $\command(t) = \command_s$ for all $t \in [t_s,\infty)$, with $\command_s \in V$ satisfying $d(\command_s)>0$. Suppose also that for any $\inputref$ on the line segment connecting $\inputref(t_s^-)$ and $\command_s$, we have $\inputref \in V_1(\command_s)$. Then, with the reference update law \eqref{equ:adjustment_2}, there exists $t_f \in [t_s,\infty)$ such that $\inputref(t) = \command(t)$ for all $t \in [t_f,\infty)$. In particular, this result does not depend on the length of the learning phase.

{\bf Proof:} Let $J_k := \|\inputref(t_k^+) - \command_s\|$. Because in the reference update law \eqref{equ:adjustment_2}, $\kappa$, determined by Algorithm~2, takes values in the interval $[0, 1]$, the sequence $\{J_k\}_{k = k_s}^{\infty}$, with $t_{k_s}$ denoting the first sample time instant after $t_s$, is non-increasing. Since $\{J_k\}_{k = k_s}^{\infty}$ is also bounded from below by $0$, by the monotone convergence theorem, $J_k$ converges as $k \to \infty$. Let $\hat{J} := \lim_{k \to \infty} J_k$. 

In particular, due to the requirement $\tilde{\inputref}(t^+) \in V_2\big(\inputref(t^-), \command_s\big)$ $= \big\{\inputref \in V \,\big|\, \|\inputref - \command_s \| \leq \max\, (\|\inputref(t^-) - \command_s \| - \lambda, 0 )\big\}$ in \eqref{equ:adjustment_2} for the reference input $\inputref(t^+)$ to be updated from $\inputref(t^-)$ to $\tilde{\inputref}(t^+)$, it must hold that $J_{k+1} \le \max(J_k -\lambda, 0)$ whenever $J_{k+1} \neq J_k$. Note that the $\lambda$ here is a positive constant. In this case, the sequence $\{J_k\}_{k = k_s}^{\infty}$ converges to $\hat{J}$ through at most a finite number of jumps. In other words, there exists $k_f \in \mathbb{N}_0$ such that $J_k = \hat{J}$ for all $k \ge k_f$. Under \eqref{equ:adjustment_2}, this also implies $\inputref(t) = \inputref(t_{k_f}^+) := \hat{\inputref}$ for all $t \in [t_{k_f}, \infty)$, i.e., the modified reference $\inputref(t)$ converges to $\hat{\inputref}$ in finite time.

We now show that $\hat{J} = 0$ and $\hat{\inputref} = \command_s$ by contradiction. 

Suppose $\hat{\inputref} \neq \command_s$. First note that the requirement $\tilde{\inputref}(t^+) \in V_1(r_s)$ in \eqref{equ:adjustment_2} for $\inputref(t^+)$ to be updated from $\inputref(t^-)$ to $\tilde{\inputref}(t^+)$ ensures $\hat{\inputref} = \inputref(t_{k_f}^+) \in V_1(r_s)$, which implies $\big(\frac{d(\hat{\inputref})}{L}\big)^\beta - \lambda \ge \big(\frac{\min\,(d(r_s),\delta)}{L}\big)^\beta - \lambda > 0$. Then, since $\inputref(t) = \hat{\inputref}$ for all $t \in [t_{k_f}, \infty)$ and $\state_\inputref(\hat{\inputref})$ is GAS by {\it (A2)}, there exists $\tau \in \{t_k\}_{k=k_f}^\infty$ such that $\|\state(\tau) - \state_\inputref(\hat{\inputref})\| \le \big(\frac{d(\hat{\inputref})}{L}\big)^\beta - \lambda$, which can also be expressed as $\big(\frac{d(\hat{\inputref})}{L}\big)^\beta - \|\state(\tau) - \state_\inputref(\hat{\inputref})\| \ge \lambda > 0$.

Then, according to \eqref{eq:alg_alpha_2}, the $\kappa$ produced by Algorithm~\ref{alg:ComputeKappa} at the sample time instant $\tau$ satisfies either $\kappa \ge \frac{(d(\hat{\inputref})/L)^\beta-  \|\state(\tau) - \state_\inputref(\hat{\inputref})\|}{\| r_s-\hat{\inputref}\|} > 0$ or $\kappa = 1$. In this case, $\tilde{\inputref} := \hat{\inputref} + \kappa (r_s-\hat{\inputref})$ satisfies either
\begin{align}
& \|\hat{\inputref} -r_s\| - \|\tilde{\inputref} -r_s\| = \|\hat{\inputref} -r_s\| - \|\hat{\inputref} + \kappa (r_s-\hat{\inputref}) -r_s\| \nonumber \\
&= \kappa\, \| \hat{\inputref} - r_s \| \ge \Big(\frac{d(\hat{\inputref})}{L}\Big)^\beta - \|\state(\tau) - \state_\inputref(\hat{\inputref})\| \ge \lambda,
\end{align}
or $\tilde{\inputref} = \hat{\inputref} + \kappa (r_s-\hat{\inputref}) = r_s$, leading to $\|\tilde{\inputref} - r_s\| = 0$. In either case, $\tilde{\inputref} = \hat{\inputref} + \kappa (r_s-\hat{\inputref}) \in V_2(\hat{\inputref}, \command_s)$. Note also that with the reference update law \eqref{equ:adjustment_2}, all of $\big\{\inputref(t_k^+)\big\}_{k = k_s}^{\infty}$ lie on the line segment connecting $\inputref(t_s^-)$ and $\command_s$, including $\hat{\inputref} = \inputref(t_{k_f}^+)$. In this case, $\tilde{\inputref} = \hat{\inputref} + \kappa (r_s-\hat{\inputref})$ also lies on the line segment connecting $\inputref(t_s^-)$ and $\command_s$. Then, by the second assumption in the proposition statement, we have $\tilde{\inputref} \in V_1(r_s)$. 

Since $\kappa > 0$ and $\tilde{\inputref} = \hat{\inputref} + \kappa (r_s-\hat{\inputref}) \in V_1(r_s) \cap V_2(\hat{\inputref}, \command_s)$, under \eqref{equ:adjustment_2}, the reference input should be updated from $\hat{\inputref}$ to $\tilde{\inputref}$ at the sample time instant $\tau \in \{t_k\}_{k=k_f}^\infty$. This contradicts our assumption that $\inputref(t)$ converges to some $\hat{\inputref} \neq \command_s$ at $t = t_{k_f}$. Since in the above we have shown that $\inputref(t)$ indeed converges to some $\hat{\inputref}$ at $t = t_{k_f}$, we can conclude that $\hat{\inputref} = \command_s$. This completes the proof.

It is clear that the above proof does not depend on the number of data points in the dataset $\mathcal{D}$. Therefore, this finite-time convergence result does not depend on the length of learning. $\blacksquare$
}

%% file: sections/04_fuel_truck_dynamics.tex
\section{Application to Fuel Truck Rollover Avoidance} \label{sec:truck_dyn}

We now consider an application of the proposed learning reference governor (LRG) to fuel truck rollover avoidance. In what follows, we introduce models to represent the roll dynamics of a fuel truck with a partially filled tank. These models were originally proposed in \cite{slosh_model,equiv_model}, where sloshing dynamics of liquid fuel in the tank are accounted for and modeled by an equivalent trammel pendulum. We summarize them here for the sake of completeness.

\subsection{Equivalent trammel pendulum model of liquid sloshing}

\begin{figure}[h!]
\begin{center}
\begin{picture}(240.0, 125.0)
\put(  -4,  -4){\epsfig{file=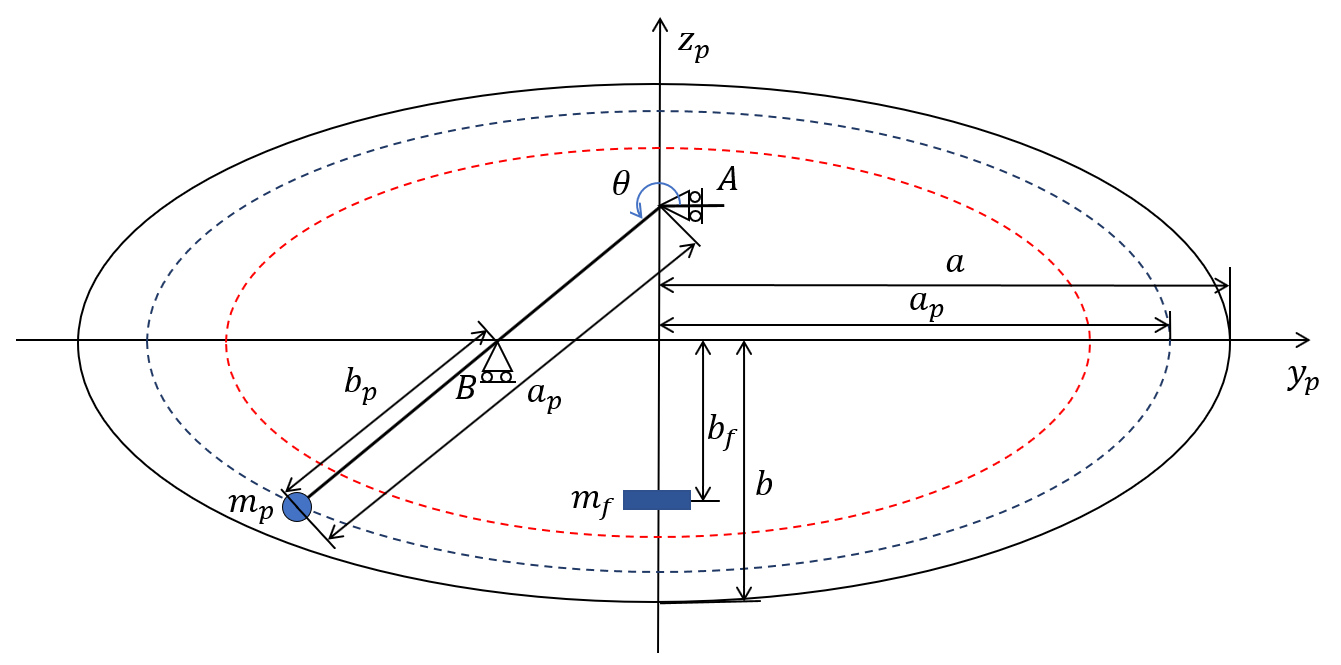,width=1\linewidth}}  
\end{picture}
\end{center}
      \caption{Diagram of the trammel pendulum model.}
      \label{fig:trammel_pend}
\end{figure}

A trammel pendulum model, shown in Fig.~\ref{fig:trammel_pend}, is used to describe the sloshing dynamics of liquid fuel in the tank. In particular, the model divides the liquid mass $m_l$ into two parts: the fixed mass $m_f$, which is fixed relative to the tank, and the pendulum mass $m_p$, which accounts for the dynamic motion of liquid. The pivot $A$ of the pendulum can move along the $z_p$ axis, and the pivot $B$ can move along the $y_p$ axis. The rod linking pivots $A$, $B$ and the pendulum mass $m_p$ is assumed to be massless and rigid, i.e., the lengths $a_p$ and $b_p$ are constants. Consequently, this pendulum system has only one degree of freedom, which can be defined by the pendulum angle $\theta$ in Fig.~\ref{fig:trammel_pend}. This angle $\theta$ is related to the tilt angle of liquid free surface and is assumed to be measured (e.g., using  a combination of wave gauges, and level and optical sensors \cite{gabl2019capturing}) for our LRG implementation. The following equation of motion for $\theta$ in the tank-fixed inertia frame can be derived \cite{equiv_model}, 
\begin{equation}
    \ddot{\theta} (a_p^2 \sin^2\theta + b_p^2 \cos^2\theta) + \frac{1}{2}\dot{\theta}^2 (a_p^2 - b_p^2) \sin 2\theta + g b_p \cos\theta = 0,
\end{equation}
where $g = 9.81$ (m/s) is the acceleration due to gravity.

For a given fuel tank with a given liquid fill ratio, the trammel pendulum parameters $m_f$, $m_p$, $a_p$ and $b_p$ are estimated according to,
\begin{align}
    m_f =&\, m_l - m_p, \\
    m_p =&\, (m_1 + m_2\Delta + m_3\Lambda + m_4\Delta^2 + m_5\Delta\Lambda \nonumber \\
    &+ m_6\Lambda^2 + m_7\Delta^3 + m_8\Delta^2\Lambda + m_9\Delta\Lambda^2)\, m_l, \\
    a_p =&\, \Lambda\, b_p, \\
    b_p =&\, (b_1 + b_2\Delta + b_3\Lambda + b_4\Delta^2 + b_5\Delta\Lambda \nonumber \\
    &+ b_6\Lambda^2 + b_7\Delta^3 + b_8\Delta^2\Lambda + b_9\Delta\Lambda^2)\, b,
\end{align}
where $\Delta$ is the liquid fill ratio, which is defined as the ratio between the height of liquid's free surface and the height of the tank, $\Lambda$ is the ratio between the tank's width and its height, i.e., $\Lambda=\frac{b}{a}$, $m_l$ is the liquid mass, and $b$ is half of the height of the tank. The coefficients $m_1$ to $m_9$ and $b_1$ to $b_9$ can be determined by fitting the dynamics of this pendulum model to liquid sloshing dynamics simulated by high-fidelity fluid simulation software (such as ANSYS Fluent).

\begin{figure}[h!]
\begin{center}
\begin{picture}(225.0, 125.0)
\put(  0,  -4){\epsfig{file=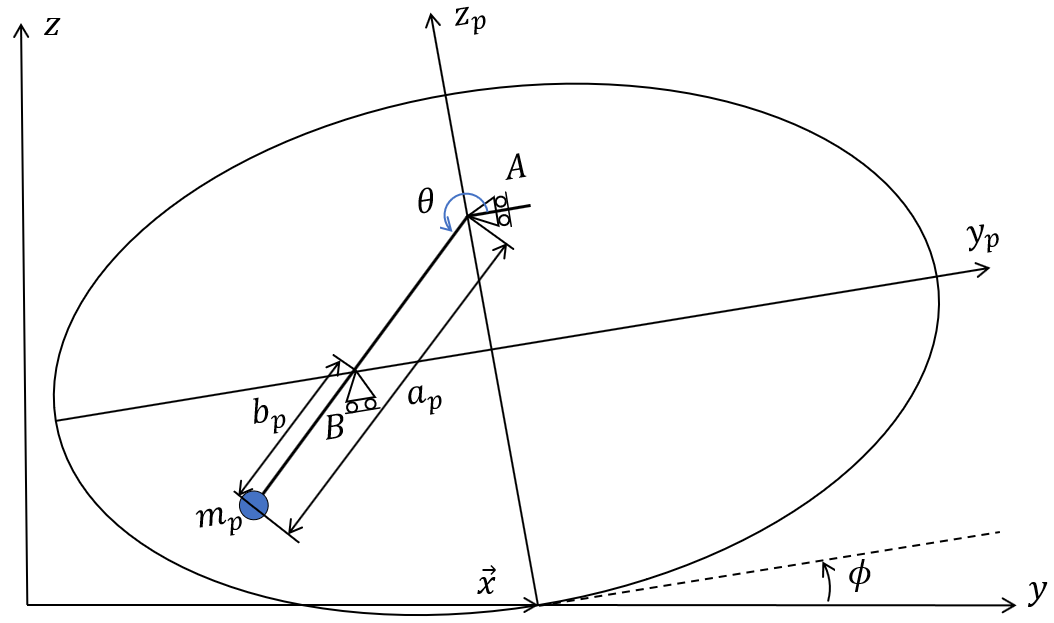,width=0.9\linewidth}}  
\end{picture}
\end{center}
      \caption{Illustration of motion of the trammel pendulum system.}
      \label{fig:trammel_pend_non_inn}
\end{figure}

Now taking into account the motion of the tank when the truck is making a turn (illustrated in Fig.~\ref{fig:trammel_pend_non_inn}), the following equations of motion of the trammel pendulum system in the tank-fixed non-inertia frame can be derived \cite{equiv_model},
\begin{equation}\label{equ:pendulum_2}
\begin{aligned}
    &\ddot{\theta} (a_p^2 \sin^2\theta + b_p^2 \cos^2\theta ) + \ddot{\phi} (a_p b_p + a_p b \sin\theta) \\
    &+ \dot{\phi}^2 \Big[ \frac{1}{2}(a_p^2 - b_p^2)\sin 2\theta - b_p b \cos\theta \Big] + \frac{1}{2}\dot{\theta}^2 (a_p^2 - b_p^2) \sin 2\theta \\
    &- \ddot{x} (a_p\sin\theta \cos\phi + b_p\cos\theta \sin\phi ) \\
    &+ g(b_p\cos\theta \cos\phi - a_p \sin\theta \sin\phi) = 0,
    \end{aligned}
\end{equation}
\begin{equation}
\begin{aligned}
    &\ddot{\phi}\Big[ (b + b_p\sin\theta)^2 + a_p^2\cos^2 \theta\Big] \\
    &+ 2\dot{\theta}\dot{\phi} \Big[ \frac{1}{2}(b_p^2 - a_p^2) \sin 2\theta + b_p b \cos\theta \Big] + \ddot{\theta} (a_p b_p + a_p b \sin\theta) \\
    &- \ddot{x} (b\cos\phi + a_p \cos\theta \sin\phi + b_p \sin\theta \cos\phi) + \dot{\theta}^2 a_p b \cos\theta \\
    &- g \Big[ (b+b_p\sin\theta) \sin\phi - a_p \cos\theta \cos\phi \Big] = 0,
\end{aligned}
\end{equation}
where $\phi$ denotes the roll angle of the tank, which is also the roll angle of the truck, and $x$ represents the translation of the bottom point of the tank in the horizontal direction.

\begin{figure}[h!]
\begin{center}
\begin{picture}(240.0, 430.0)
\put(  12,  210){\epsfig{file=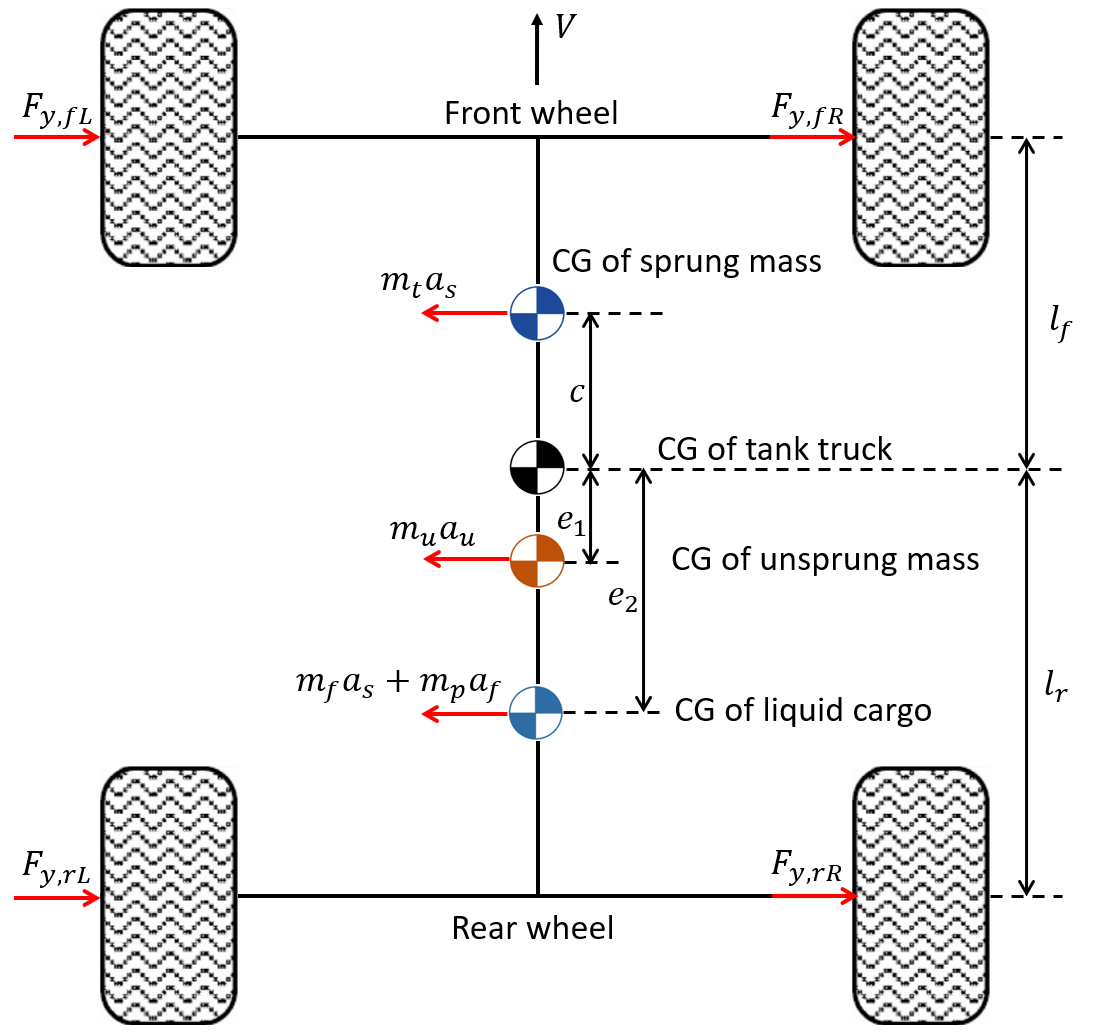,width=0.925\linewidth}}  
\put(  5,  -10){\epsfig{file=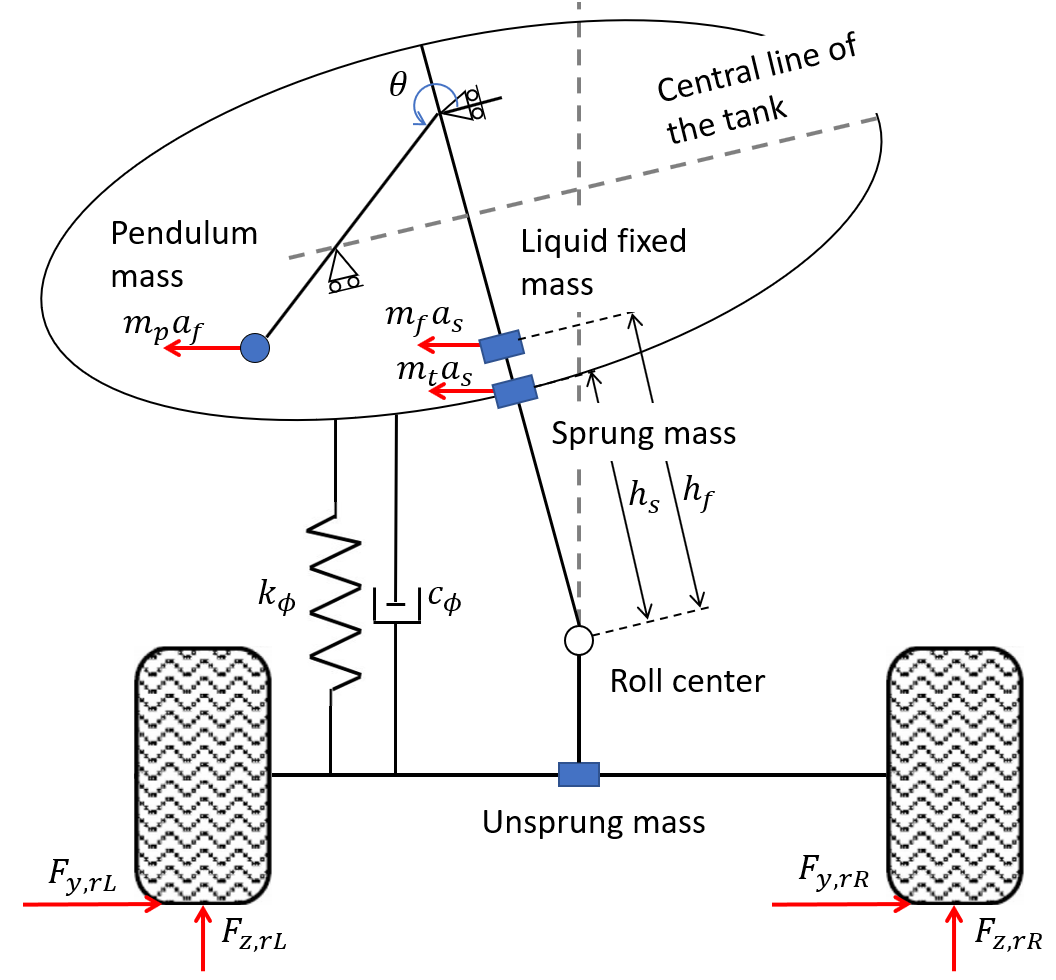,width=0.925\linewidth}}  
\end{picture}
\end{center}
      \caption{Top view and back view of the tank truck dynamic model.}
      \label{fig:top_back_view}
\end{figure}

\subsection{Tank truck dynamic model}

The model representing the tank truck dynamics is illustrated in Fig.~\ref{fig:top_back_view}. Firstly, considering forces acting on the truck in the lateral direction and according to Newton's second law, in the earth-fixed inertia frame we have,
\begin{equation} \label{eq:newton_2nd_law}
    (m_t + m_f) a_s + m_u a_u + m_p a_f = F_{y,f} + F_{y,r},
\end{equation}
where $m_t$ is the sprung mass of the tank truck without load, $m_u$ is the unsprung mass of the truck, $m_f$ and $m_p$ are defined as above. $a_s$, $a_u$, and $a_f$ are the lateral acceleration of the sprung mass, the unsprung mass, and the pendulum mass $m_p$, respectively. $F_{y,f} = F_{y,fL} + F_{y,fR}$ and $F_{y,r} = F_{y,rL} + F_{y,rR}$ are the front tire cornering force and rear tire cornering force, respectively.

The lateral accelerations of the sprung mass and the unsprung mass can be expressed as \cite{equiv_model},
\begin{equation}
    \begin{aligned}
    a_s &= V(\dot{\beta} + r) - h_s \ddot{\phi} + c\dot{r}, \\
    a_u &= V(\dot{\beta} + r) - e_1\dot{r},
    \end{aligned}
\end{equation}
where $V$ is the vehicle's driving speed and is treated as a parameter, $\beta$ is the vehicle's slip angle, $r$ is the vehicle's yaw rate, $h_s$ (shown in Fig.~\ref{fig:top_back_view}) is the distance between the roll center and center of gravity (CG) of the sprung mass, $c$ (shown in Fig.~\ref{fig:top_back_view}) is the longitudinal distance between the CG of the sprung mass and that of the tank truck, and $e_1$ (shown in Fig.~\ref{fig:top_back_view}) is the longitudinal distance between the CG of the unsprung mass and that of the tank truck. The lateral acceleration of the pendulum mass can be expressed as \cite{equiv_model},
\begin{equation}
    a_f = \ddot{x} - \mathcal{D}_1 \ddot{\phi} - \mathcal{D}_2 \ddot{\theta} + 2 \mathcal{D}_3 \dot{\phi}\dot{\theta} + \mathcal{D}_4 \dot{\phi}^2 + \mathcal{D}_5 \dot{\theta}^2,
\end{equation}
where $\ddot{x}$ is defined as in \eqref{equ:pendulum_2} and satisfies $\ddot{x} = V \dot{\beta}$, and
\begin{equation}\label{eq:D1toD5}
    \begin{aligned}
    \mathcal{D}_1 &= b\cos\phi + a_p \cos\theta\sin\phi + b_p \sin\theta \cos\phi, \\
    \mathcal{D}_2 &= a_p\sin\theta \cos\phi + b_p\cos\theta \sin\phi, \\
    \mathcal{D}_3 &= a_p\sin\theta \sin\phi - b_p\cos\theta\cos\phi, \\
    \mathcal{D}_4 &= b\sin\phi - a_p \cos\theta \cos\phi + b_p\sin\theta \sin\phi, \\
    \mathcal{D}_5 &= - a_p \cos\theta \cos\phi + b_p \sin\theta \sin\phi. 
    \end{aligned}
\end{equation}

\begin{figure}[h!]
\begin{center}
\begin{picture}(220.0, 210.0)
\put(  6,  -4){\epsfig{file=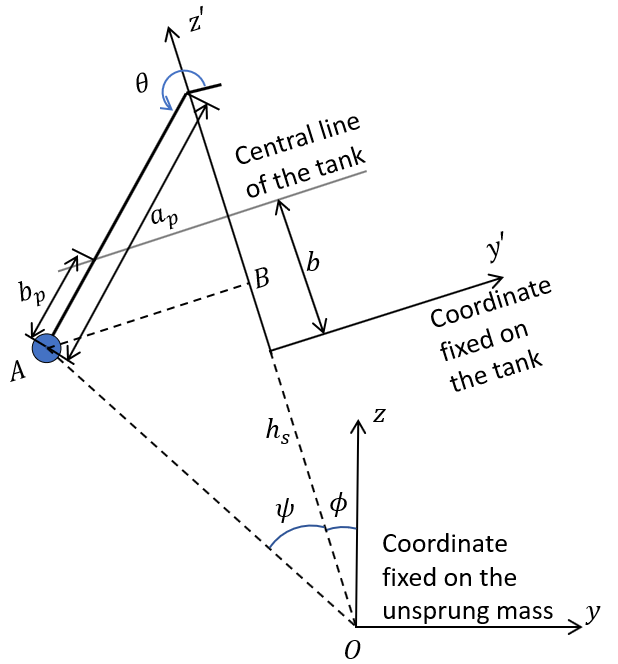,width=0.8\linewidth}}  
\end{picture}
\end{center}
      \caption{Illustration of the rotational dynamics of the trammel pendulum system about the roll center.}
      \label{fig:rot_roll_axis}
\end{figure}

The roll angle of the pendulum mass about the roll center, $\angle$, can be expressed as (see Fig.~\ref{fig:rot_roll_axis}), 
\begin{equation}
    \angle = \phi + \psi,
\end{equation}
where $\phi$ is defined as above and $\psi$ satisfies
\begin{equation}
    \psi = \arctan \frac{\text{AB}}{\text{BO}} = \arctan \frac{-a_p \cos\theta}{h_s + b + b_p\sin\theta}.
\end{equation}
Then, the roll rate and roll angular acceleration in the earth-fixed inertia frame can be expressed as follows \cite{equiv_model},
\begin{equation}\label{equ:angle}
    \dot{\angle} = \dot{\phi} + \dot{\psi}, \quad\quad \ddot{\angle} = \ddot{\phi} + \ddot{\psi},
\end{equation}
where
\begin{equation}
    \dot{\psi} = \frac{\mathcal{E}_2 \dot{\theta}}{\mathcal{E}_1}, \quad\quad \ddot{\psi} = \frac{\mathcal{E}_4\ddot{\theta} - \mathcal{E}_5 \dot{\theta}^2}{\mathcal{E}_3},
\end{equation}
in which 
\begin{equation} \label{eq:AtoB}
    \begin{aligned}
    \mathcal{E}_1 =&\, (h_s + b + b_p\sin\theta)^2 + (a_p \cos\theta)^2, \\
    \mathcal{E}_2 =&\, a_p \Big[ (h_s + b)\sin\theta + b_p \Big], \\
    \mathcal{E}_3 =&\, \mathcal{E}_1^2, \quad \mathcal{E}_4 = \mathcal{E}_1 a_p \Big[(h_s + b)\sin\theta + b_p \Big], \\
    \mathcal{E}_5 =&\, 2a_p \cos\theta \Big[ (h_s + b) \sin\theta + b_p \Big] \\
    & \times \Big[b_p (h_s + b + b_p \sin\theta) - a_p^2 \sin\theta \Big] \\
    & - \mathcal{E}_1 a_p (h_s + b) \cos\theta.
    \end{aligned}
\end{equation}

Furthermore, according to Newton's second law for rotation and using the expressions in \eqref{equ:angle}, the vehicle's roll moment balance and yaw moment balance can be expressed as follows \cite{equiv_model},
\begin{equation} \label{eq:moment_balance_1}
\begin{aligned}
    & I_z \dot{r} - I_{xz} \ddot{\phi} - m_f a_s e_2 - m_p a_f e_2 - I_{xzp} (\ddot{\phi} + \ddot{\psi}) + I_{zzp} \dot{r} \\
    & - I_{xyp} (\dot{\phi} + \dot{\psi})^2 - I_{yzp}r (\dot{\phi} + \dot{\psi}) = F_{y,f} l_f - F_{y,r} l_r, 
\end{aligned}
\end{equation}
\begin{equation} \label{eq:moment_balance_2}
\begin{aligned}
    & I_x \ddot{\phi} - I_{xz} \dot{r} + m_t h_s V(\dot{\beta} + r) + m_f h_f a_s + (h_s + b) m_p a_f \\
    & + I_{xxp} (\ddot{\phi} + \ddot{\psi}) - I_{xzp} \dot{r} + I_{xyp} r (\dot{\phi} + \dot{\psi}) + I_{yzp} r^2 \\
    &= -k_\phi \phi - c_\phi \dot{\phi} + \phi \Big(m_t h_s g + m_f h_f g + (h_s +b) m_p g \Big) \\
    & - m_p g a_p \cos\theta,
\end{aligned}
\end{equation}
where $l_f$ (resp. $l_r$) is the longitudinal distance between the CG of the tank truck and the front wheel axle (resp. the rear wheel axle), $e_2$ (shown in Fig.~\ref{fig:top_back_view}) is the longitudinal distance between the CG of the tank truck and that of the liquid tank, $h_f$ (shown in Fig.~\ref{fig:top_back_view}) is the vertical distance between the CG of the liquid fixed mass $m_f$ and the roll center, $k_\phi$ is the suspension roll stiffness, and $c_\phi$ is the suspension roll damping. Moreover, $I_{xxp}$ and $I_{zzp}$ are the moments of inertia of the pendulum mass $m_p$ about the $x$-axis and $z$-axis, respectively; $I_{xyp}$, $I_{xzp}$, and $I_{yzp}$ are the products of inertia of the pendulum mass about the $x$- and $y$-axes, the $x$- and $z$-axes, and the $y$- and $z$-axes, respectively; and $I_x$, $I_z$, and $I_{xz}$ are determined according to the parallel axis theorem of the moment of inertia as follows \cite{equiv_model},
\begin{equation}
    \begin{aligned}
    I_x &= I_{xxs} + I_{xxf} + m_t h_s^2 \\
    I_z &= I_{zzs} + I_{zzu} + I_{zzf} + m_t c^2 + m_u e_1^2, \\
    I_{xz} &= I_{xzs} + I_{xzf} + m_t h_s c,
    \end{aligned}
\end{equation}
where $I_{xxs}$ and $I_{zzs}$ (resp. $I_{xxf}$ and $I_{zzf}$) are the moments of inertia of the sprung mass $m_t$ (resp. the liquid fixed mass $m_f$) about the $x$-axis and $z$-axis, respectively; $I_{zzu}$ is the moment of inertia of the unsprung mass $m_u$ about the $z$-axis; and $I_{xzs}$ and $I_{xzf}$ are the product of inertia of the sprung mass and that of the liquid fixed mass about the $x$- and $z$-axes.

The tire cornering forces $F_{y,f}$ and $F_{y,r}$ are determined according to the magic formula \cite{rajamani2011vehicle},
\begin{equation}
    F_{y,i} = \mathcal{D} \sin \bigg ( \mathcal{C} \arctan \Big( \mathcal{B} \alpha_i - \mathcal{E} \big( \mathcal{B}\alpha_i - \arctan \mathcal{B}\alpha_i \big) \Big) \bigg),
\end{equation}
with $i \in \{f,r\}$, where $\mathcal{B}$, $\mathcal{C}$, $\mathcal{D}$, and $\mathcal{E}$ are constant parameters to be fitted, and $\alpha_f$ and $\alpha_r$ are the front and rear tire sideslip angles, computed as
\begin{equation}\label{equ:sideslip}
    \begin{aligned}
    &\alpha_f = \delta_f - \arctan \bigg( \frac{V \beta + r l_f}{V} \bigg), \\
    &\alpha_r = \delta_r - \arctan \bigg( \frac{V \beta - r l_r}{V} \bigg), 
    \end{aligned}
\end{equation}
where $\delta_f$ is the front wheel steering angle, and $\delta_r$ is the rear wheel steering angle and is assumed to be $0$ in this paper (corresponding to a truck with only front wheel steering).

To sum up, equations \eqref{equ:pendulum_2}-\eqref{equ:sideslip} define a $6$-order system with the vehicle's roll angle $\phi$, roll rate $\dot{\phi}$, slip angle $\beta$, yaw rate $r$, the pendulum angle $\theta$, and angular velocity $\dot{\theta}$ as its states and with the front wheel steering angle $\delta_f$ as its scalar input.

We choose to use the load transfer ratio (LTR) to represent rollover constraints \cite{Bencatel_rollover}, which is defined as:
\begin{equation}
    \text{LTR} := \frac{F_{z,R} - F_{z,L}}{mg},
\end{equation}
where $F_{z,R}$ and $F_{z,L}$ are the total vertical forces on the right-side tires and on the left-side tires, respectively, and $mg = (m_t + m_u + m_l)g$ is the total weight of the vehicle. The LTR measures how much of the vehicle vertical load is concentrated on one side of the vehicle. In particular, the absolute value of LTR being greater than $1$ implies that the tires on one side of the vehicle may have been off the ground. Following \cite{LTR_cal}, the LTR is estimated as,
\begin{equation}
    \text{LTR} \approx - \frac{2}{mg W} (k_\phi \phi + c_\phi \dot{\phi}),
\end{equation}
where $W$ is the width of the vehicle.


%% file: sections/05_results.tex
\section{Results} \label{sec:results}

In this section, we present the results of the simulation study. Firstly, the trammel pendulum model is verified against the FLUENT simulation results in \cite{slosh_model}. Then, we present and compare the simulation results of the tank truck dynamic model under no load, solid load, and liquid load scenarios. Finally, we apply learning reference governor (LRG) to protect the tank truck from rollover accidents, and the results during learning process and after learning with different rollover tests are reported.

\input{sections/05_01_model_validation.tex}

\input{sections/05_02_model_simulation.tex}

\input{sections/05_03_learning_results.tex}

\input{sections/05_04_multispeed_extension}

%% file: sections/05_01_model_validation.tex
\subsection{Validation of modeling liquid sloshing using the trammel pendulum} \label{sec:vali_res}

The fuel sloshing dynamics are represented by an equivalent trammel pendulum model. The parameter values of the trammel pendulum model for generating our simulation results are taken from \cite{slosh_model}, which correspond to a typical fuel tank in the market.

\begin{figure}[H]
\begin{center}
\begin{picture}(240.0, 90.0)
\put(  -6,  -8){\epsfig{file=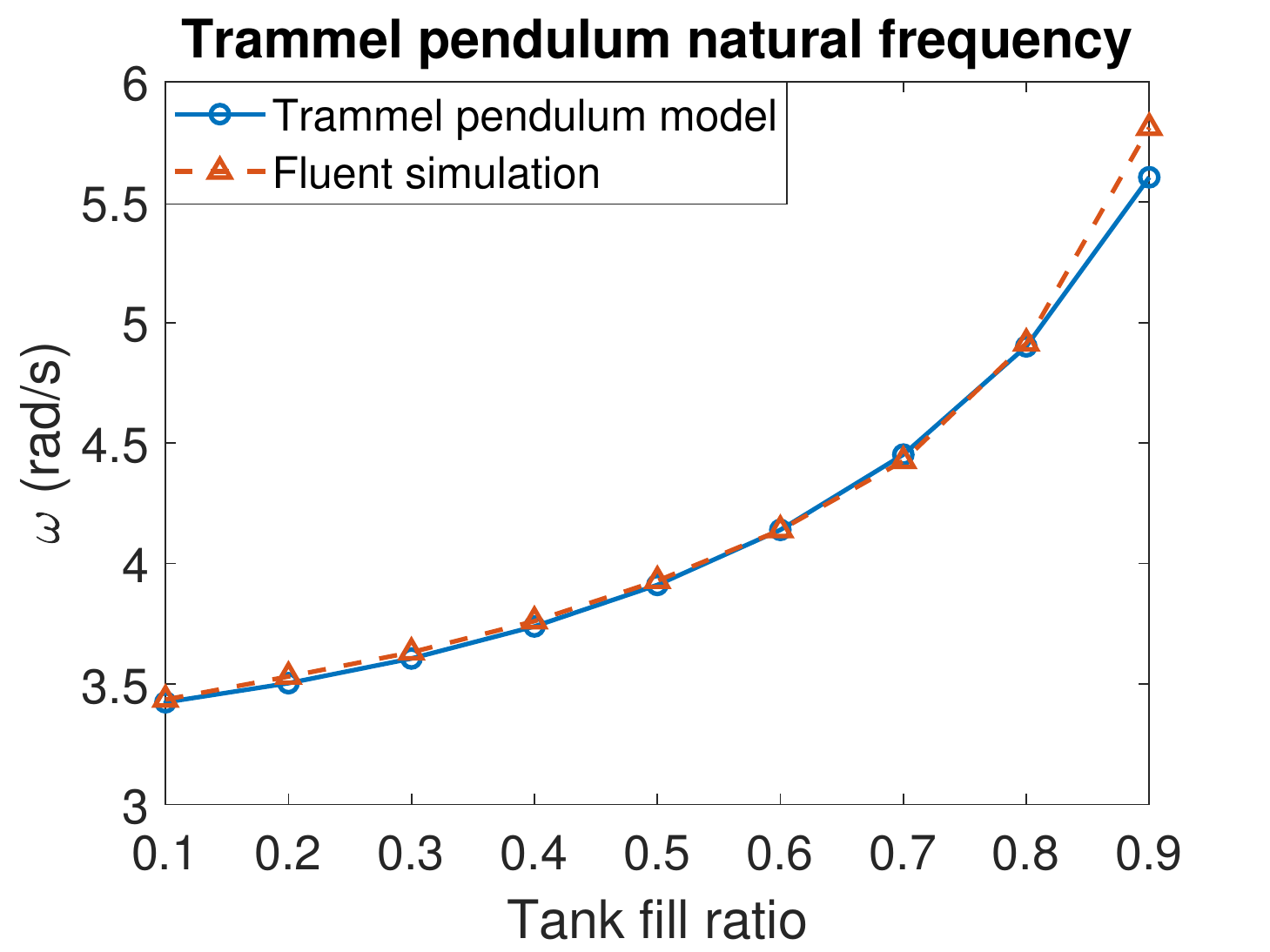,height=.195\textwidth}}  
\put(  121,  -8){\epsfig{file=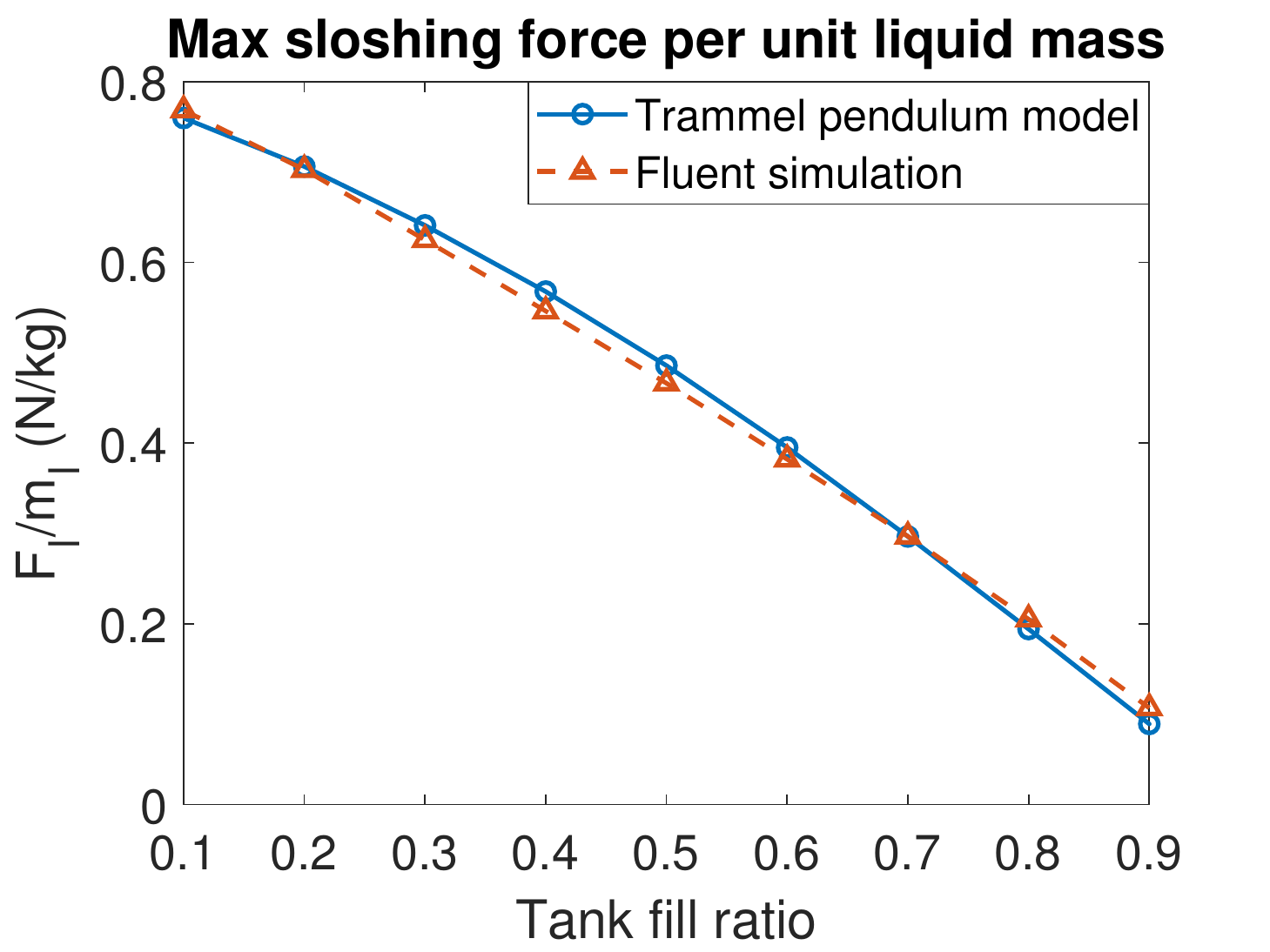,height=.195\textwidth}}  
\put( -5, 86.5){\small{(a)}}
\put( 120, 86.5){\small{(b)}}
\end{picture}
\end{center}
      \caption{(a) Natural frequency of the trammel pendulum and (b) forces exerted by the pendulum on the tank as functions of tank fill ratio compared with FLUENT simulation results presented in \cite{slosh_model}.}
      \label{fig:sloshing_validation}
\end{figure}

To verify the fuel sloshing model, we focus on the natural frequency of the fuel sloshing dynamics and the force exerted by the fuel on the tank. The natural frequency of the trammel pendulum is shown in Fig.~\ref{fig:sloshing_validation} (a). Higher tank fill ratio results in higher frequency of the trammel pendulum, and this reasonably accurately matches the FLUENT simulation results in \cite{slosh_model}. Next, we consider the maximum force exerted on the tank due to fuel sloshing. In the trammel pendulum model, this force is calculated as \cite{slosh_model},

\begin{equation}
    F_l = \max_t \; m_p a_p (-\dot{\theta}\sin\theta + \ddot{\theta}\cos\theta )
\end{equation}

Shown in Fig.~\ref{fig:sloshing_validation} (b) is the maximum force exerted by the trammel pendulum on the tank per unit liquid mass as we vary the tank fill ratio. This also matches the FLUENT simulation results in \cite{slosh_model}.

%% file: sections/05_02_model_simulation.tex
\subsection{Simulation results of tank truck dynamics} \label{res:res_sim}

A series of simulations has been performed based on the tank truck dynamic model described in Section~\ref{sec:truck_dyn} under different load conditions. Some parameters used in the simulations are shown in Table~\ref{tab:truck_sim_params}.

\begin{table}[h!]
\centering
\begin{tabular}{ |c||c| } 
 \hline
 Parameters & Values \\ 
 \hline
 \hline
 $V$ & $25$ m/s  \\ 
 $m_u$ & $300$ kg \\
 $h_s$ & $0.8580$ m \\
 $l_f$ & $1.160$ m \\
 $l_r$ & $1.750$ m \\
 $I_{xxs}$ & $1280$ kg/m$^2$ \\
 $I_{zzs}$ & $2800$ kg/m$^2$ \\
 $I_{xzs}$ & 0 \\
 $c_\phi$ & $7471$  N$\cdot$m$\cdot$s/rad \\
 $k_\phi$ & $95707$ N$\cdot$m \\
 \hline
\end{tabular}
\caption{Tank truck simulation parameters.}
\label{tab:truck_sim_params}
\end{table}

The first scenario we considered is the truck running without any load, i.e., $m_t = 1700$ (kg), $m_l = 0$ (kg). We apply constant steering inputs of 0.02 (rad) and 0.05 (rad) to the system (these steering angles correspond to the actual steering of the front wheels).

\begin{figure}[h!]
\begin{center}
\begin{picture}(245.0, 190.0)
\put(  -6,  -8){\epsfig{file=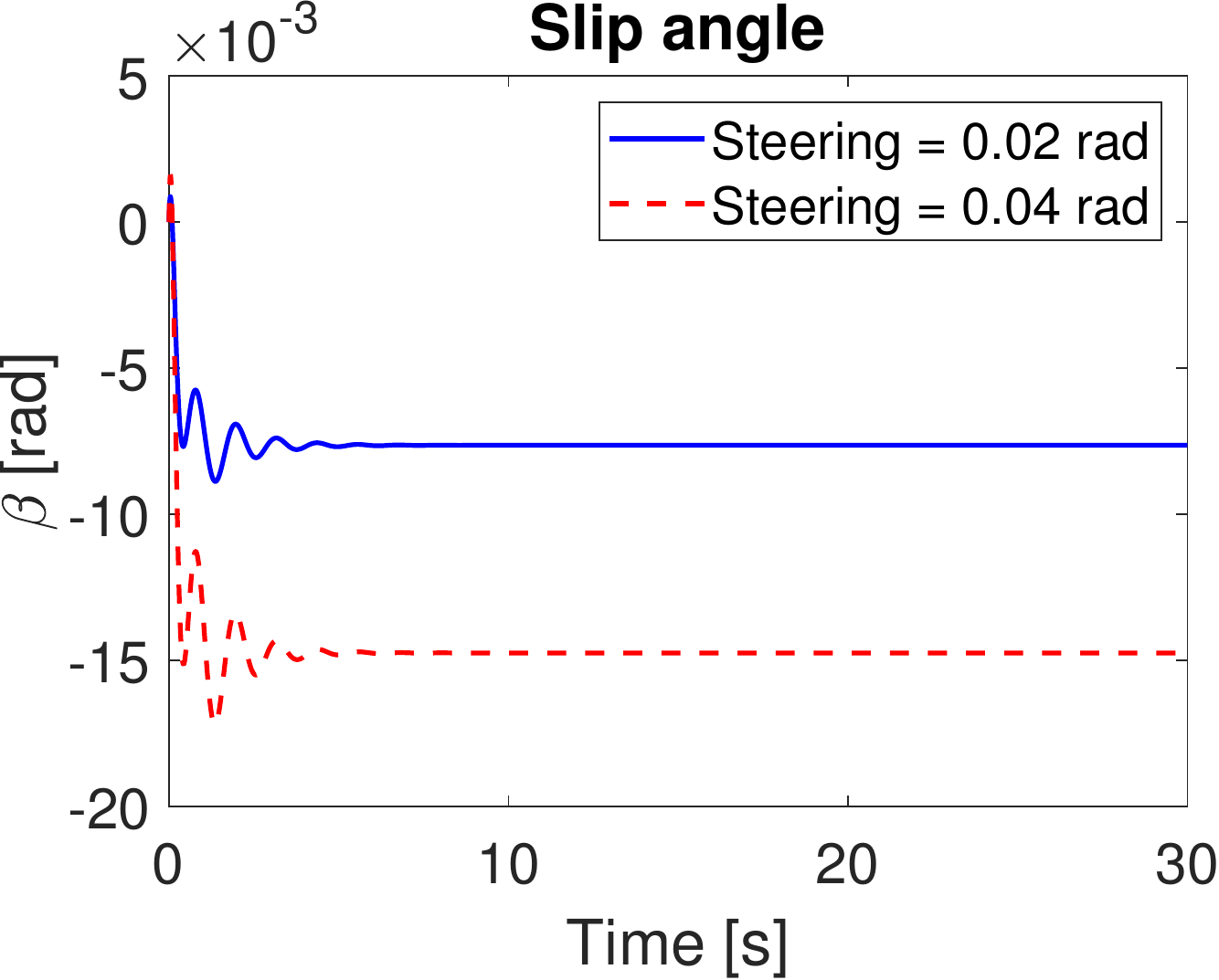,height=.195\textwidth}}  
\put(  125,  -8){\epsfig{file=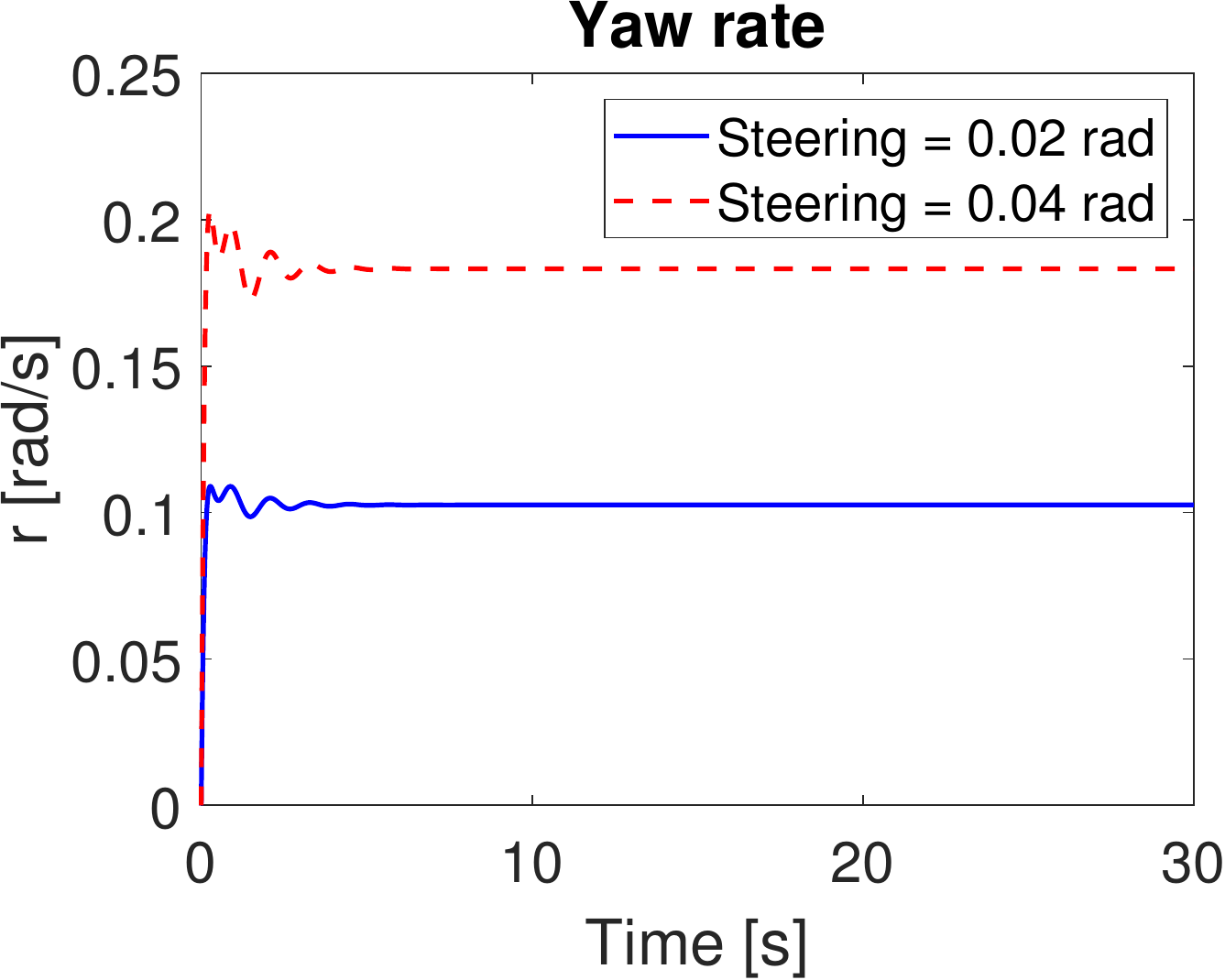,height=.195\textwidth}}  
\put(  -6,  95){\epsfig{file=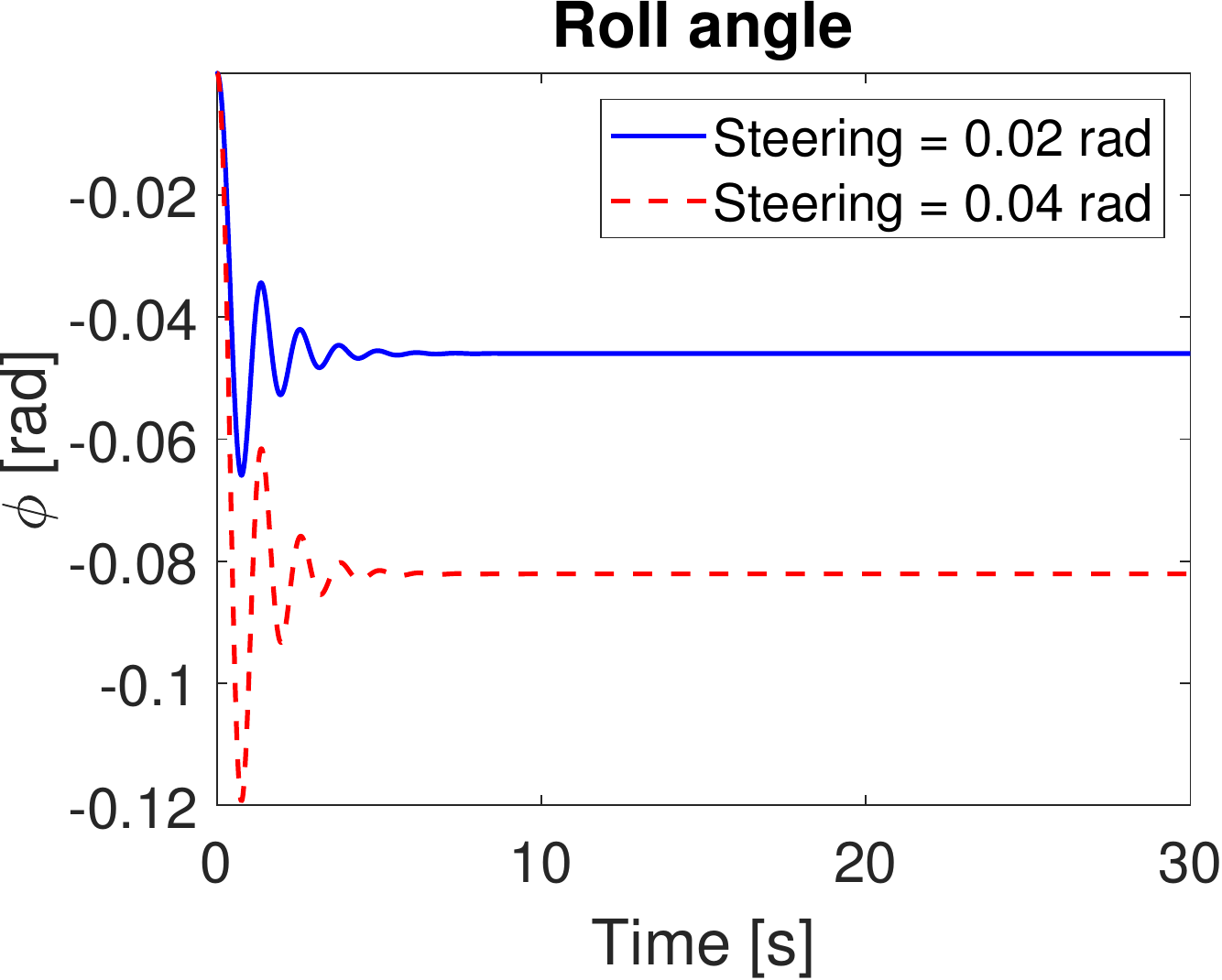,height=.195\textwidth}}  
\put(  125,  95){\epsfig{file=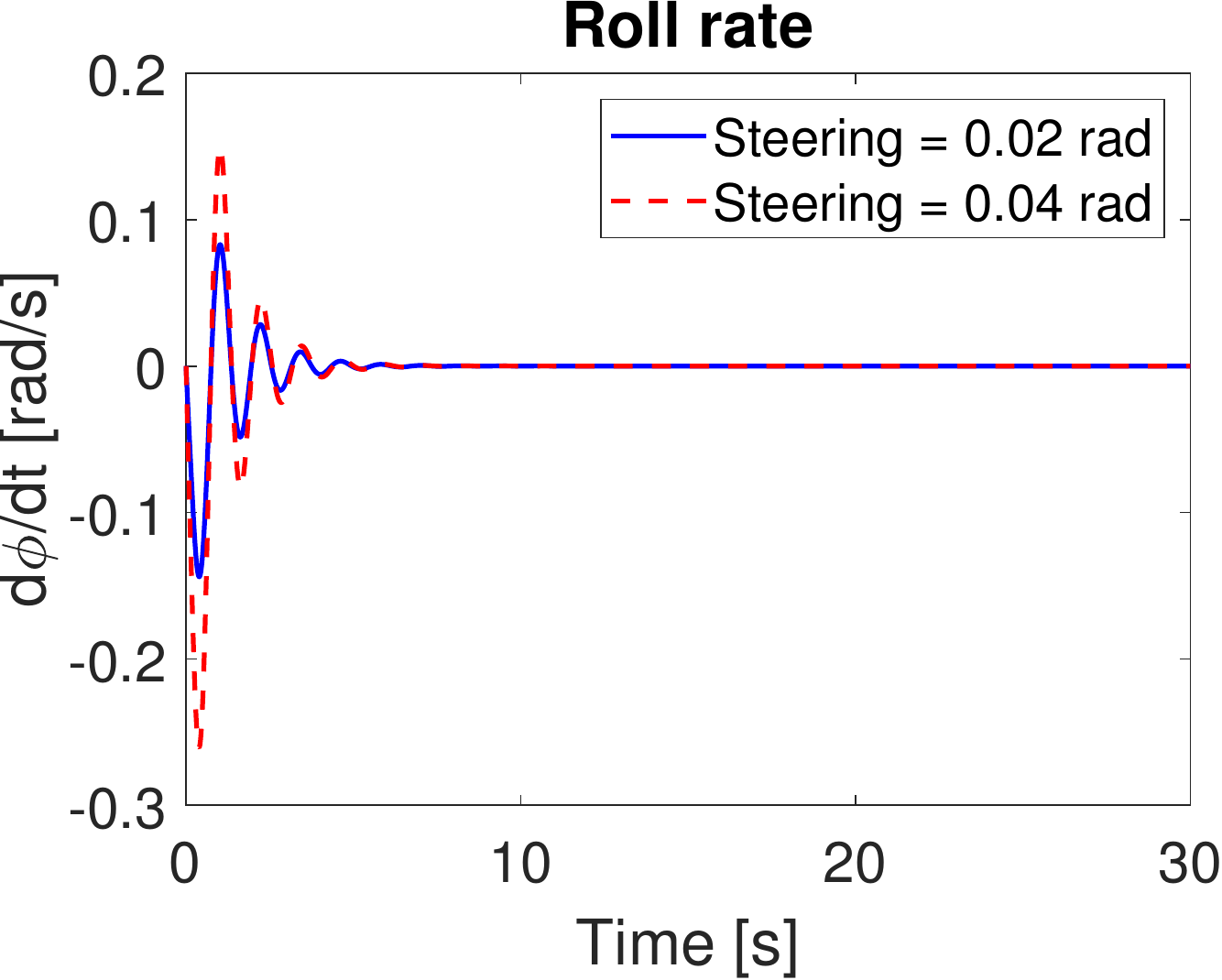,height=.195\textwidth}}  

\end{picture}
\end{center}
      \caption{The roll angle, roll rate, slip angle, and yaw rate responses for step commands in the no load scenario ($m_t = 1700$ (kg) and $m_l = 0$ (kg)).}
      \label{fig:tank_truck_dyn_scen_1}
\end{figure}

From Fig.~\ref{fig:tank_truck_dyn_scen_1}, we can observe convergence of states to an equilibrium under different steering wheel angles. Note also that larger steering wheel angle results in larger roll angle, slip angle and yaw rate at steady state. 

\begin{figure}[h!]
\begin{center}
\begin{picture}(245.0, 190.0)
\put(  -6,  -8){\epsfig{file=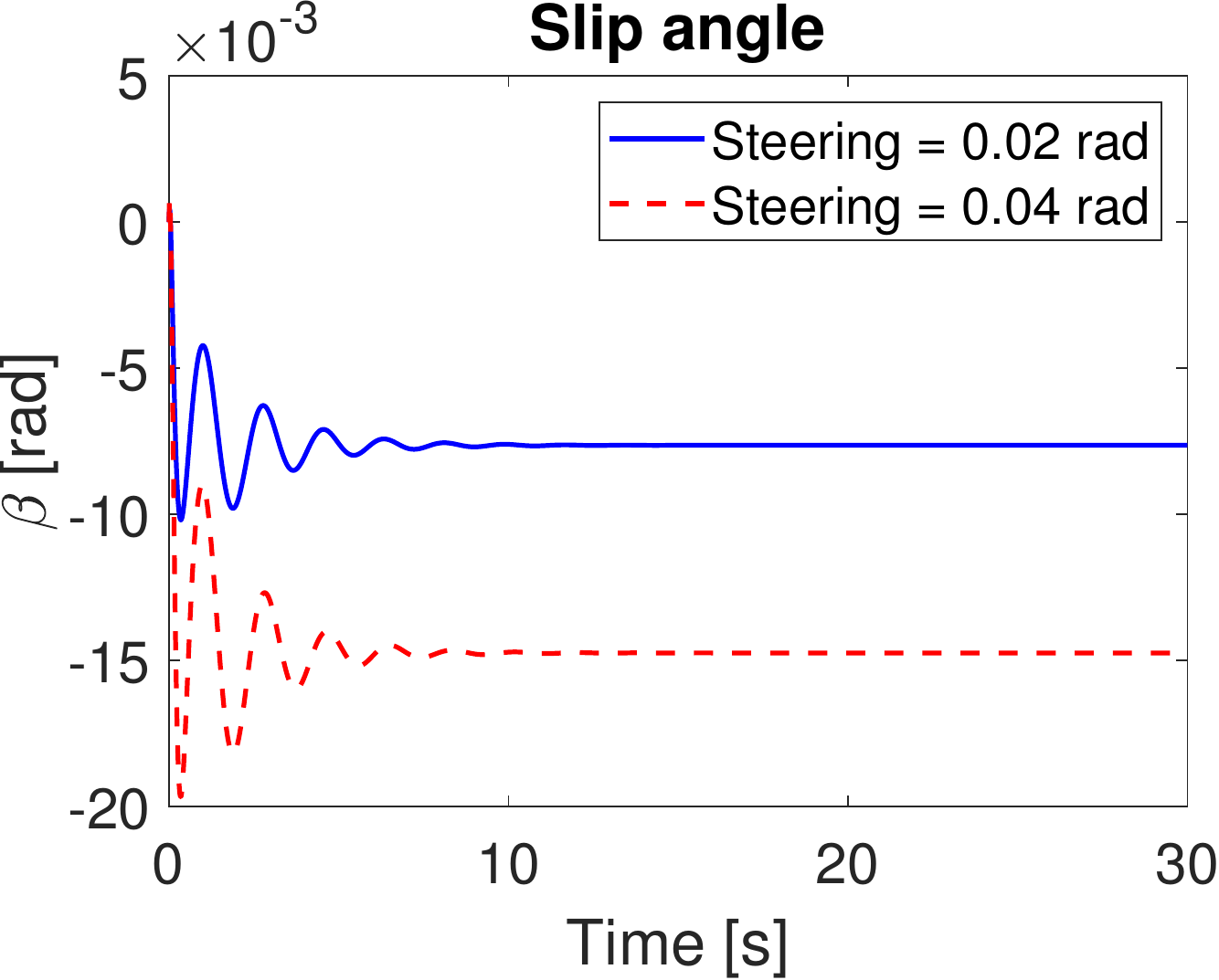,height=.195\textwidth}}  
\put(  125,  -8){\epsfig{file=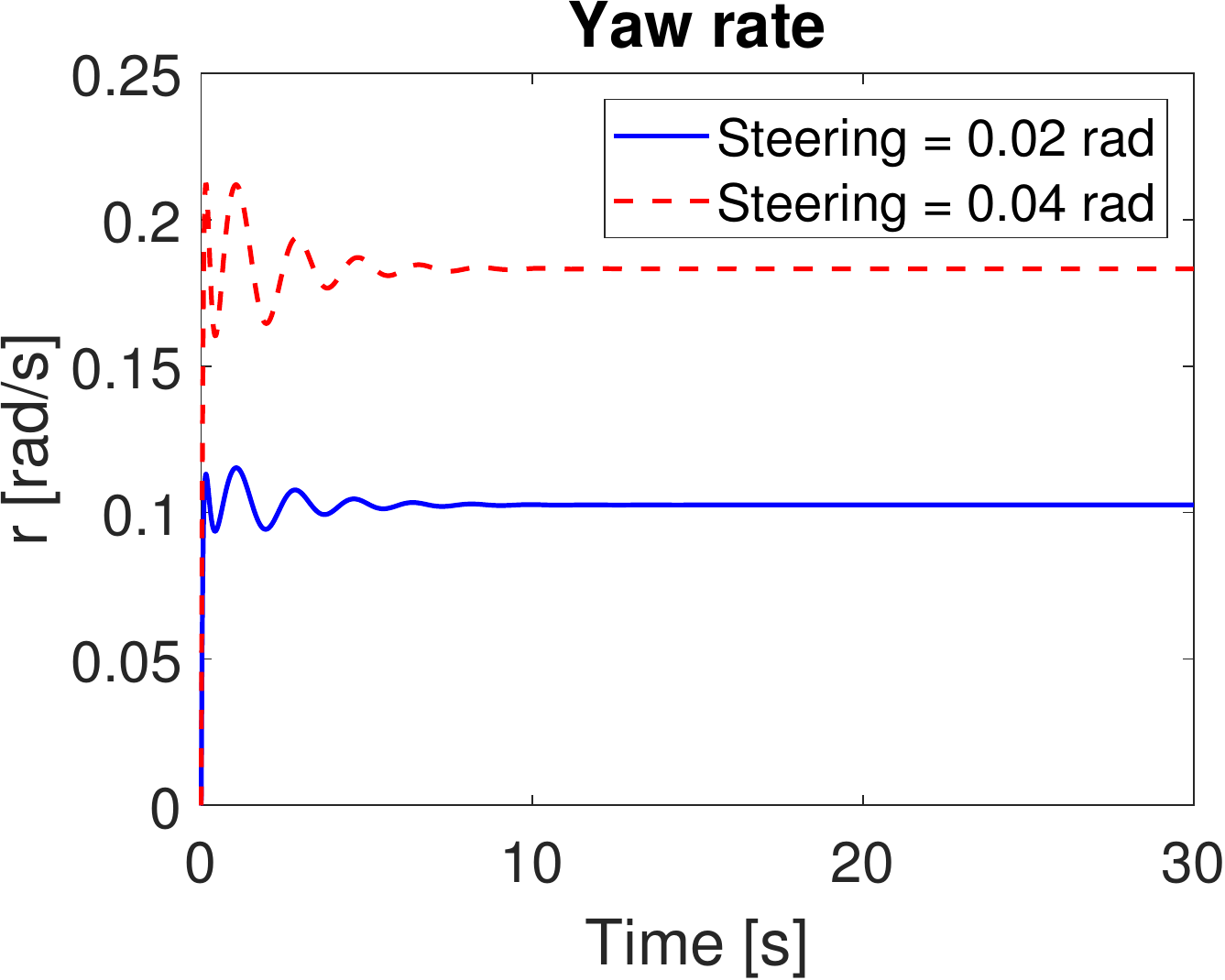,height=.195\textwidth}}  
\put(  -6,  95){\epsfig{file=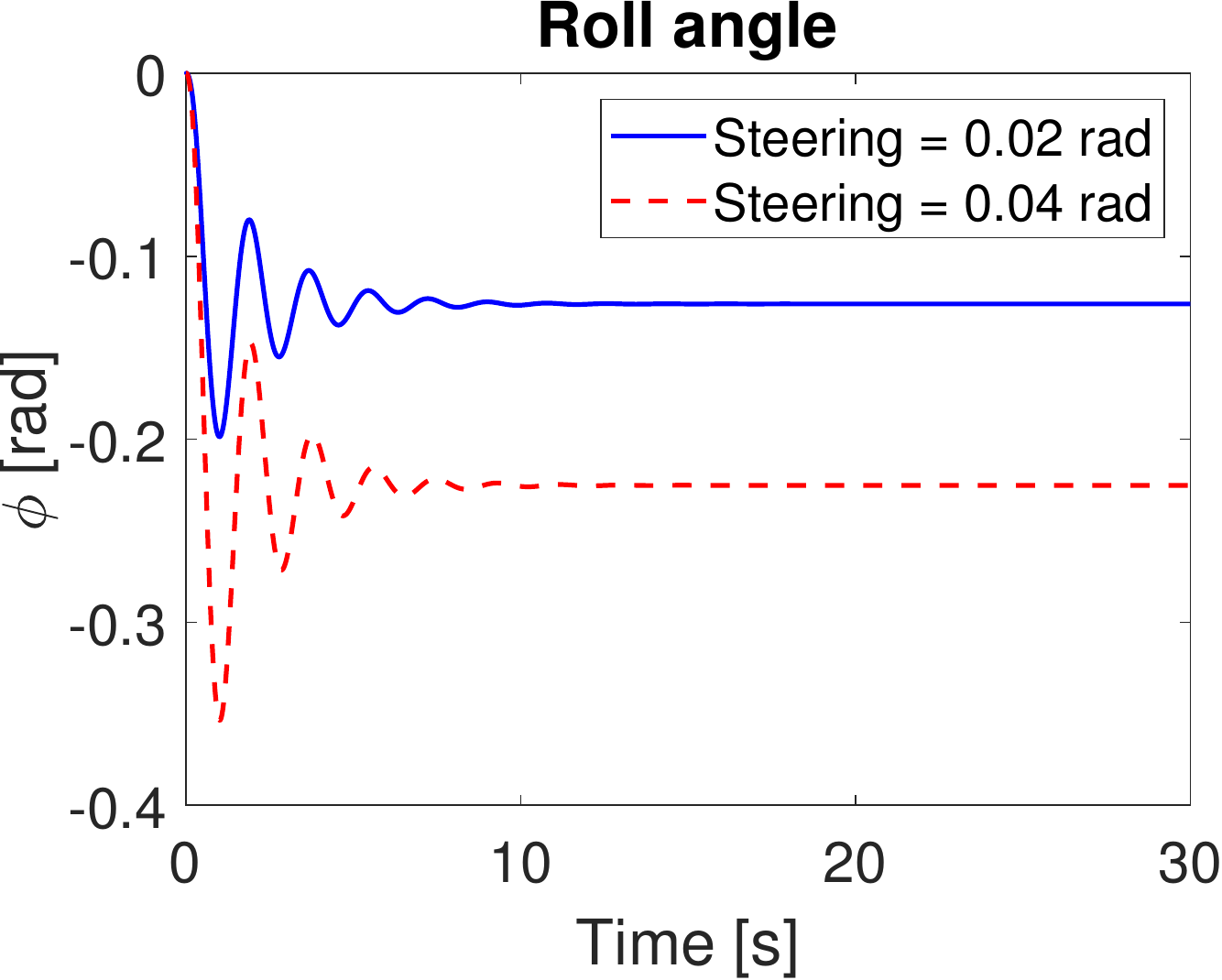,height=.195\textwidth}}  
\put(  125,  95){\epsfig{file=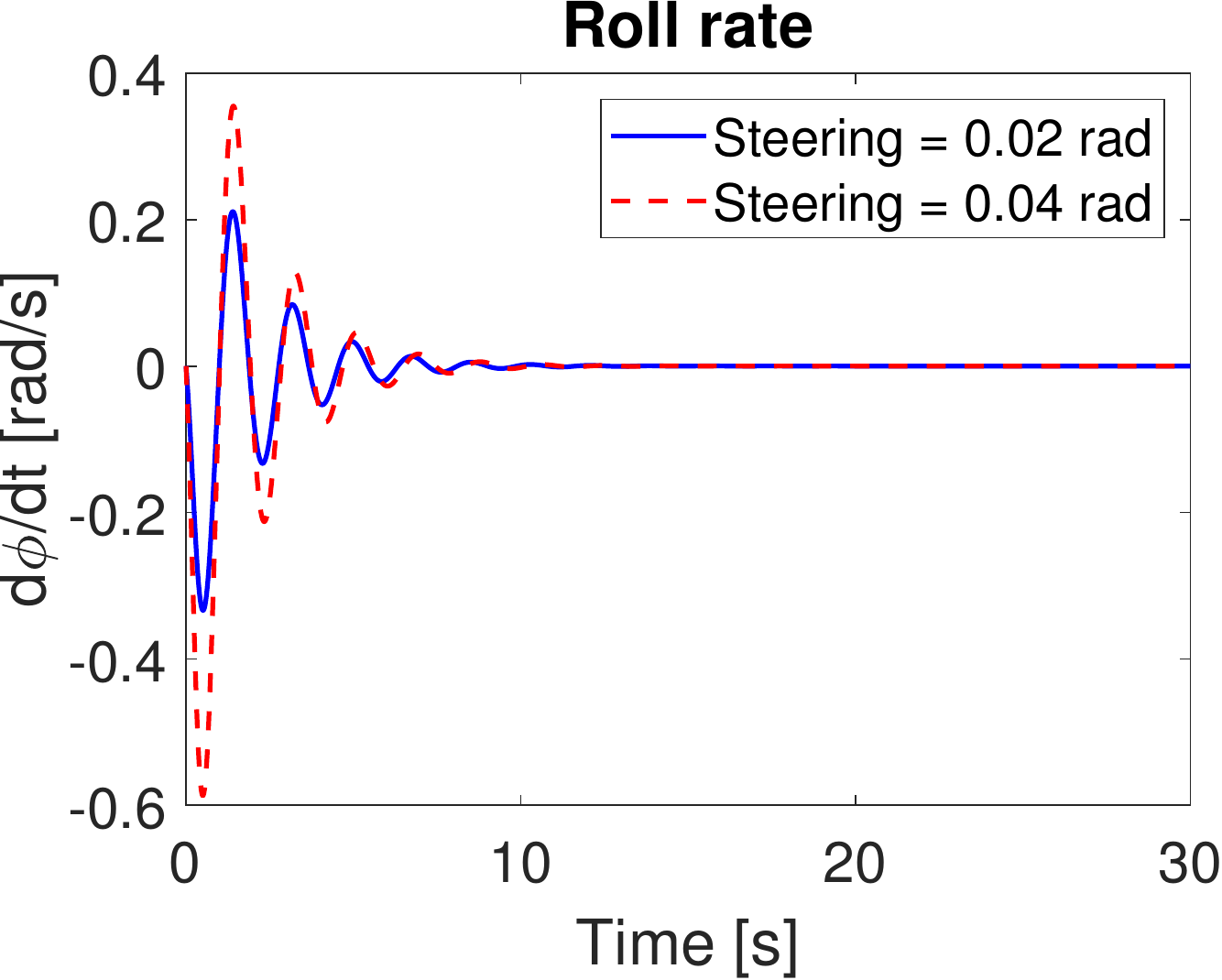,height=.195\textwidth}}  

\end{picture}
\end{center}
      \caption{The roll angle, roll rate, slip angle, and yaw rate responses for step commands in the solid load scenario ($m_t = 3700$ (kg) and $m_l = 0$ (kg)).}
      \label{fig:tank_truck_dyn_scen_2}
\end{figure}

The second scenario we consider is the case where there is $2000$ (kg) solid load only, and the load weight is added to the sprung mass, i.e., $m_t = 3700$ (kg), $m_l = 0$ (kg). Constant steering commands are applied and the results are shown in Fig.~\ref{fig:tank_truck_dyn_scen_2}.

Comparing Fig.~\ref{fig:tank_truck_dyn_scen_1} and Fig.~\ref{fig:tank_truck_dyn_scen_2}, the solid load case exhibits larger overshoot in roll rate and higher propensity for rolling over. Note that in both the no load and the solid load cases, the response of the pendulum angle $\theta$ and the pendulum angular velocity $\dot{\theta}$ are not shown, because in both cases the pendulum mass $m_p$ is 0 (kg).

\begin{figure}[h!]
\begin{center}
\begin{picture}(245.0, 290.0)
\put(  -6,  -8){\epsfig{file=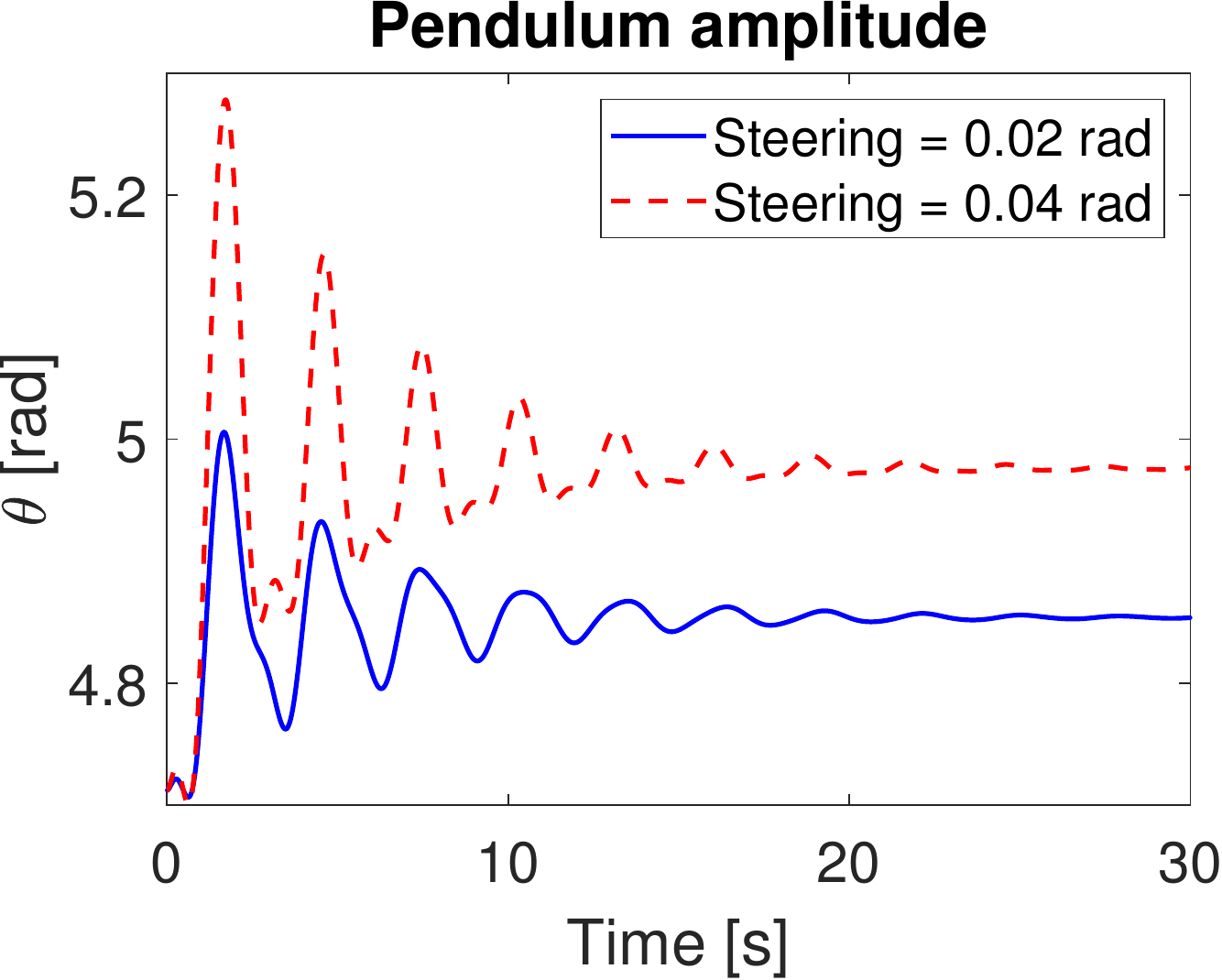,height=.195\textwidth}}  
\put(  125,  -8){\epsfig{file=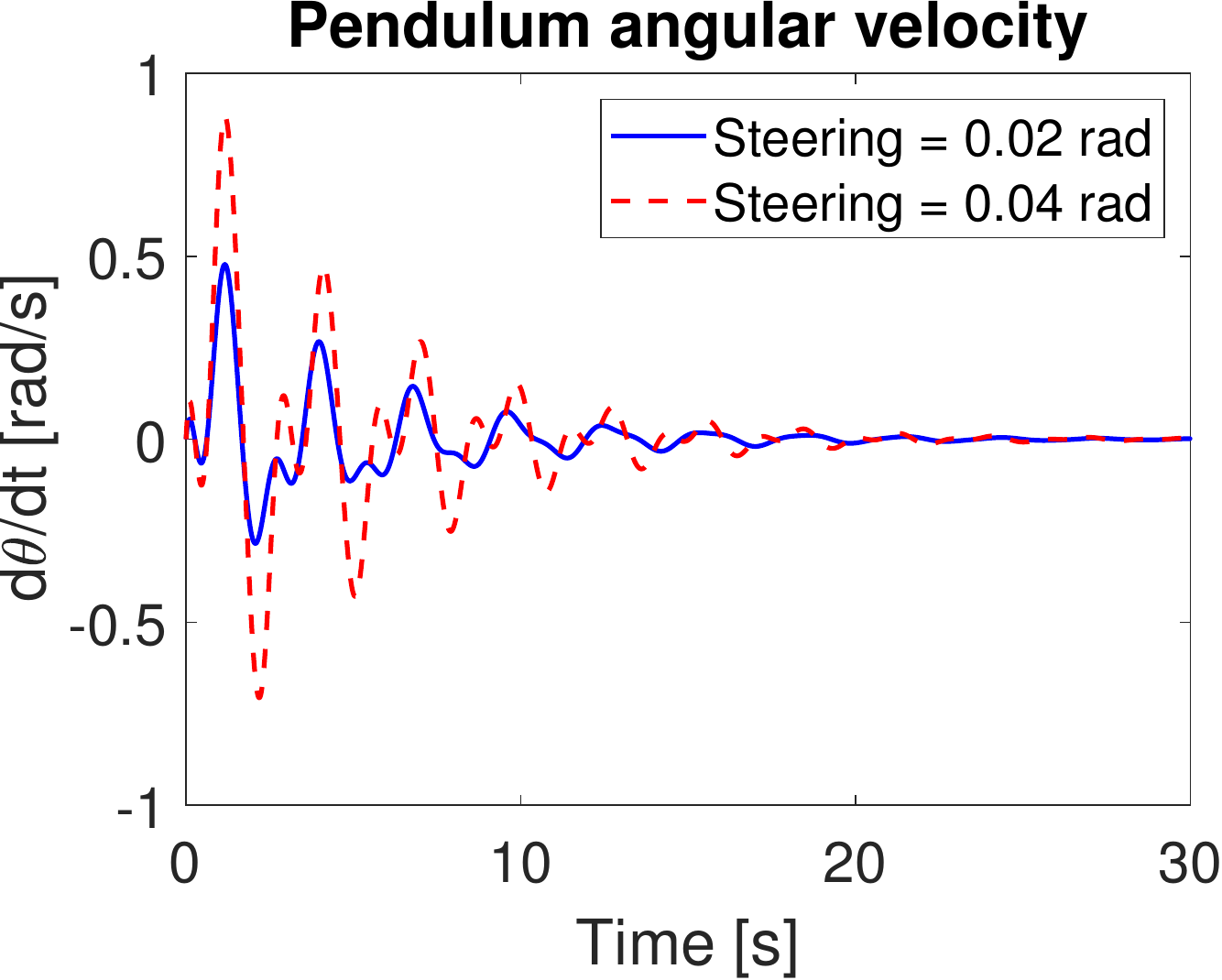,height=.195\textwidth}}  
\put(  -6,  95){\epsfig{file=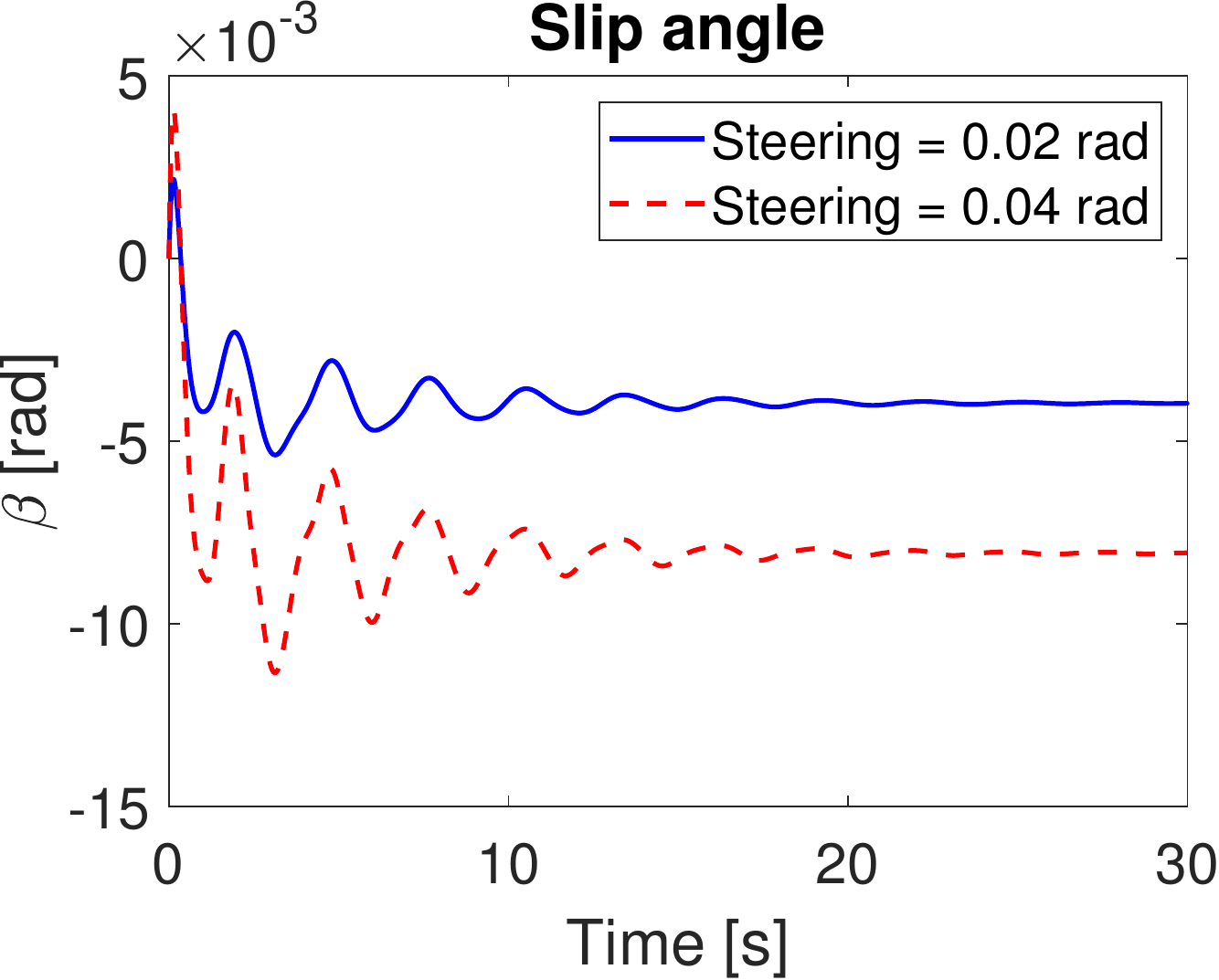,height=.195\textwidth}}  
\put(  125,  95){\epsfig{file=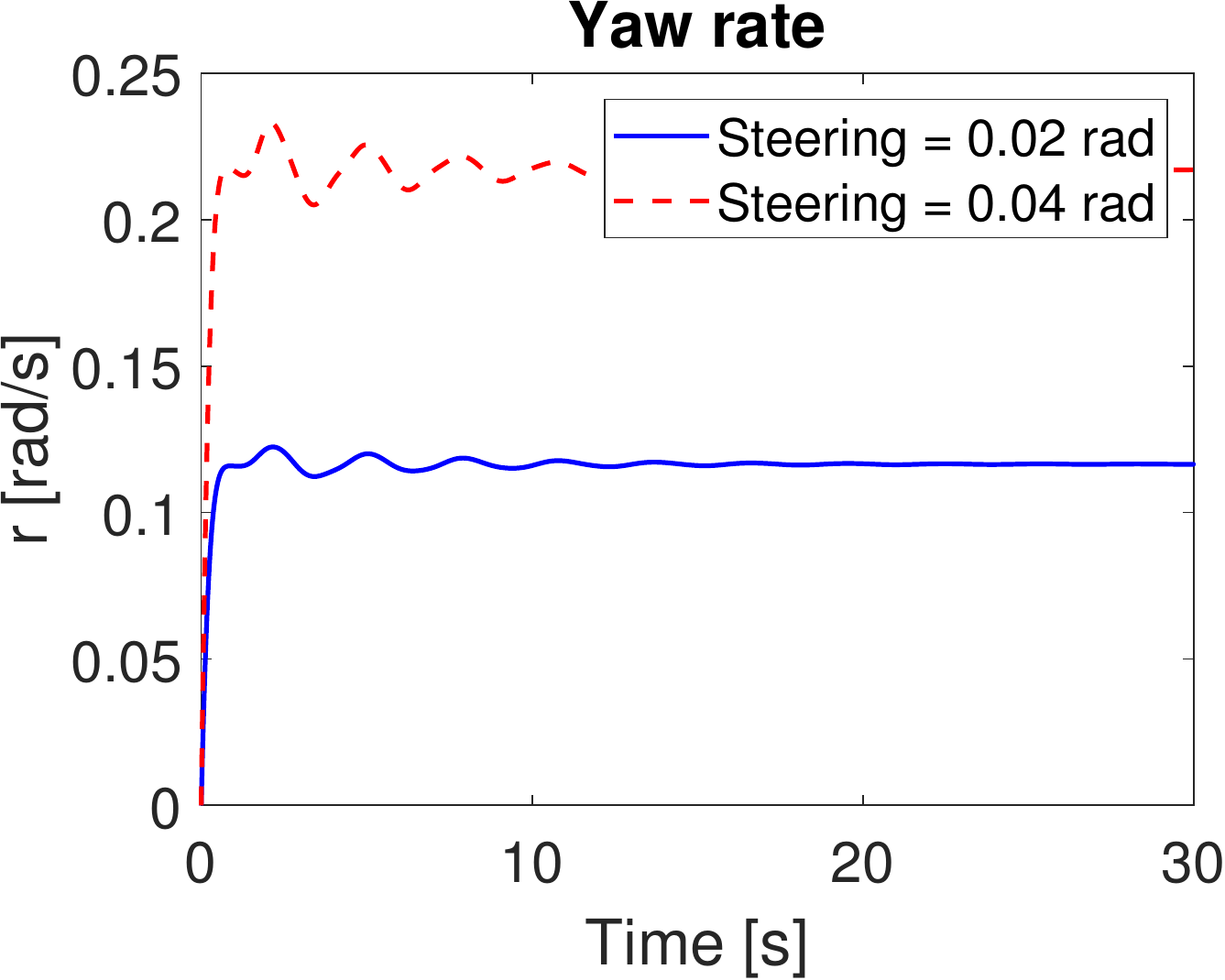,height=.195\textwidth}}  
\put(  -6,  197){\epsfig{file=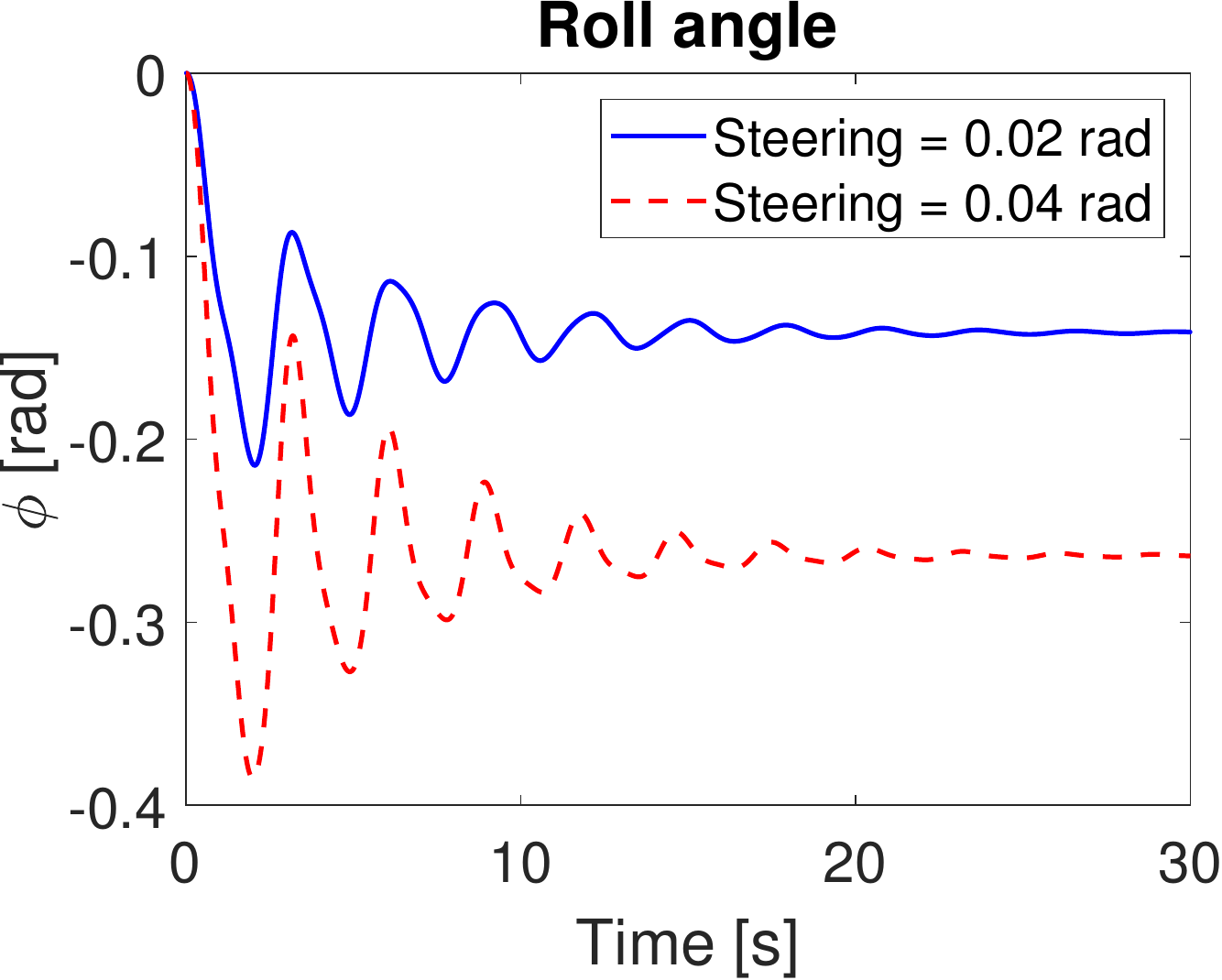,height=.195\textwidth}}  
\put(  125,  197){\epsfig{file=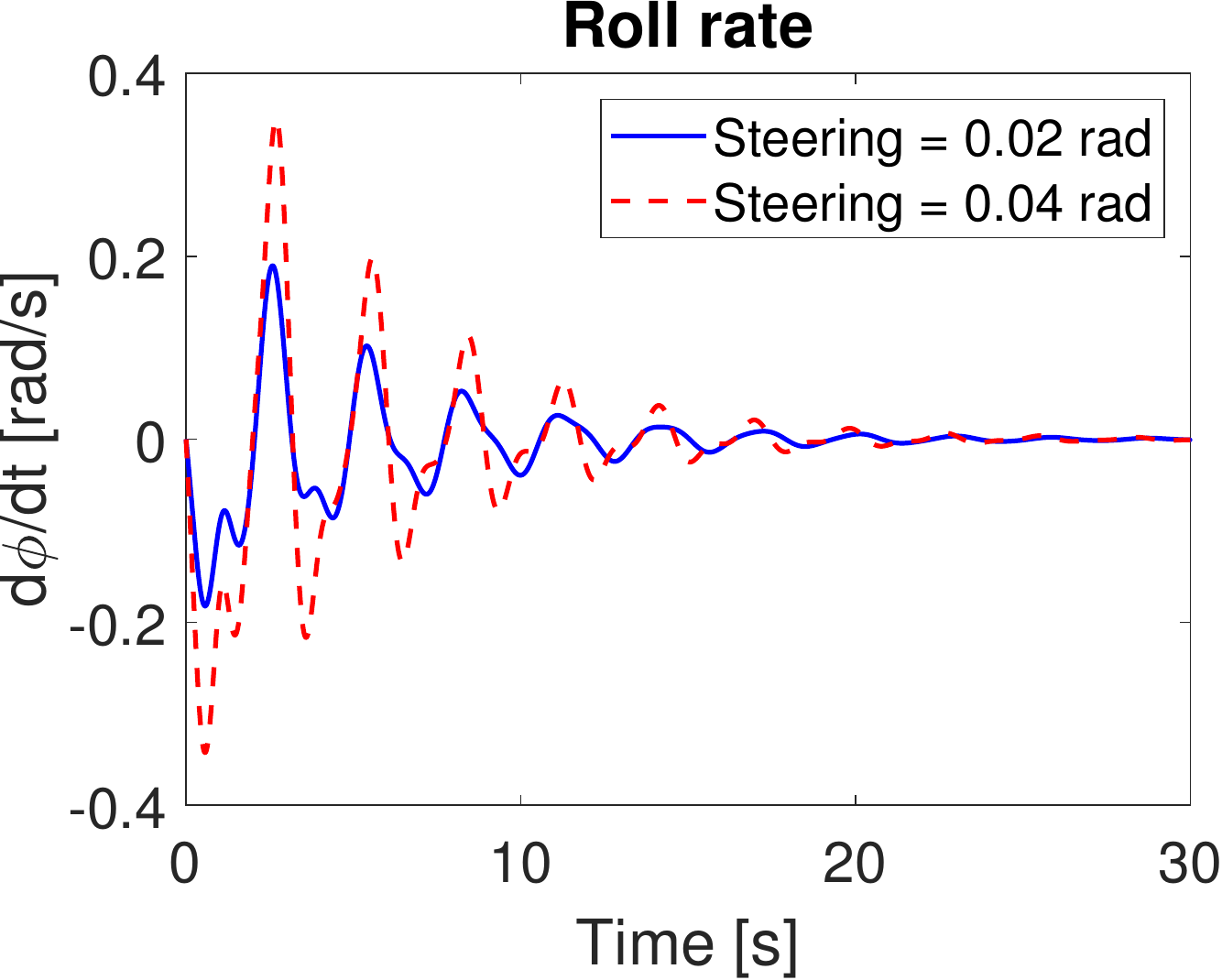,height=.195\textwidth}}  

\end{picture}
\end{center}
      \caption{The roll angle, roll rate, slip angle, yaw rate, pendulum angle, and pendulum angular velocity responses for step commands in the liquid load scenario ($m_t = 1700$ (kg) and $m_l = 2000$ (kg)).}
      \label{fig:tank_truck_dyn_scen_3}
\end{figure}

Next, we consider another scenario where the truck carries a circular liquid tank with radius 1 (m), and the tank is 50\% filled. The mass of the liquid is $m_l = 2000$ (kg). The response is shown in Fig.~\ref{fig:tank_truck_dyn_scen_3}. Due to the effect of liquid fuel sloshing, it takes longer for states to converge to the equilibrium.

\begin{figure}[h!]
\begin{center}
\begin{picture}(245.0, 190.0)
\put(  58,  -8){\epsfig{file=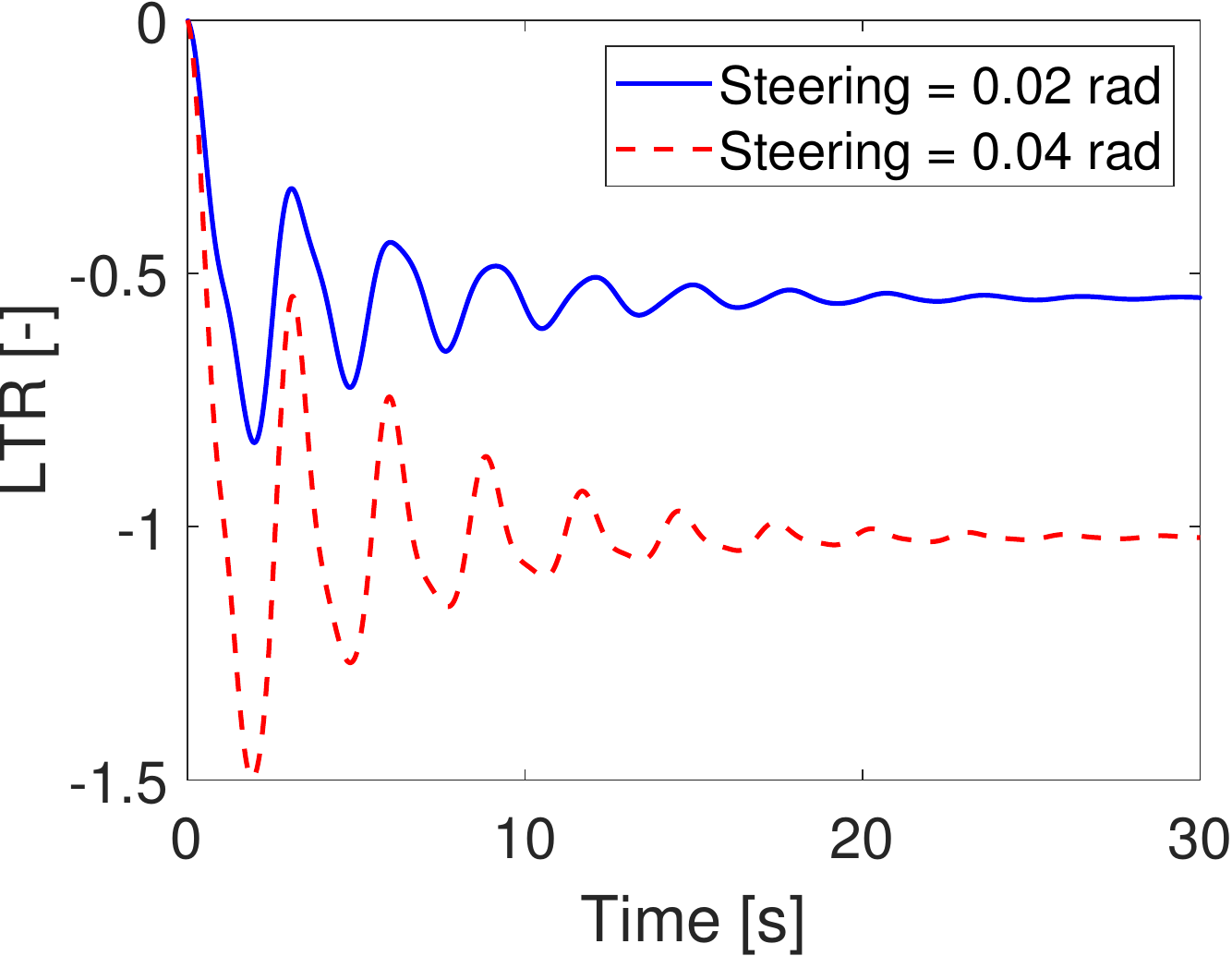,height=.195\textwidth}}  
\put(  -6,  95){\epsfig{file=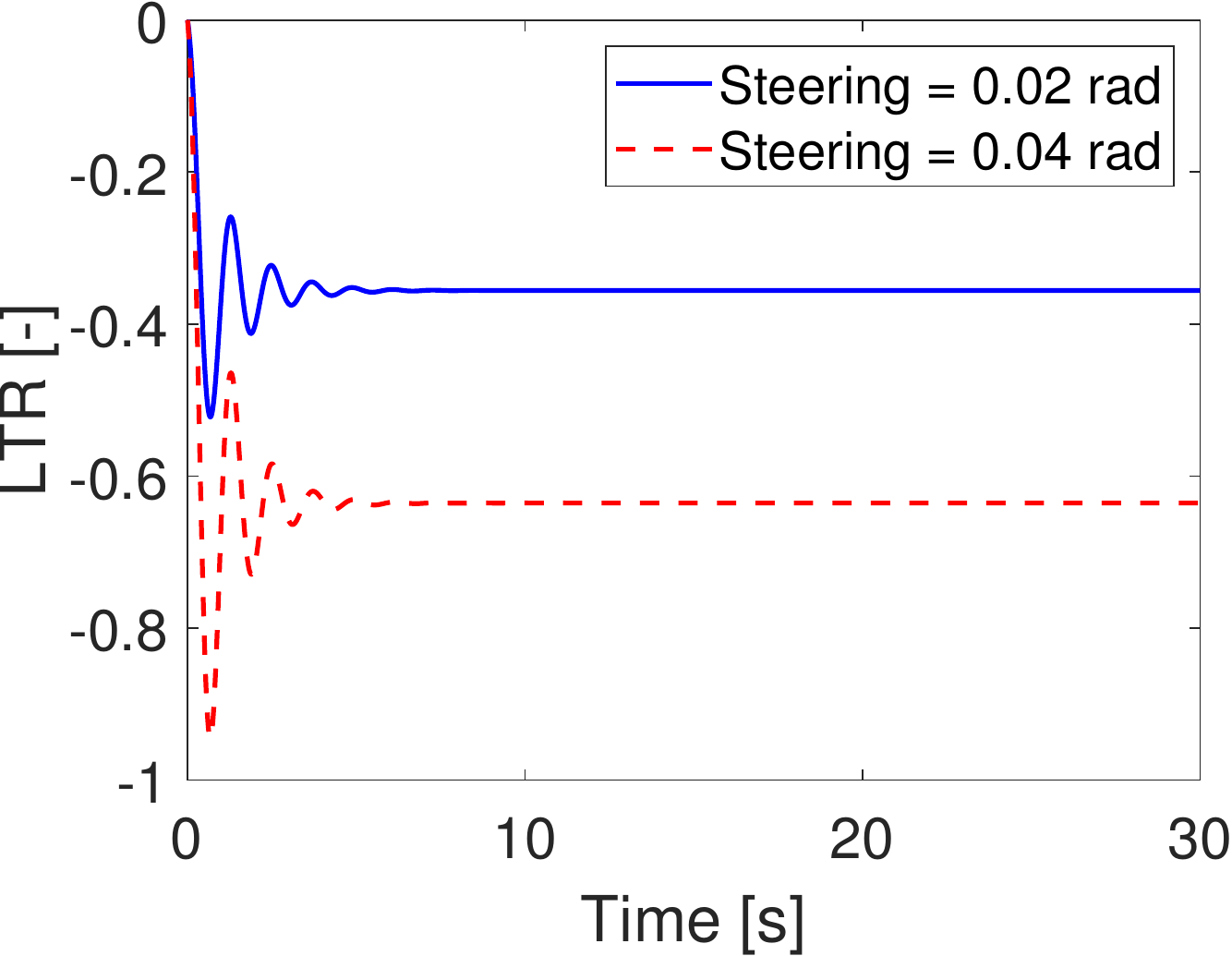,height=.195\textwidth}}  
\put(  125,  95){\epsfig{file=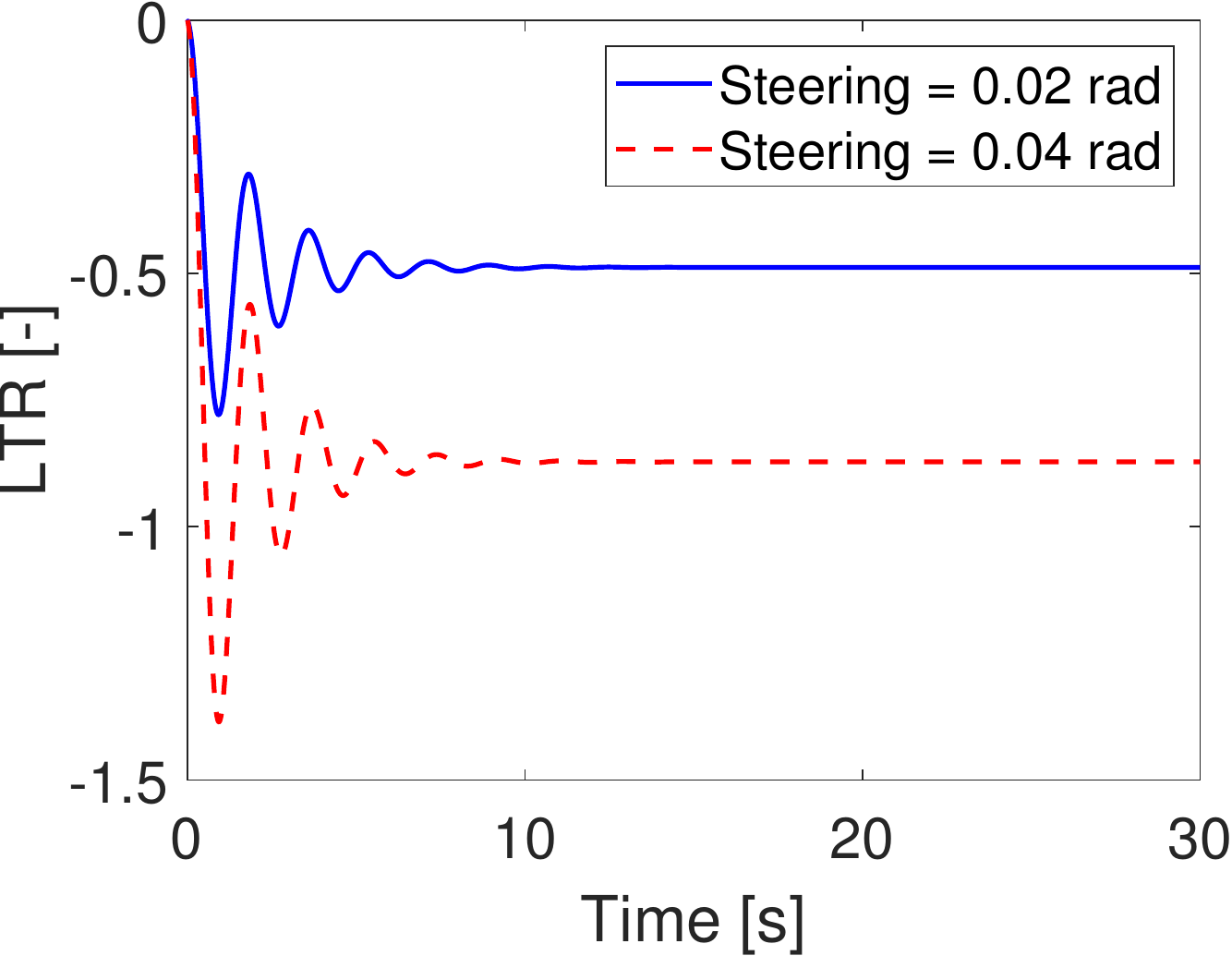,height=.195\textwidth}}  
\end{picture}
\end{center}
      \caption{The LTR responses of the no load scenario (top left), solid load scenario (top right), and the liquid load scenario (bottom).}
      \label{fig:tank_truck_dyn_ltr}
\end{figure}

The resulting load transfer ratio (LTR) responses of all three scenarios are shown in Fig.~\ref{fig:tank_truck_dyn_ltr}, which indicates that the vehicle with the liquid load has higher propensity to rolling over under the same steering command.

%% file: sections/05_03_learning_results.tex
\subsection{Applying learning reference governor (LRG) to the tank truck} \label{sec:res_learning}

The liquid load scenario is chosen to demonstrate the effectiveness of our learning algorithm described in Section~\ref{sec:learning_rg}. The tank is a circular tank ($a = b = 1$ (m)); the sprung mass is $m_t = 1700$ (kg); the liquid load is $m_l = 2000$ (kg); and the tank fill ratio is $\Delta = 0.5$. Other parameters are the same as listed in Table~\ref{tab:truck_sim_params}.

{\color{black}
To implement LRG, the H\"older constants $L$ and $\beta$ in \eqref{eq:lipschitz} are required. With $\beta > 1$, the estimated $\overline{D}$ by the LRG in \eqref{equ:upper_D} can be sensitive to small changes in $(\inputref, \delta\inputref, \delta\state)$ due to the nature of the exponent $1/\beta$. This sensitivity to small changes can lead to longer learning time for LRG to achieve its maximal aggressiveness. As a result, in the application of tanker truck, we assume $\beta = 1$ and estimate $L$ in \eqref{eq:lipschitz}, which corresponds to estimating the Lipschitz constant of $D$. The estimation of a Lipschitz constant is a common problem in optimization. For example, \cite{strongin2013global} and \cite{pinter2013global} use sampled or ordered evaluation points to construct an under-estimate of a Lipschitz constant, and a Lipschitz constant is obtained by multiplying the under-estimate value by a factor greater than 1. References \cite{de1981estimation} and \cite{wood1996estimation} use order statistics to estimate a Lipschitz constant for univariate functions. Reference \cite{nugroho2020nonlinear} proposes an interval-based global optimization algorithm to numerically estimate a Lipschitz constant for arbitrary nonlinear functions. In this paper, points in the space of $(\inputref, \delta\inputref, \delta\state)$ are sampled and the corresponding derivatives of $D$ are numerically estimated. Specifically, $80$ points were sampled and from them, an estimate of $L$ is inferred, which gives $L \geq 0.28$. We therefore consider choices of $L = 0.3$ and $L = 0.5$ to illustrate that a conservative estimate of $L$ is sufficient for our learning algorithm. 
}


Note that in the simulations, a constant ratio $k_{\delta_f}$ between the steering wheel angle and the forward tires steering angles is assumed. That is, the control input is the steering wheel angle $SW$ and
\begin{equation}
    \delta_f = k_{\delta_f} SW
\end{equation}
where $k_{\delta_f} = 1/20$.

Based on the LTR, the rollover constraints are imposed as
\begin{equation} \label{eq::vehicle_rollover_constraint_def}
    -\text{LTR}_{\lim} \le \text{LTR} \le \text{LTR}_{\lim}.
\end{equation}
Note that $|\text{LTR}| > 1$ corresponds to wheels lifting off, so we set $\text{LTR}_{\lim} = 1$ in this example.

During the learning phase, the commands $\command_n$ can be randomly generated. However, to better visualize how the learning algorithm gradually pushes the system to its mobility limits, in this example we use a repeated profile for $\command_n$, switching between $\pm 50$ (deg) with a duration of 20 (sec).

\begin{figure}[h!]
\begin{center}
\begin{picture}(240.0, 100.0)
\put(  -6,  -8){\epsfig{file=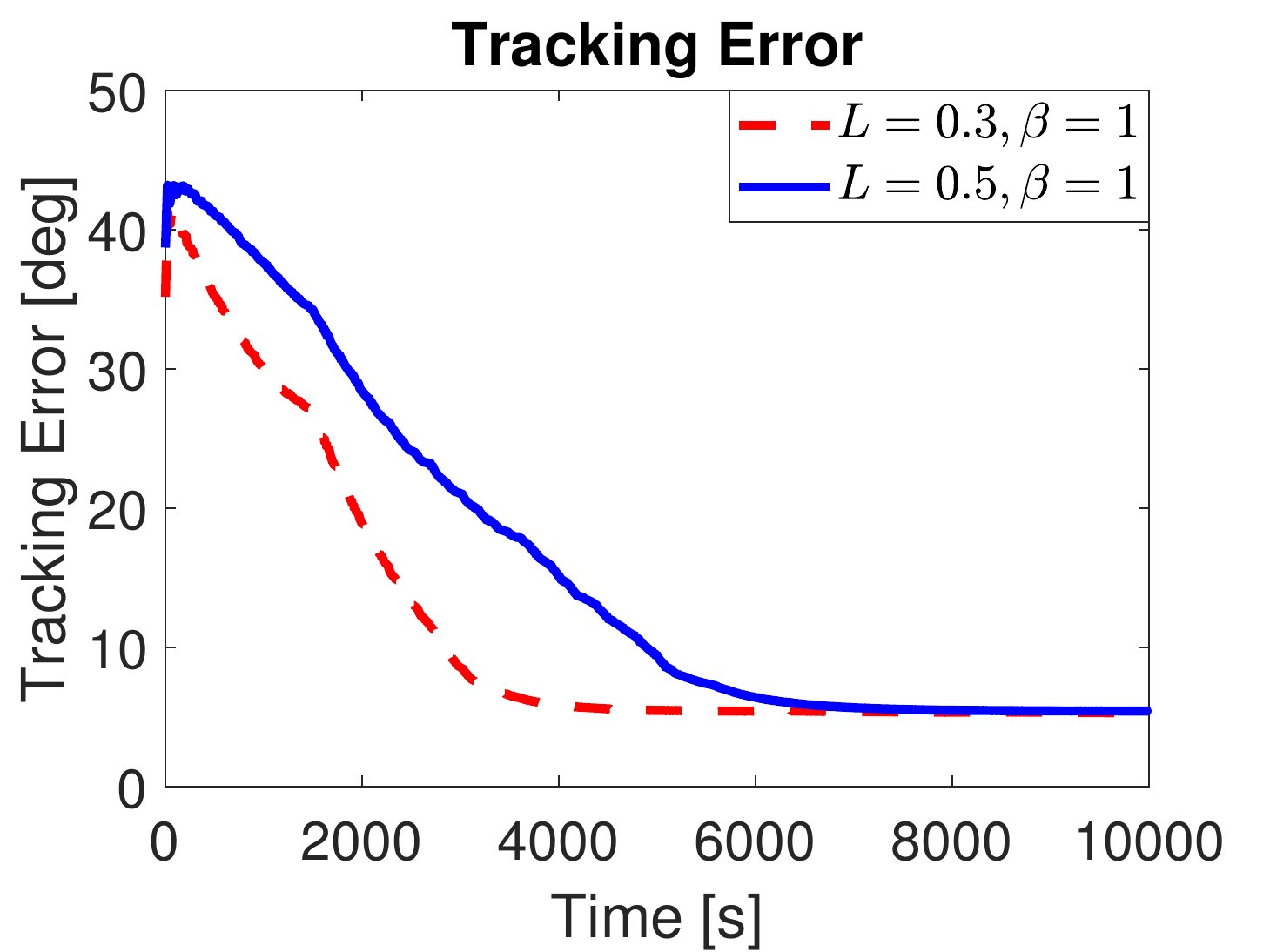,height=.195\textwidth}}  
\put(  125,  -8){\epsfig{file=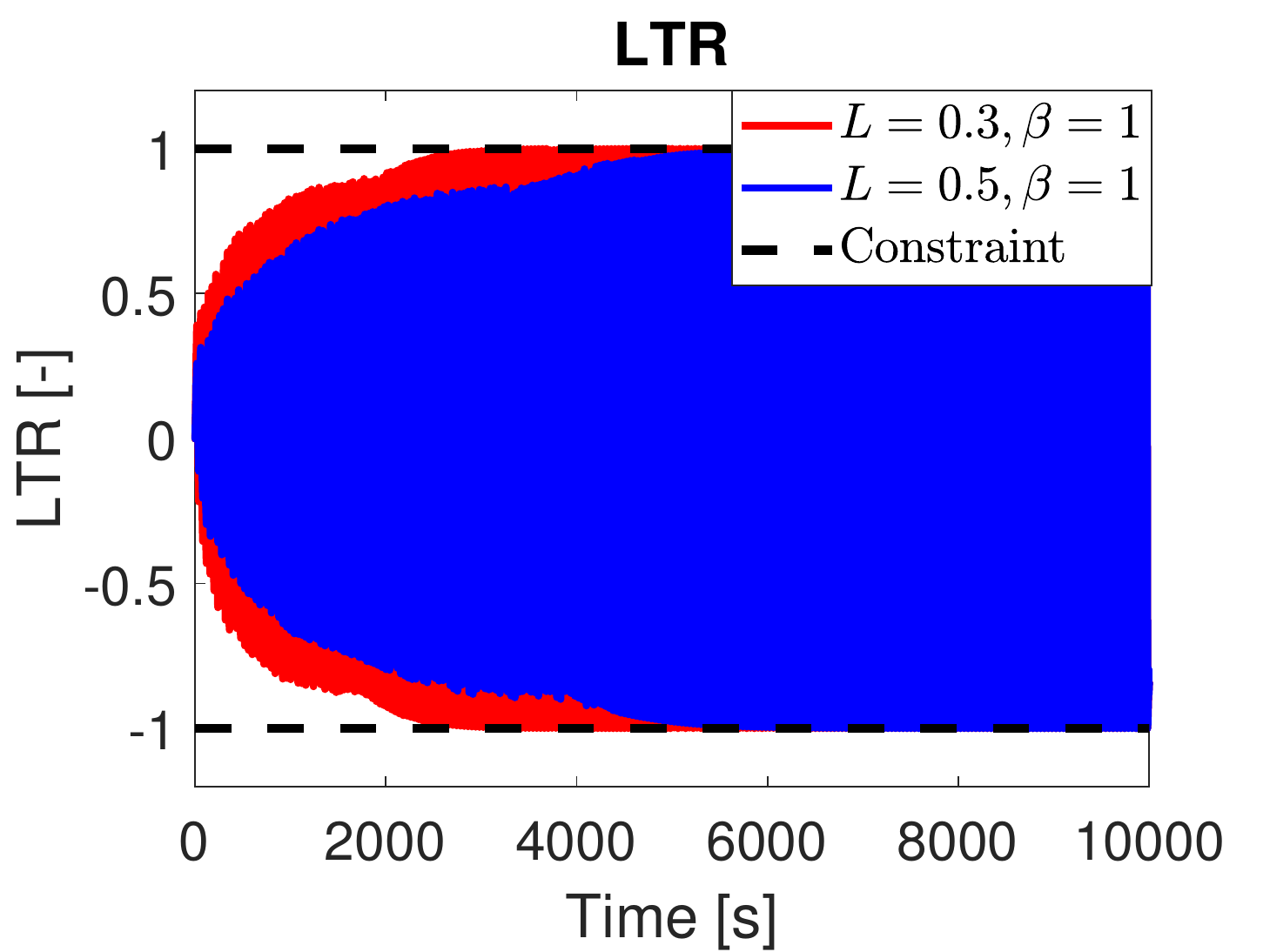,height=.195\textwidth}}  
\put( 14, 86.5){\small{(a)}}
\put( 144, 86.5){\small{(b)}}
\end{picture}
\end{center}
      \caption{Learning algorithm application to fuel truck rollover avoidance based on the tank truck dynamic model developed: (a) Tracking error profile during learning; (b) LTR profile during learning.}
      \label{fig:results_linear_carsim}
\end{figure}

The learning progress is illustrated in Fig.~\ref{fig:results_linear_carsim}. In Fig.~\ref{fig:results_linear_carsim}(a), the tracking error is defined as the average of $|\command(t) - \inputref(t)|$ over a past time window of 1000 (sec). At the beginning of learning, the tracking error is relatively high, and LRG operates conservatively with significant modifications of the command. The tracking error gradually decreases as the learning progresses and converges to a low value. Fig.~\ref{fig:results_linear_carsim}(b) shows the LTR response during learning. As the learning progresses, the vehicle gains the ability to operate with maneuvers that cause the LTR response to reach the constraint boundary but without violating it. Note that constraints are satisfied during the entire learning process. 

\begin{figure}[h]
\begin{center}
\begin{picture}(240.0, 100.0)
\put(  -6,  -8){\epsfig{file=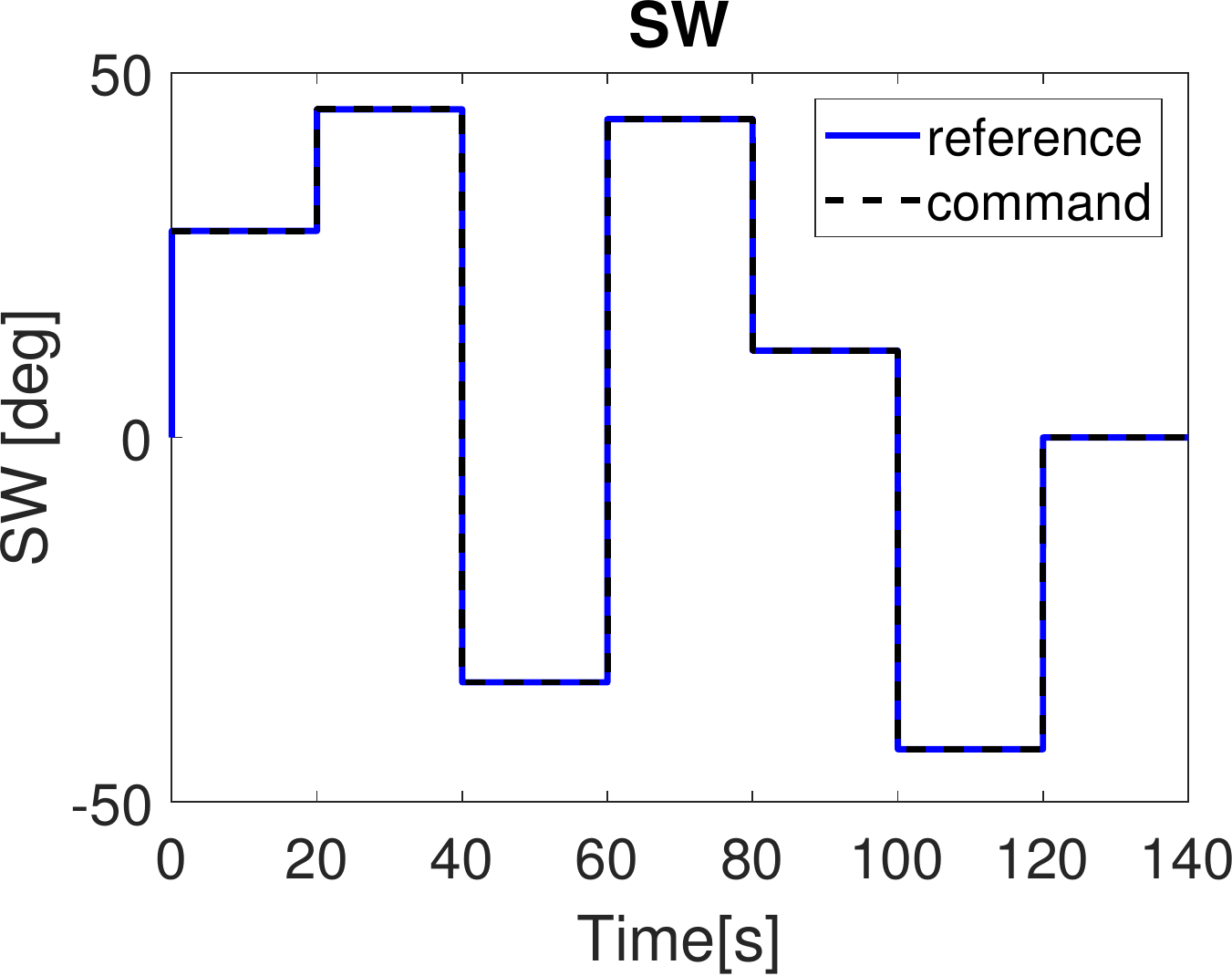,height=.195\textwidth}}  
\put(  125,  -8){\epsfig{file=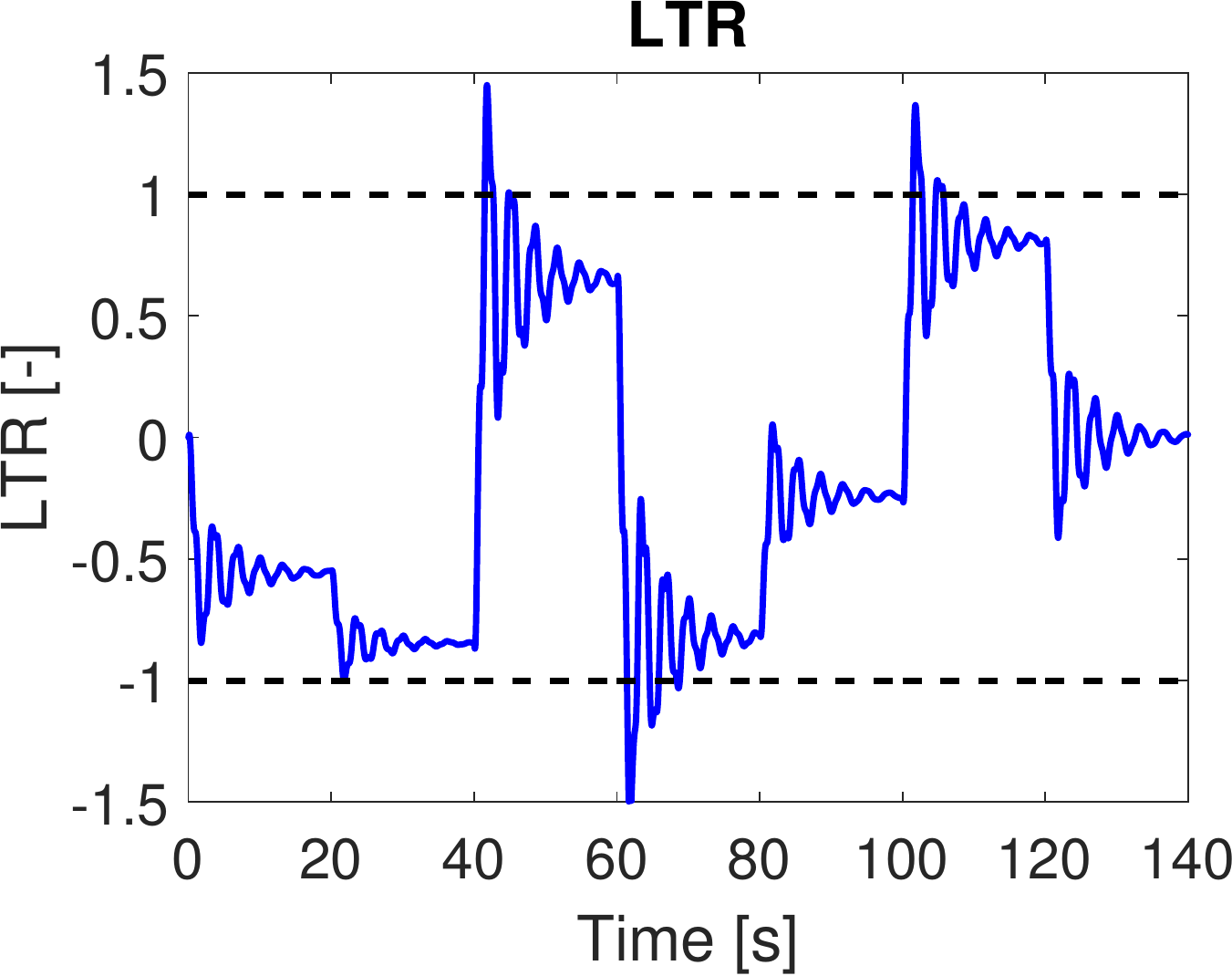,height=.195\textwidth}}  
\put( 14, 88){\small{(a)}}
\put( 144, 88){\small{(b)}}
\end{picture}
\end{center}
      \caption{Vehicle responses to step commands without LRG.}
      \label{fig:validation_noRG}
     \vspace{0.2in}
\end{figure}

After the learning is completed, the vehicle response to step commands is examined. Fig.~\ref{fig:validation_noRG} shows the response of the system without LRG with steering wheel command directly applied to the fuel truck system as shown in Fig.~\ref{fig:validation_noRG}(a), and constraint violations in the resulting LTR response in Fig.~\ref{fig:validation_noRG}(b). 

\begin{figure}[h]
\begin{center}
\begin{picture}(240.0, 100.0)
\put(  -6,  -8){\epsfig{file=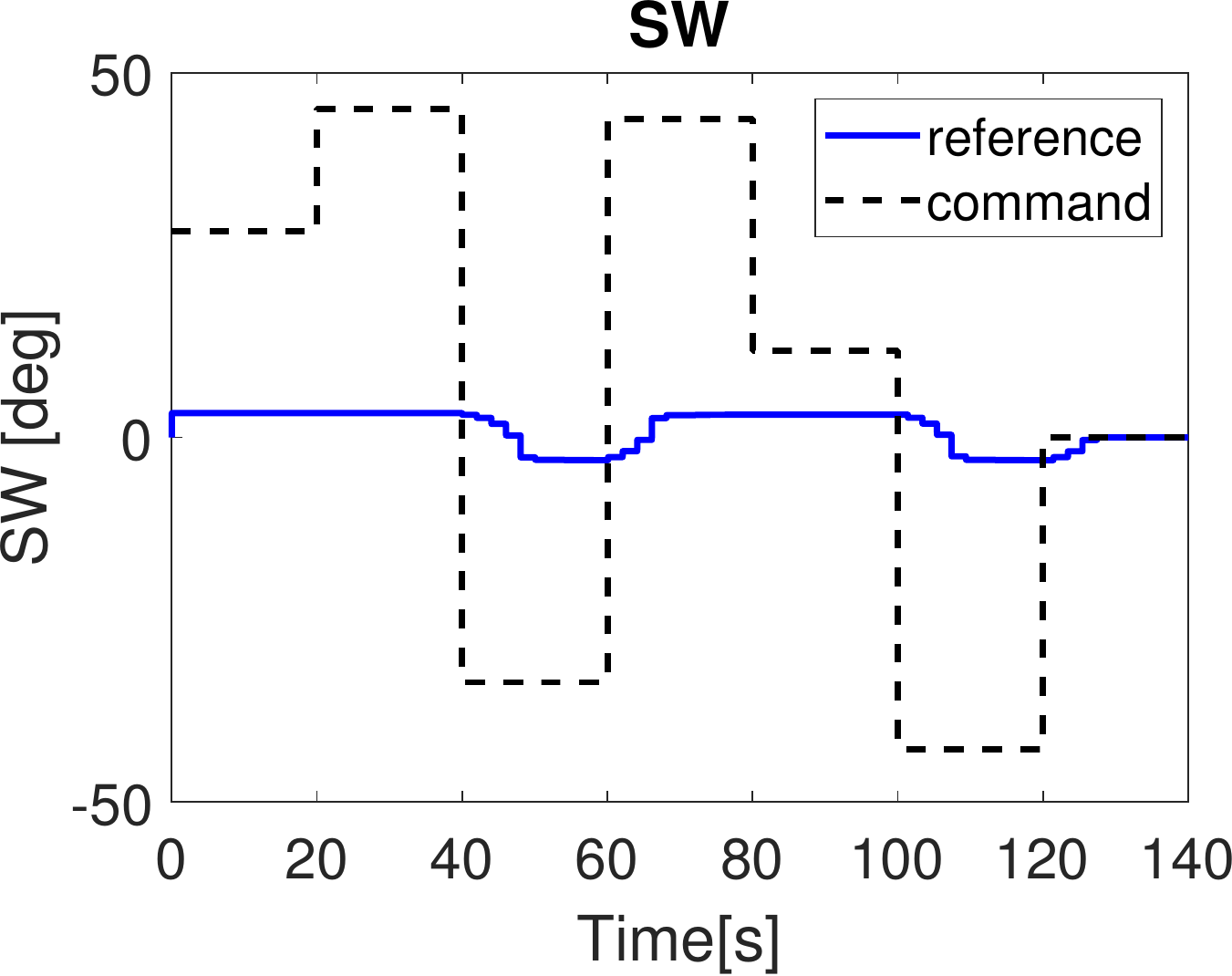,height=.195\textwidth}}  
\put(  125,  -8){\epsfig{file=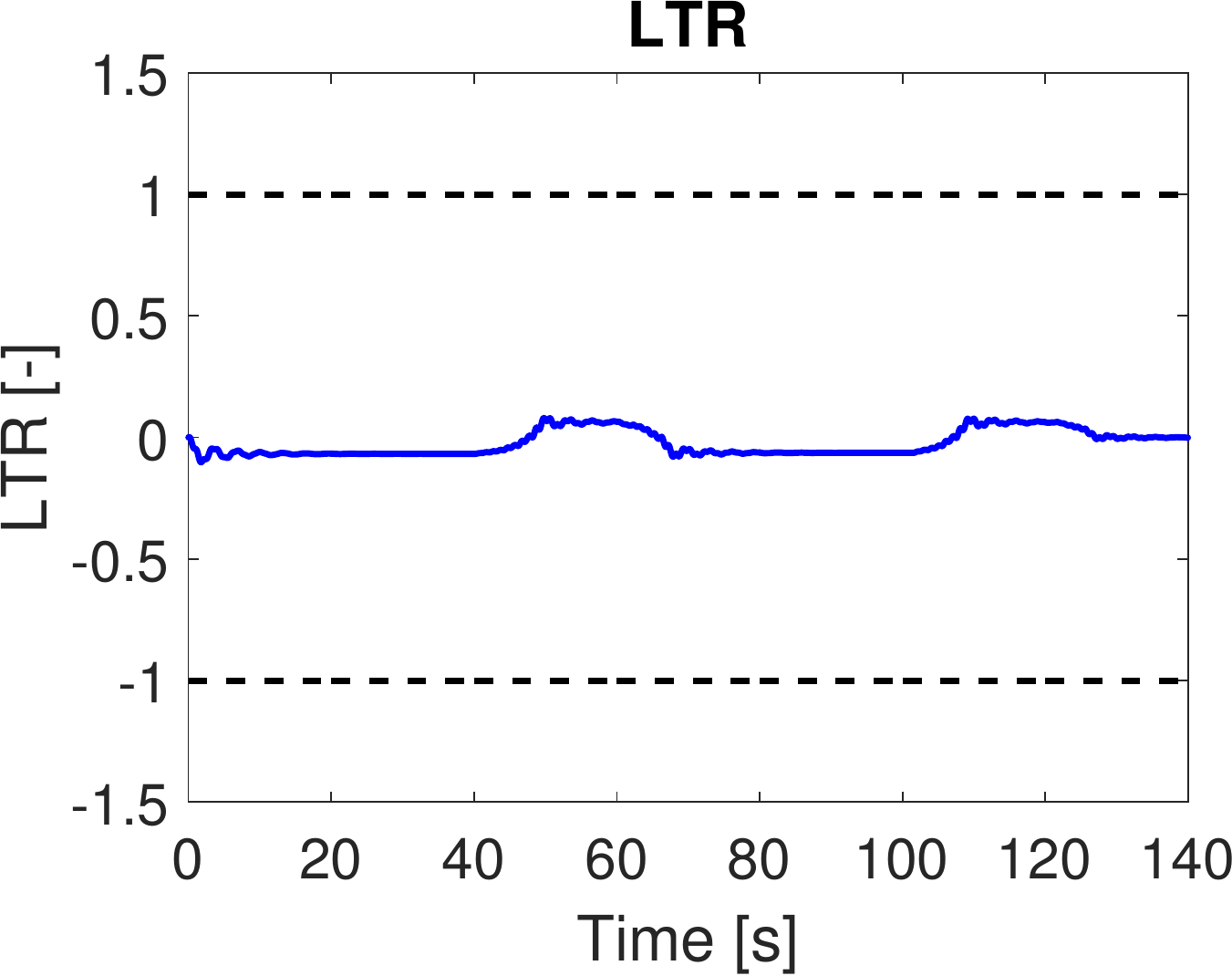,height=.195\textwidth}}  
\put( 14, 88){\small{(a)}}
\put( 144, 88){\small{(b)}}
\end{picture}
\end{center}
      \caption{Vehicle responses to step commands with LRG before learning.}
      \label{fig:validation_RG_beg}
\end{figure}

\begin{figure}[h]
\begin{center}
\begin{picture}(240.0, 100.0)
\put(  -6,  -8){\epsfig{file=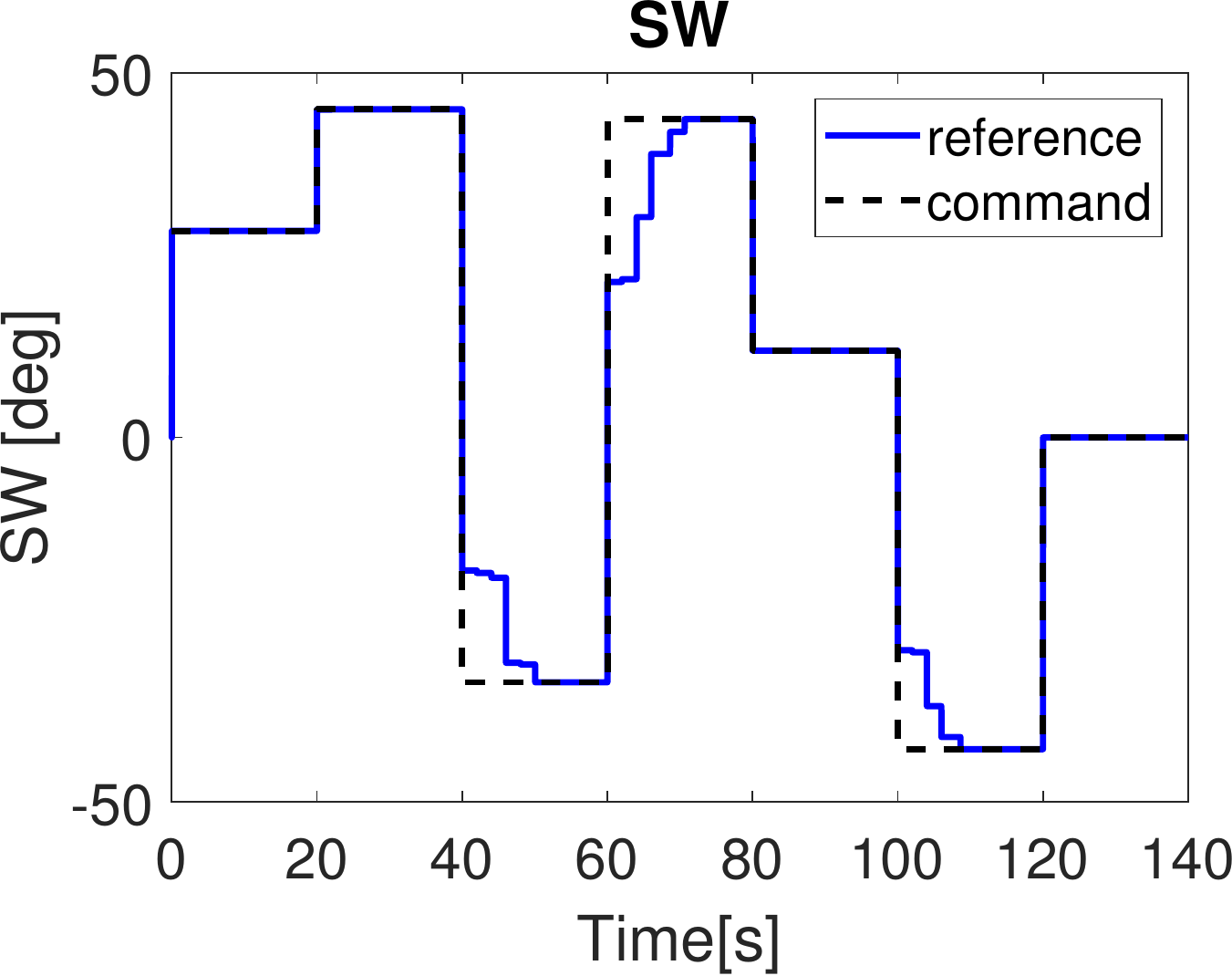,height=.195\textwidth}}  
\put(  125,  -8){\epsfig{file=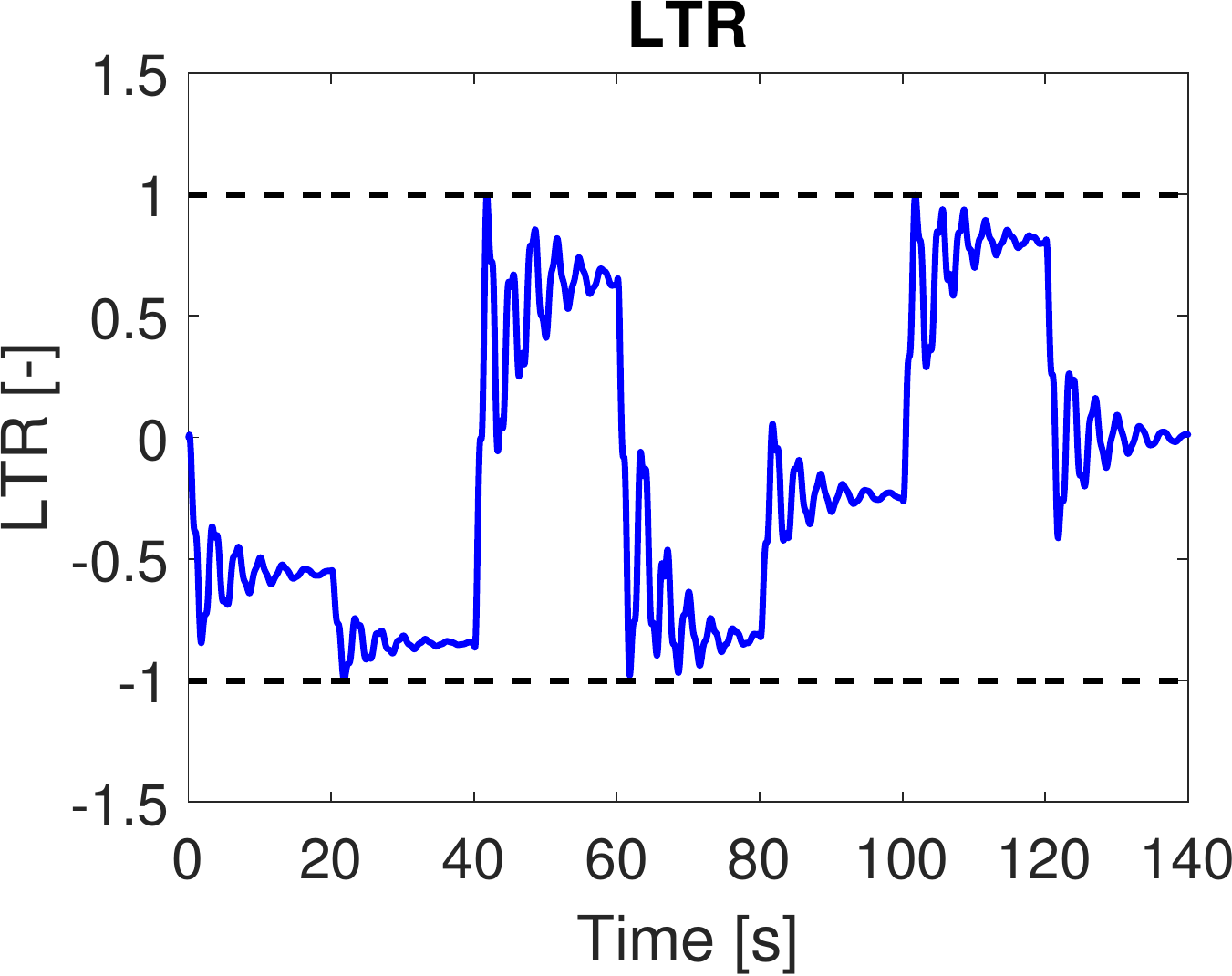,height=.195\textwidth}}  
\put( 14, 88){\small{(a)}}
\put( 144, 88){\small{(b)}}
\end{picture}
\end{center}
      \caption{Vehicle responses to step commands with LRG after learning.}
      \label{fig:validation_RG}
\end{figure}

Figs.~\ref{fig:validation_RG_beg},~\ref{fig:validation_RG} show the response with LRG before learning and after learning respectively. Before learning, LRG operates conservatively permitting almost no changes in response to the command. After learning is completed, LRG is significantly less conservative and allows the modified reference input to converge to the original command
within a relatively short time period for the same step commands.

\begin{figure}[h!]
\begin{center}
\begin{picture}(240.0, 100.0)
\put(  -6,  -8){\epsfig{file=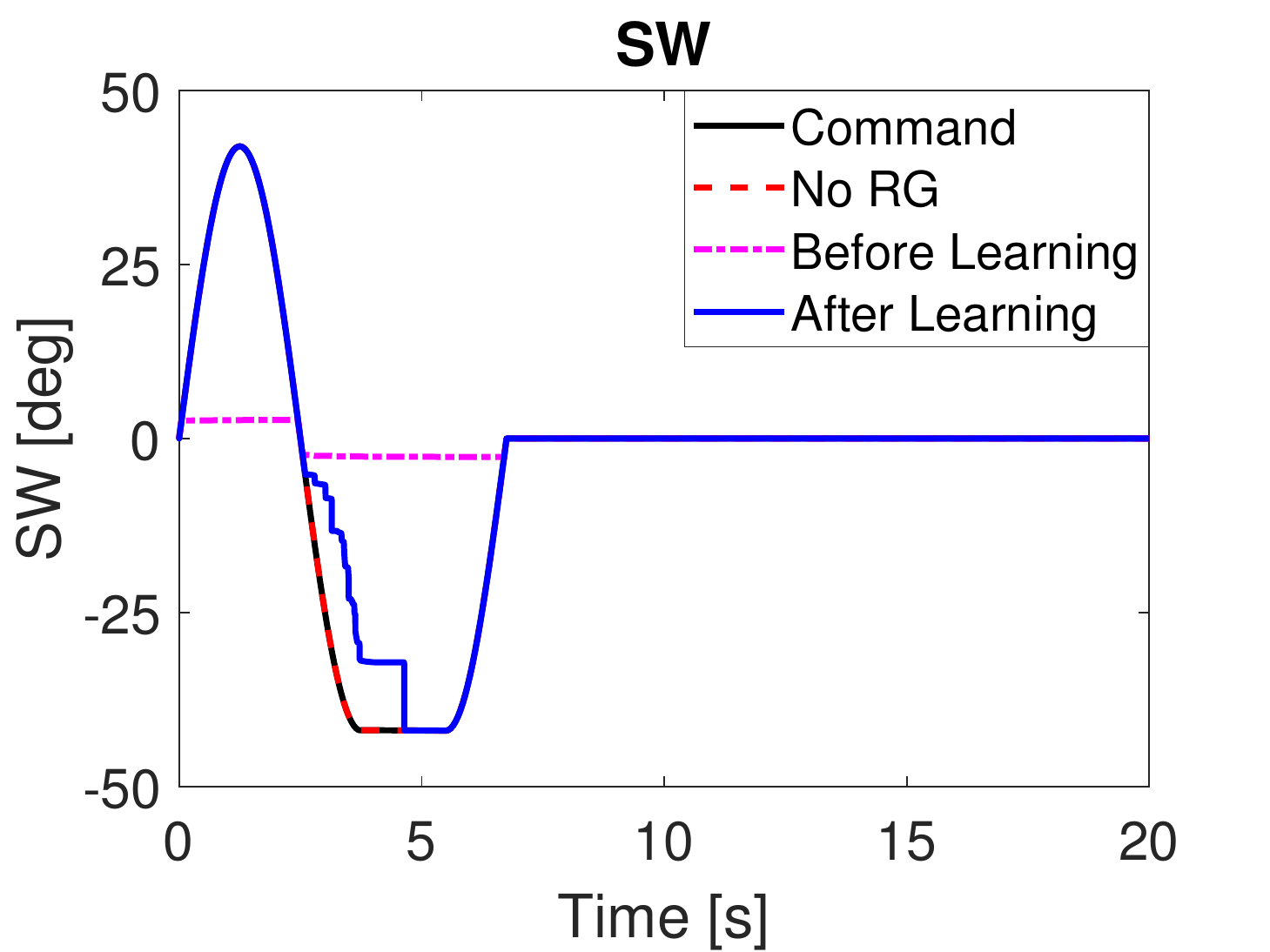,height=.195\textwidth}}  
\put(  121,  -8){\epsfig{file=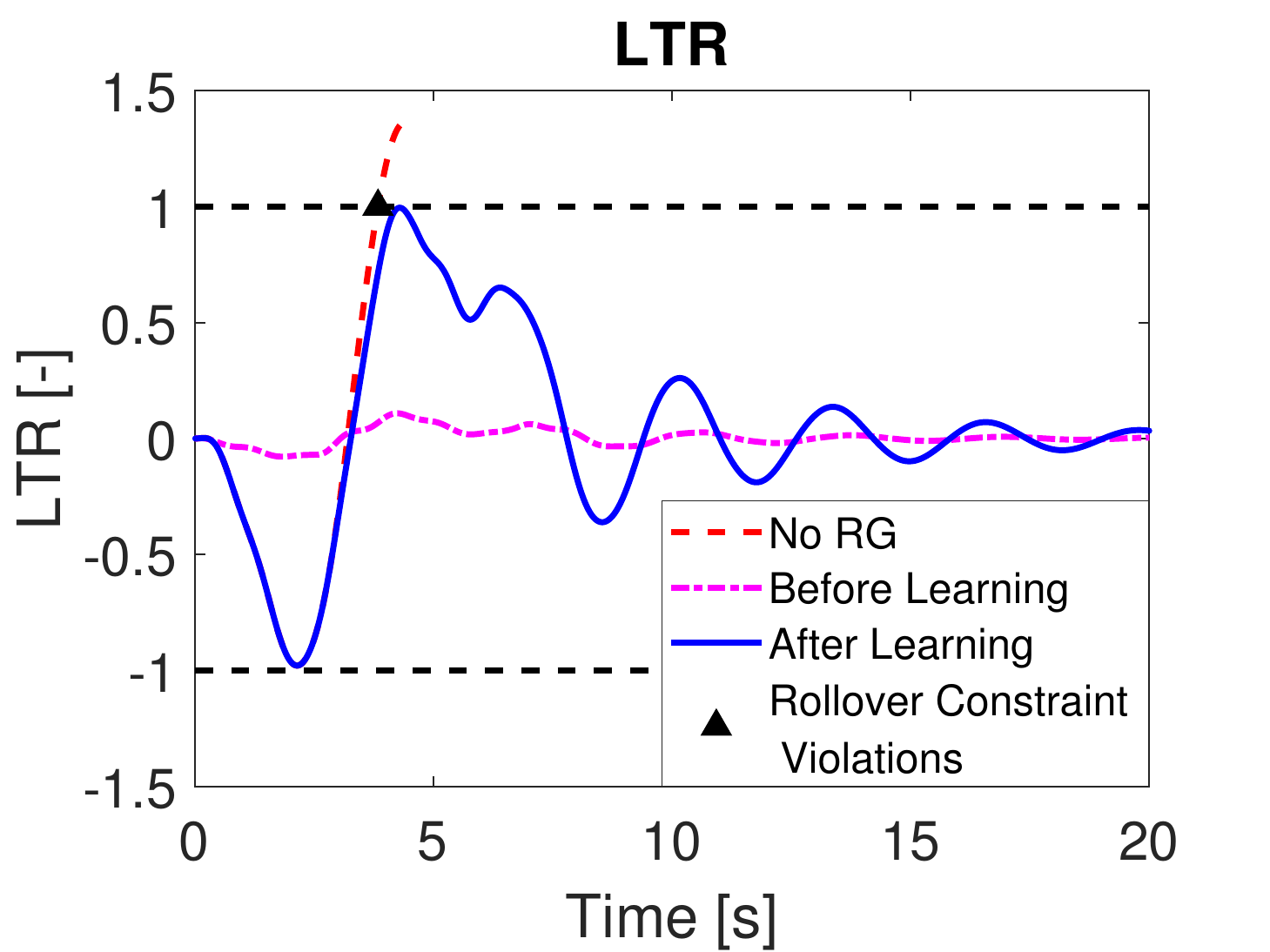,height=.195\textwidth}}  
\put( 14, 86.5){\small{(a)}}
\put( 144, 86.5){\small{(b)}}
\end{picture}
\end{center}
      \caption{(a) Steering angle and (b) LTR responses for the sine-and-dwell test.}
      \label{fig:sin_and_dwell}
\end{figure}

\begin{figure}[h!]
\begin{center}
\begin{picture}(245.0, 290.0)
\put(  -6,  -8){\epsfig{file=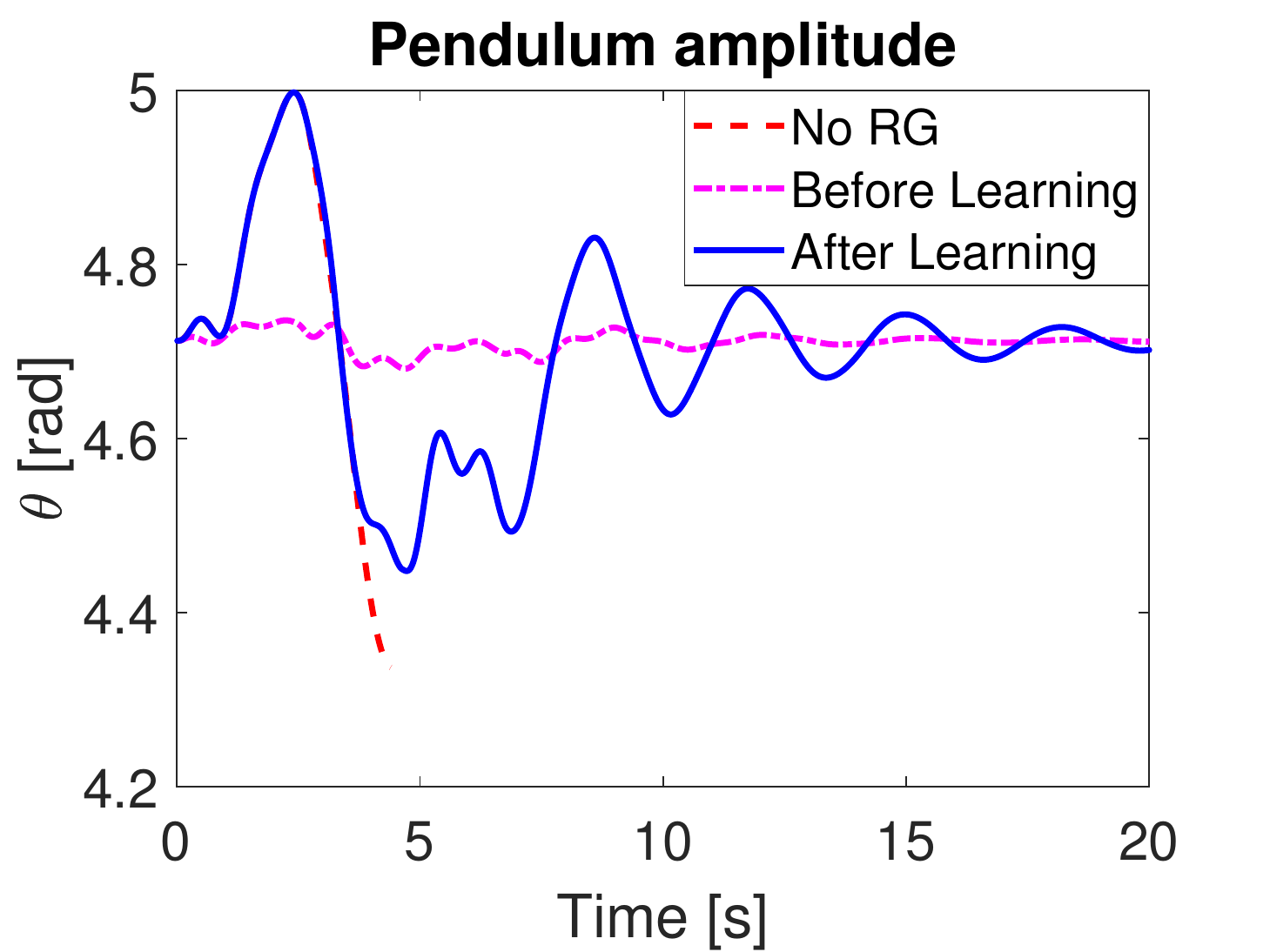,height=.195\textwidth}}  
\put(  122,  -8){\epsfig{file=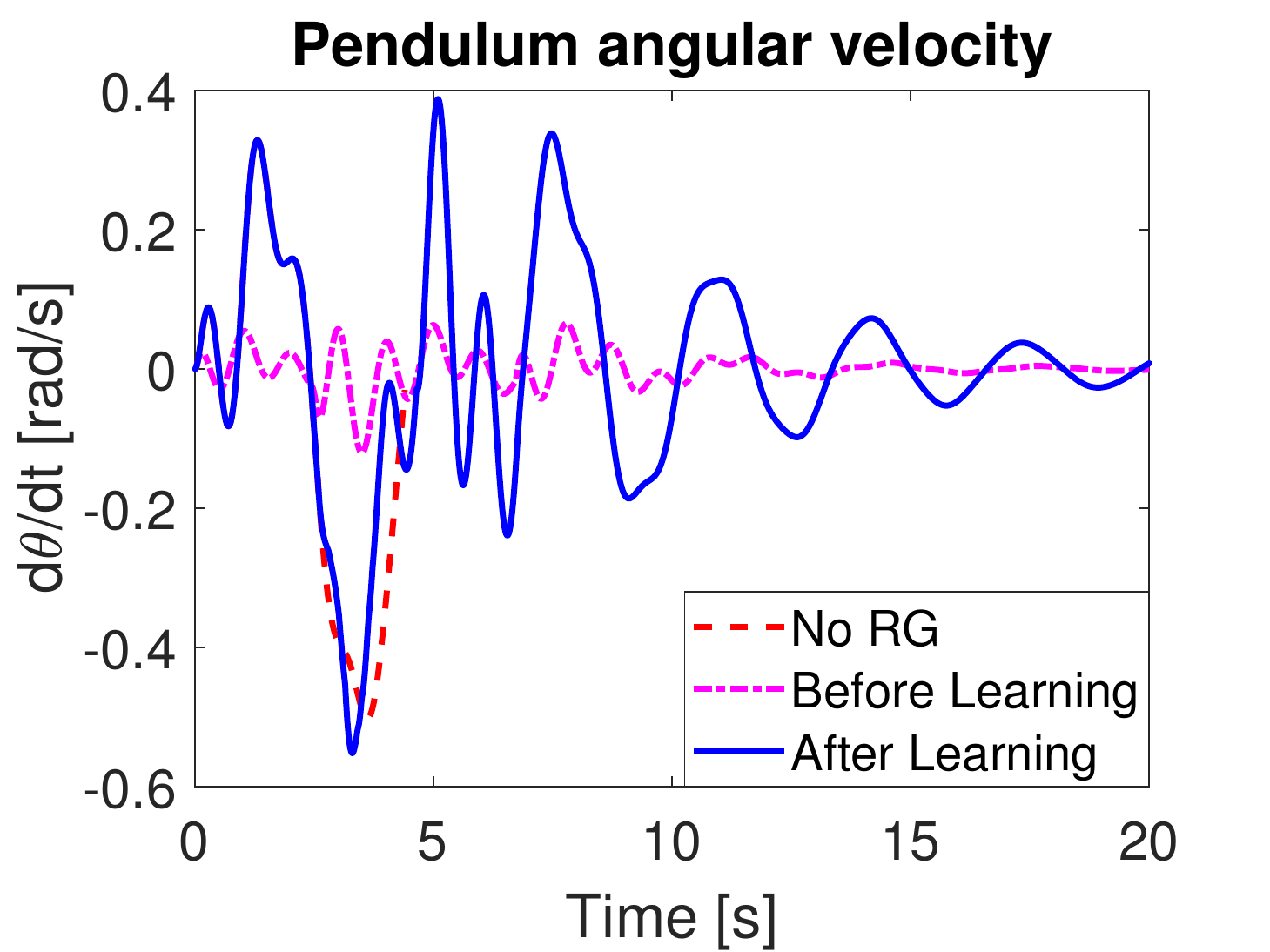,height=.195\textwidth}}  
\put(  -6,  95){\epsfig{file=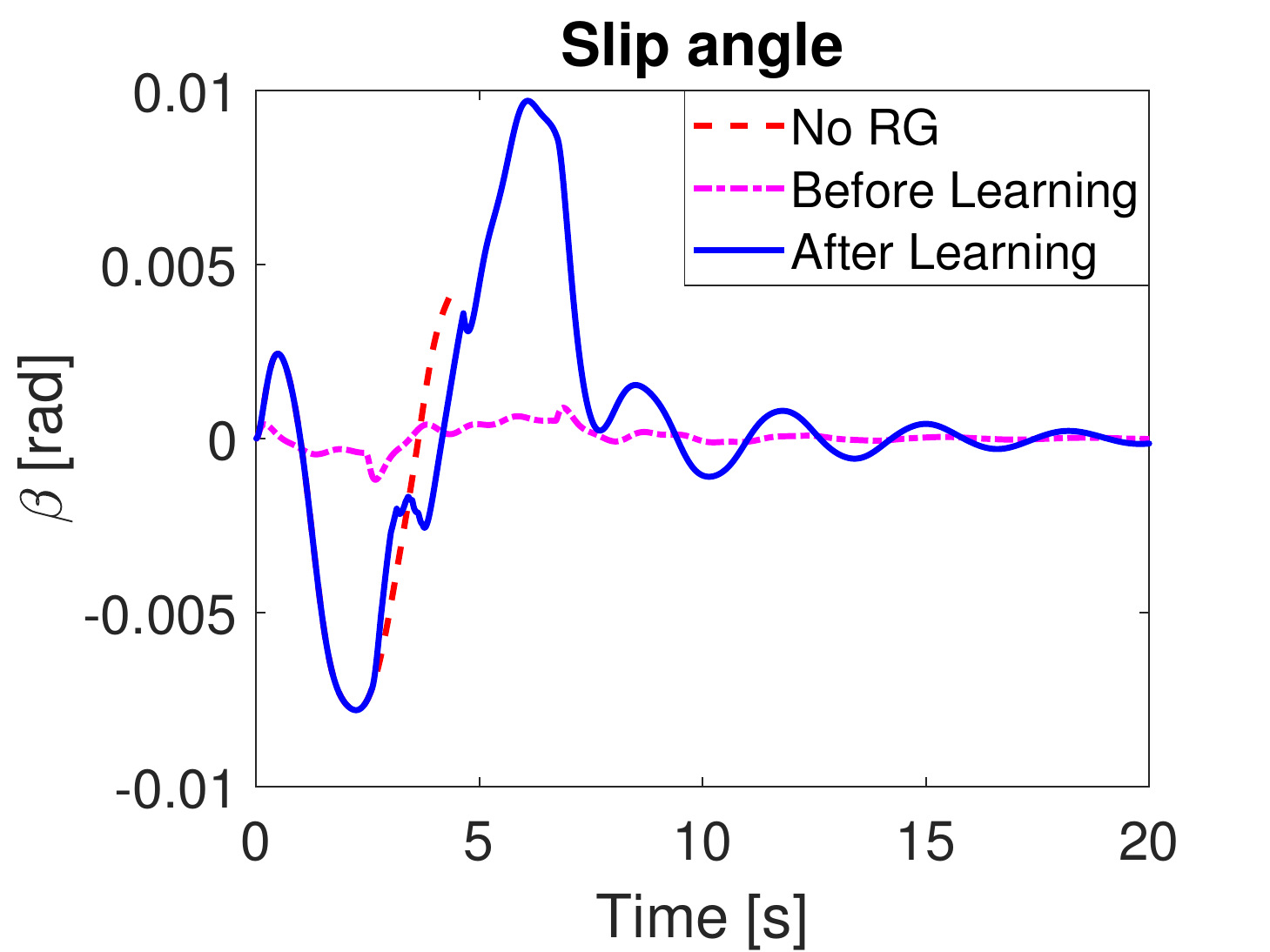,height=.195\textwidth}}  
\put(  122,  95){\epsfig{file=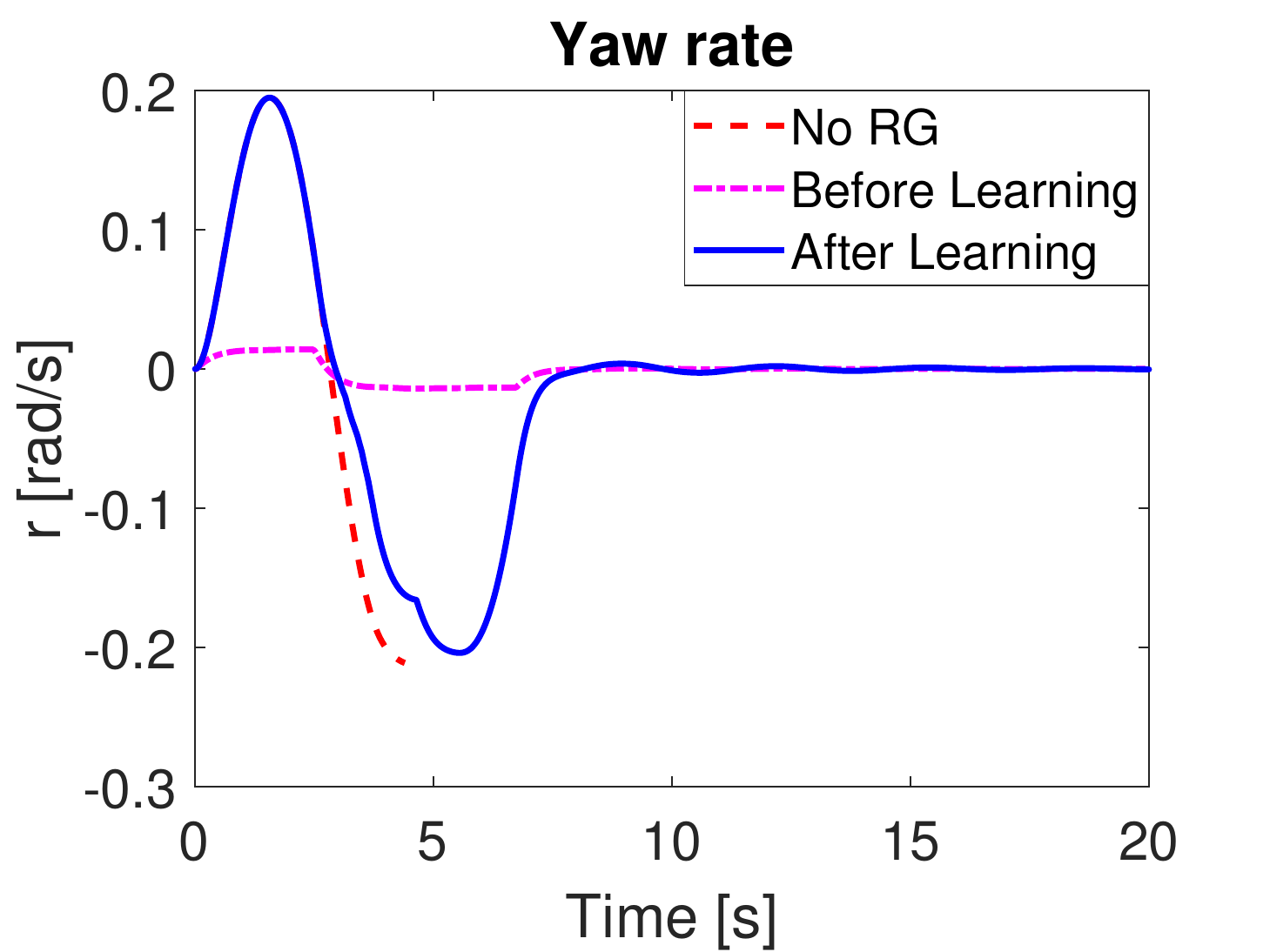,height=.195\textwidth}}  
\put(  -6,  197){\epsfig{file=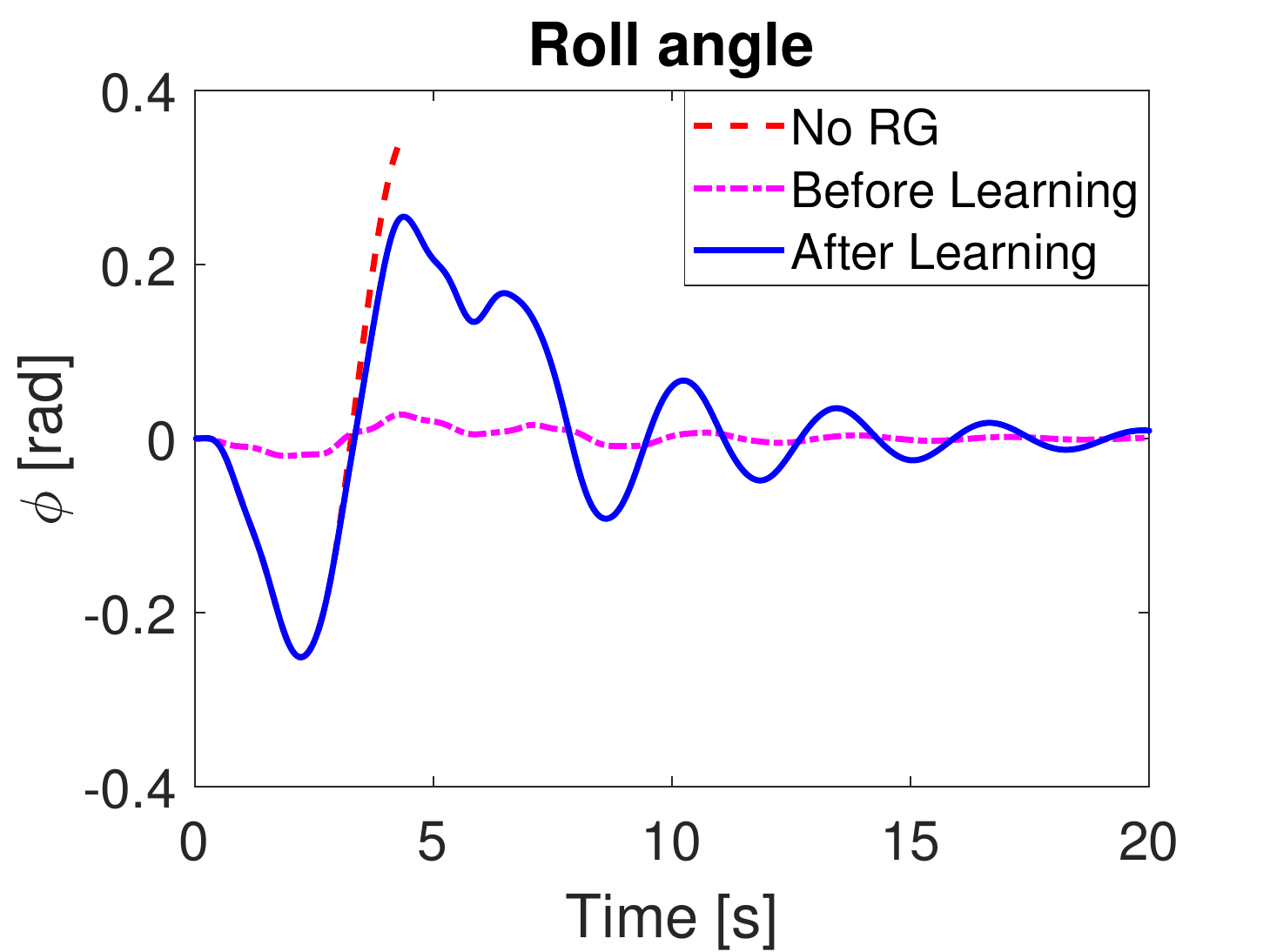,height=.195\textwidth}}  
\put(  122,  197){\epsfig{file=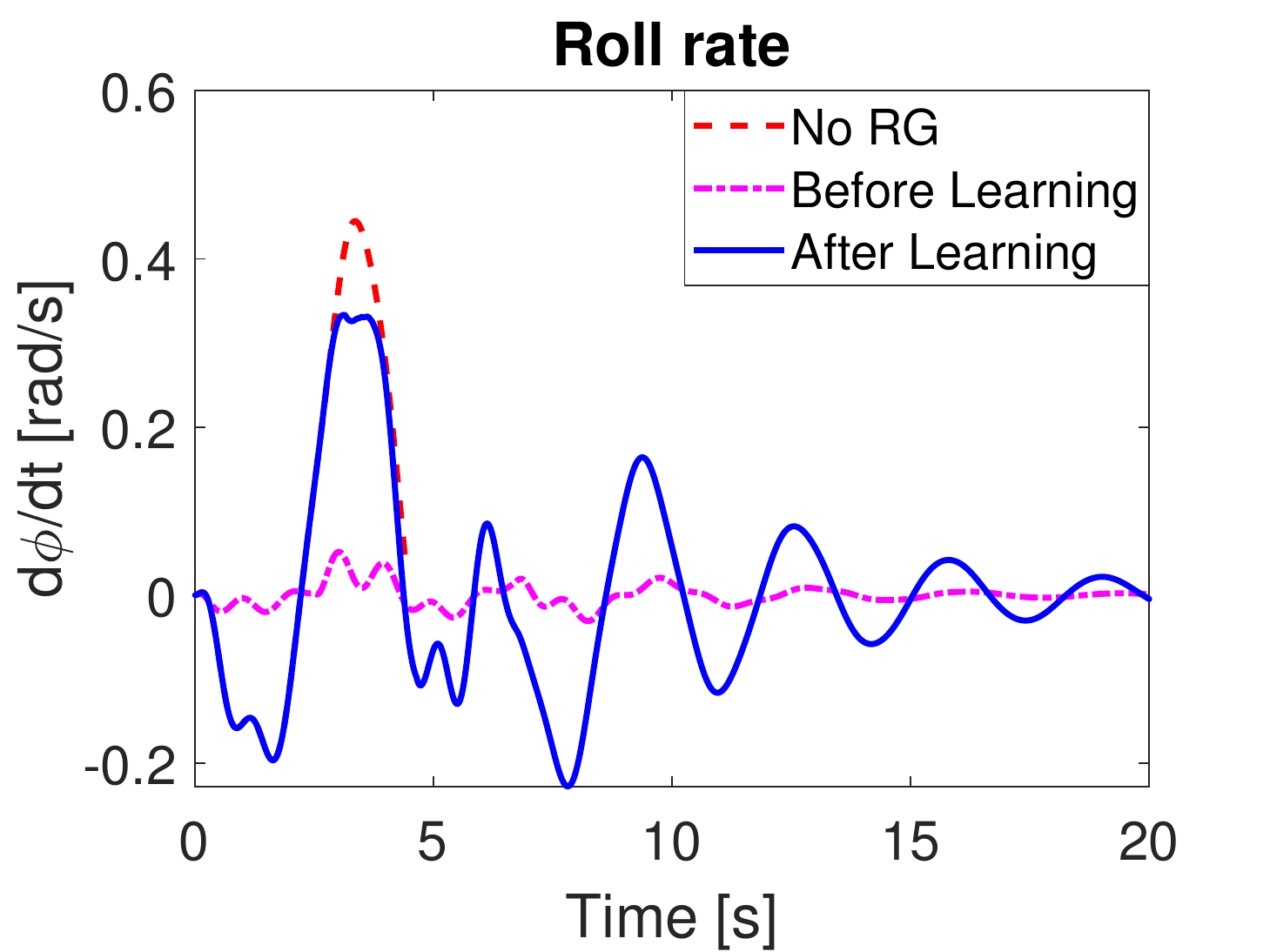,height=.195\textwidth}}  

\end{picture}
\end{center}
      \caption{The roll angle, roll rate, slip angle, yaw rate, pendulum angle, and pendulum angular velocity responses for the sine-and-dwell test.}
      \label{fig:state_traj_sine_and_dwell}
\end{figure}

We next consider vehicle response to the standard sine-and-dwell steering angle profile. The corresponding steering angle, LTR, and states responses are shown in Fig.~\ref{fig:sin_and_dwell} and Fig.~\ref{fig:state_traj_sine_and_dwell}.

Without LRG, the sine-and-dwell steering angle profile causes constraint violations, which may lead to the fuel truck rolling over. By augmenting LRG, rollover accidents can be avoided. Before learning, LRG makes relatively large modifications to the steering commands and results in conservative responses. After learning is completed, LRG is able to enforce the constraints with
significantly less command modifications.

%% file: sections/05_04_multispeed_extension.tex
\subsection{Applying LRG to variable load and speed scenarios} 

In previous section, LRG is applied to the fuel truck that has specified constant fill ratio and velocity. When the fuel truck's fill ratio and velocity change, the LRG needs to be re-trained to estimate $D$, or it can be trained at different load conditions and speeds from the beginning, and $\overline{D}$ in \eqref{equ:upper_D} can be made a function of these additional parameters. 

\begin{figure}[h]
\begin{center}
\begin{picture}(240.0, 90.0)
\put(  -6,  -8){\epsfig{file=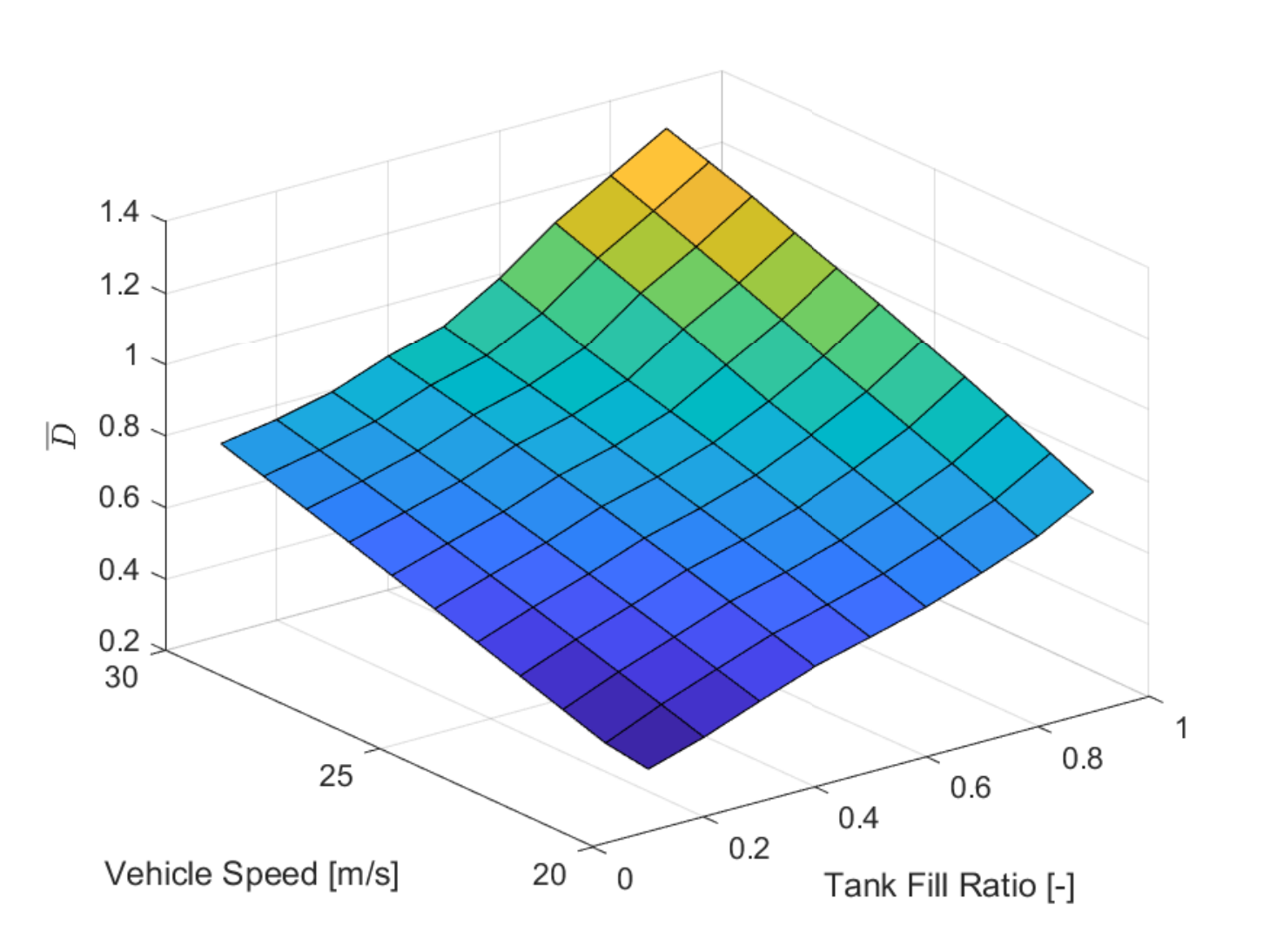,height=.195\textwidth}}  
\put(  121,  -8){\epsfig{file=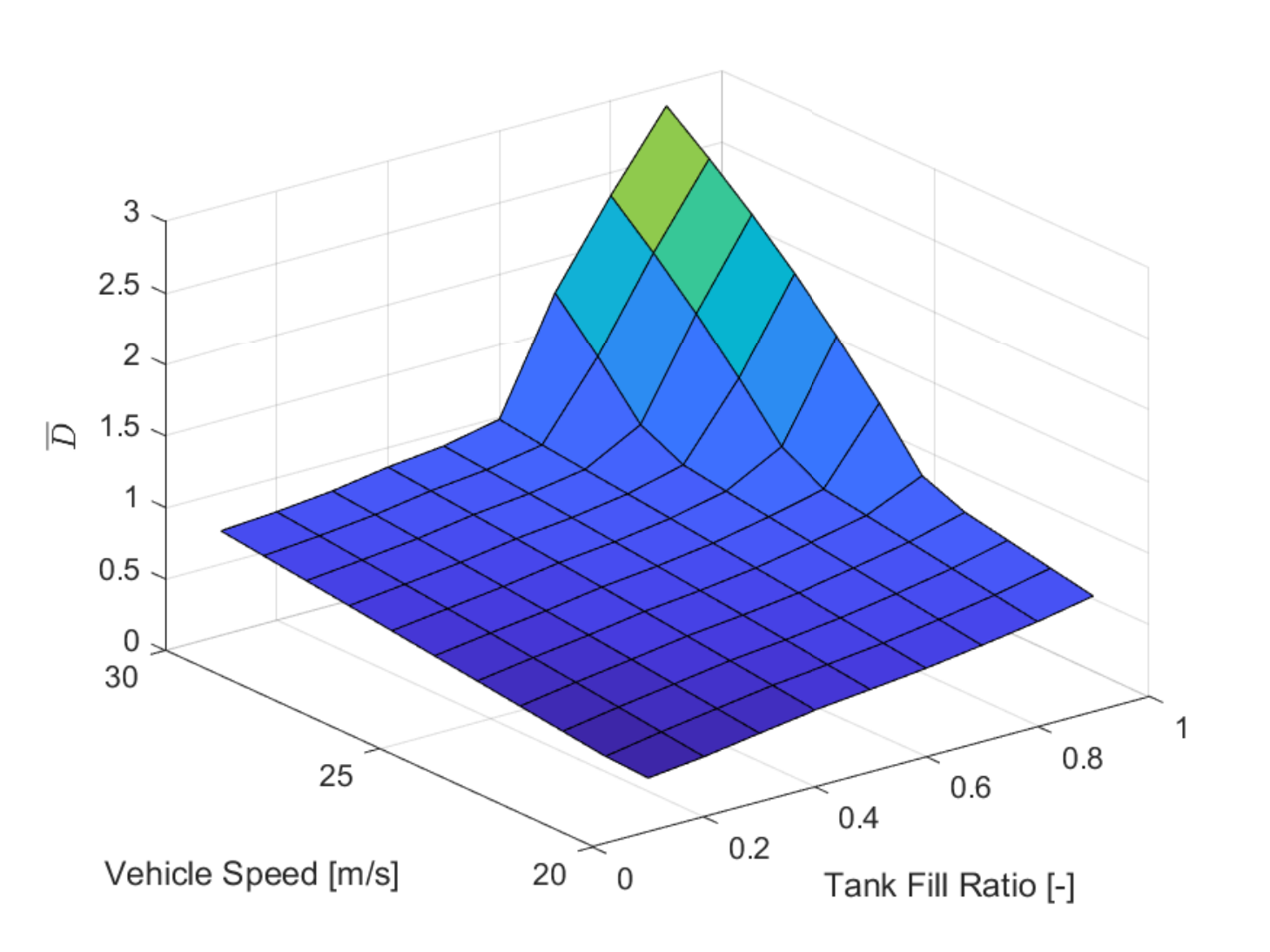,height=.195\textwidth}}  
\put( 14, 86.5){\small{(a)}}
\put( 138, 86.5){\small{(b)}}
\end{picture}
\end{center}
      \caption{Estimated $\overline{D}$ values at different speeds and fill ratios with (a) $\inputref = -25$ (deg), $\delta\inputref = 25$ (deg), $\delta\state = 0$ and (b) $\inputref = 0$ (deg), $\delta\inputref = -25$ (deg), $\delta\state = 0$.}
      \label{fig:mesh_plot_multi_speed_load}
     \vspace{0.2in}
\end{figure}

With the latter approach followed, two cross-sections of the resulting $\overline{D}$ are shown in Fig.~\ref{fig:mesh_plot_multi_speed_load}. In Fig.~\ref{fig:mesh_plot_multi_speed_load}(a), for the same $\delta\inputref$, higher vehicle speed and larger tank fill ratio will result in higher value of $\overline{D}$, which reflects the fact that the truck is easier to rollover. In Fig.~\ref{fig:mesh_plot_multi_speed_load}(b), $\overline{D}$ increases significantly when the tank fill ratio approaches 1 and the vehicle speed approaches 30 (m/s). Note that $\overline{D}$ is constructed using \eqref{equ:upper_D} based on collected measurements and $L, \beta$. This means $\overline{D}(\inputref, \delta\inputref, \delta\state)$ will be close to $D(\inputref, \delta\inputref, \delta\state)$ when there are measurements near $(\inputref, \delta\inputref, \delta\state)$. On the other hand, $\overline{D}(\inputref, \delta\inputref, \delta\state)$ relies on $L$ and $\beta$ to extrapolate values and can be conservative when all measurements are far from $(\inputref, \delta\inputref, \delta\state)$. The significantly higher value of $\overline{D}$ shown in Fig.~\ref{fig:mesh_plot_multi_speed_load}(b) is due to the fact that when the tank fill ratio and vehicle speed are large (e.g., $V = 30$ (m/s) and $\Delta = 0.9$), $\delta\inputref = -25$ (deg) results in constraint violations. As no measurements were collected around the point $(\inputref = 0, \delta\inputref=-25, \delta\state = 0)$, LRG relies on a conservative estimate of $\overline{D}$ at these loads and speeds.

After $\overline{D}$ is made dependent on tank fill ratio and vehicle speed, the sine-and-dwell test is performed again with the truck accelerating or decelerating. Following \cite{yang2016truck} and guidelines in \cite{harwood2003review}, the acceleration rate is selected as 1 ($\text{m/s}^2$), and the deceleration rate is chosen as -3 ($\text{m/s}^2$). The tank fill ratio is 0.5 and is not changing as we perform the sine-and-dwell test. 

\begin{figure}[h]
\begin{center}
\begin{picture}(245.0, 290.0)
\put(  -6,  -8){\epsfig{file=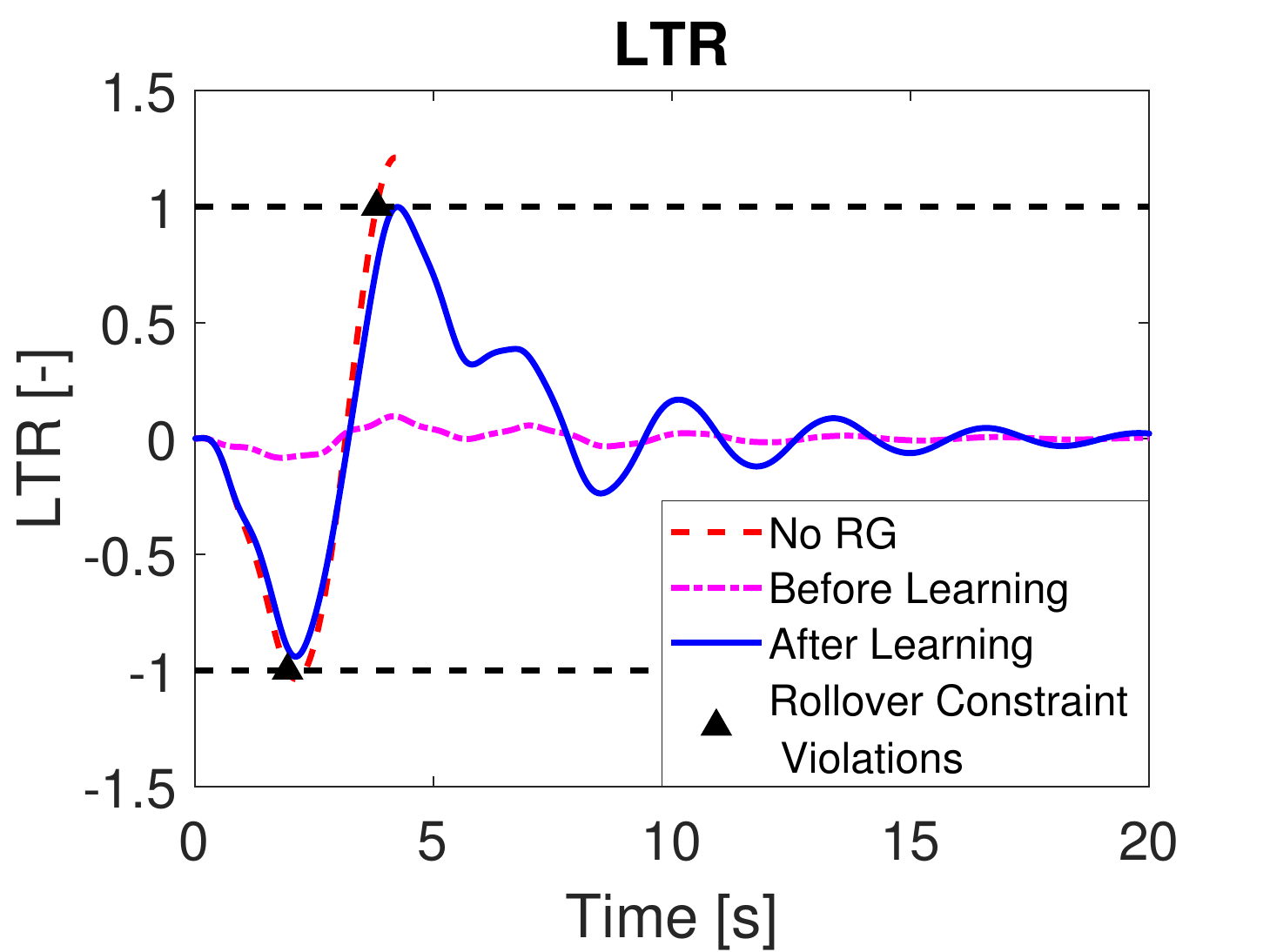,height=.195\textwidth}}  
\put(  -6,  95){\epsfig{file=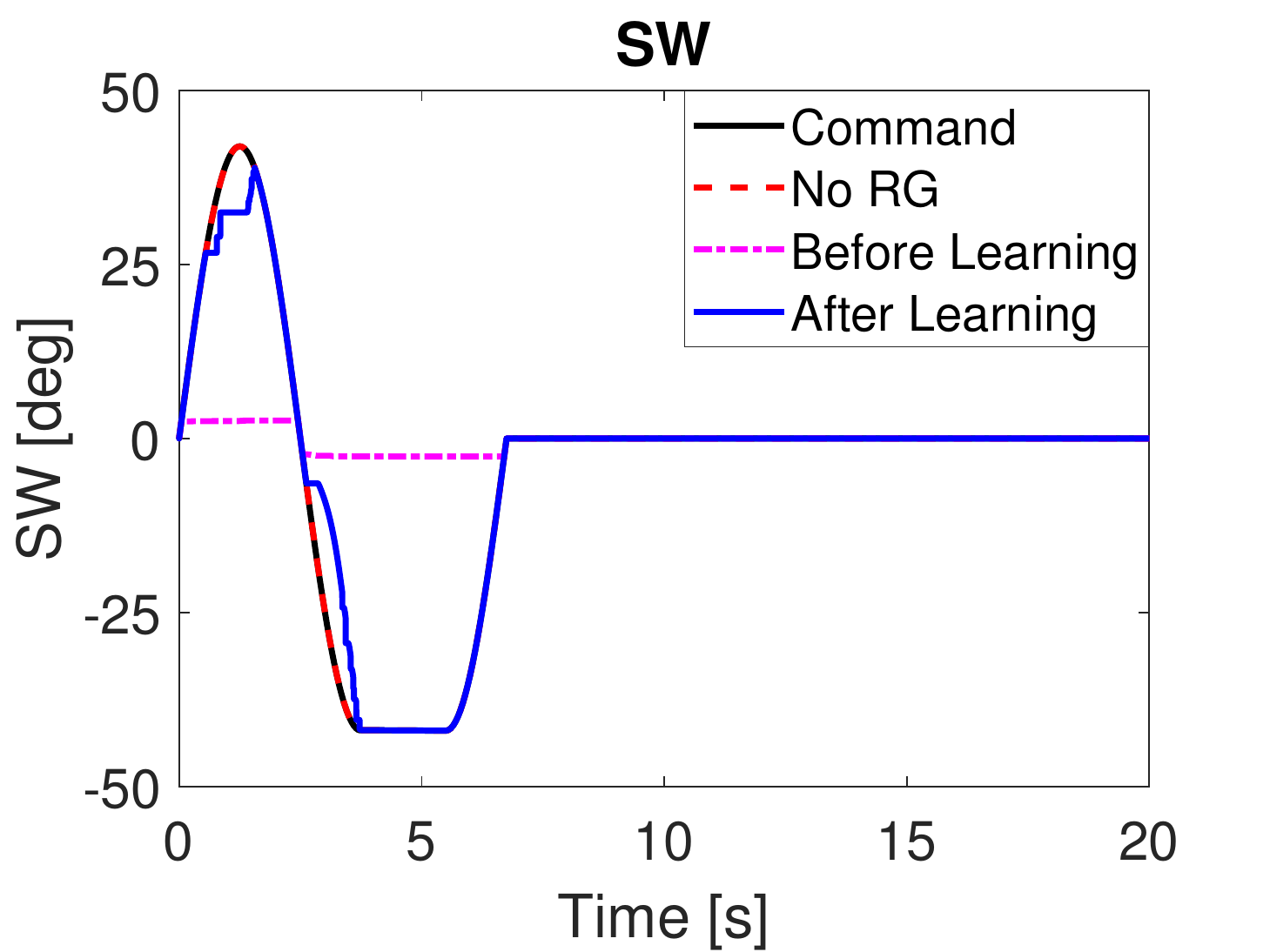,height=.195\textwidth}}  
\put(  -6,  195) {\epsfig{file=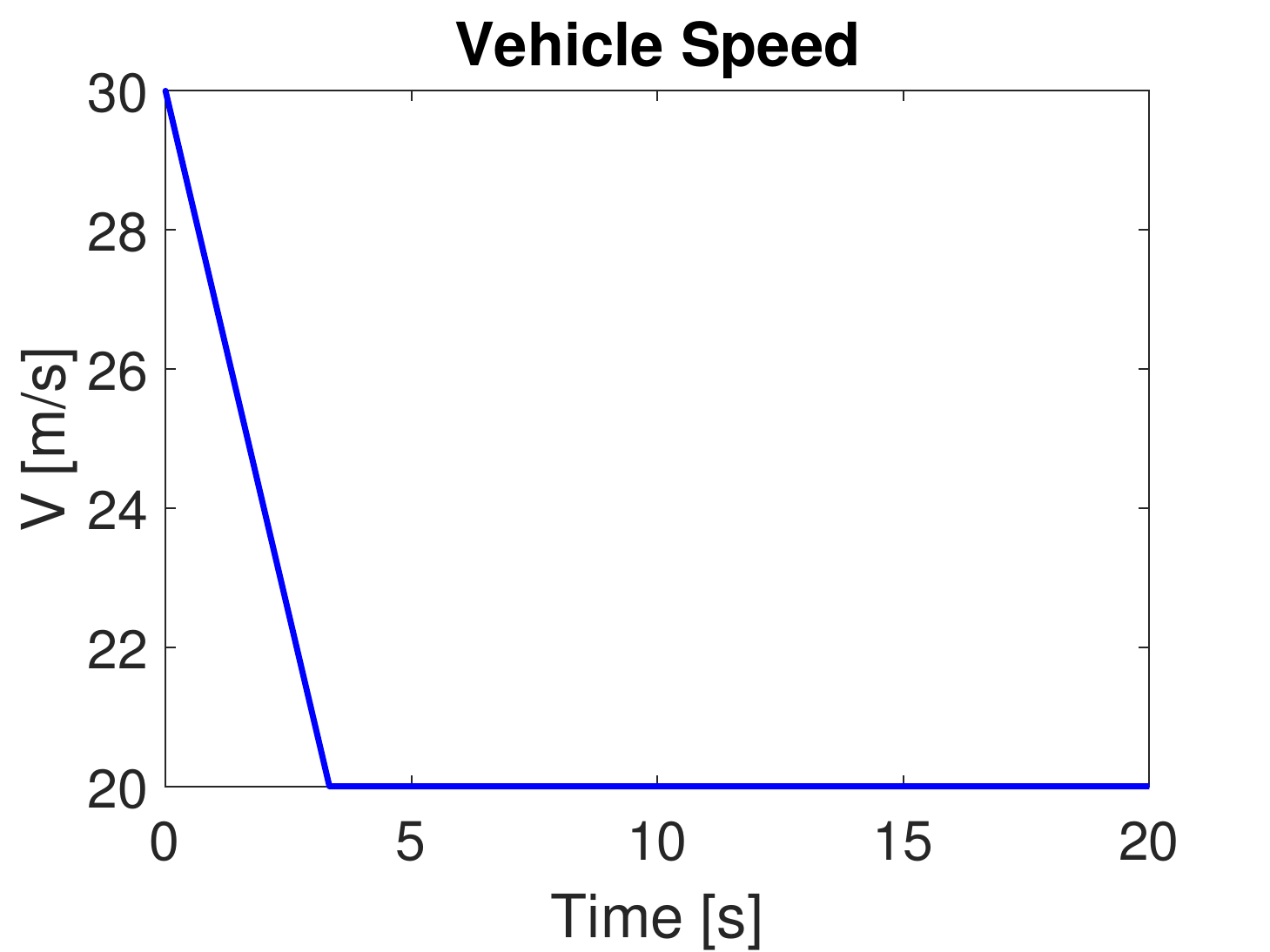,height=.195\textwidth}}  
\put(  125,  -8){\epsfig{file=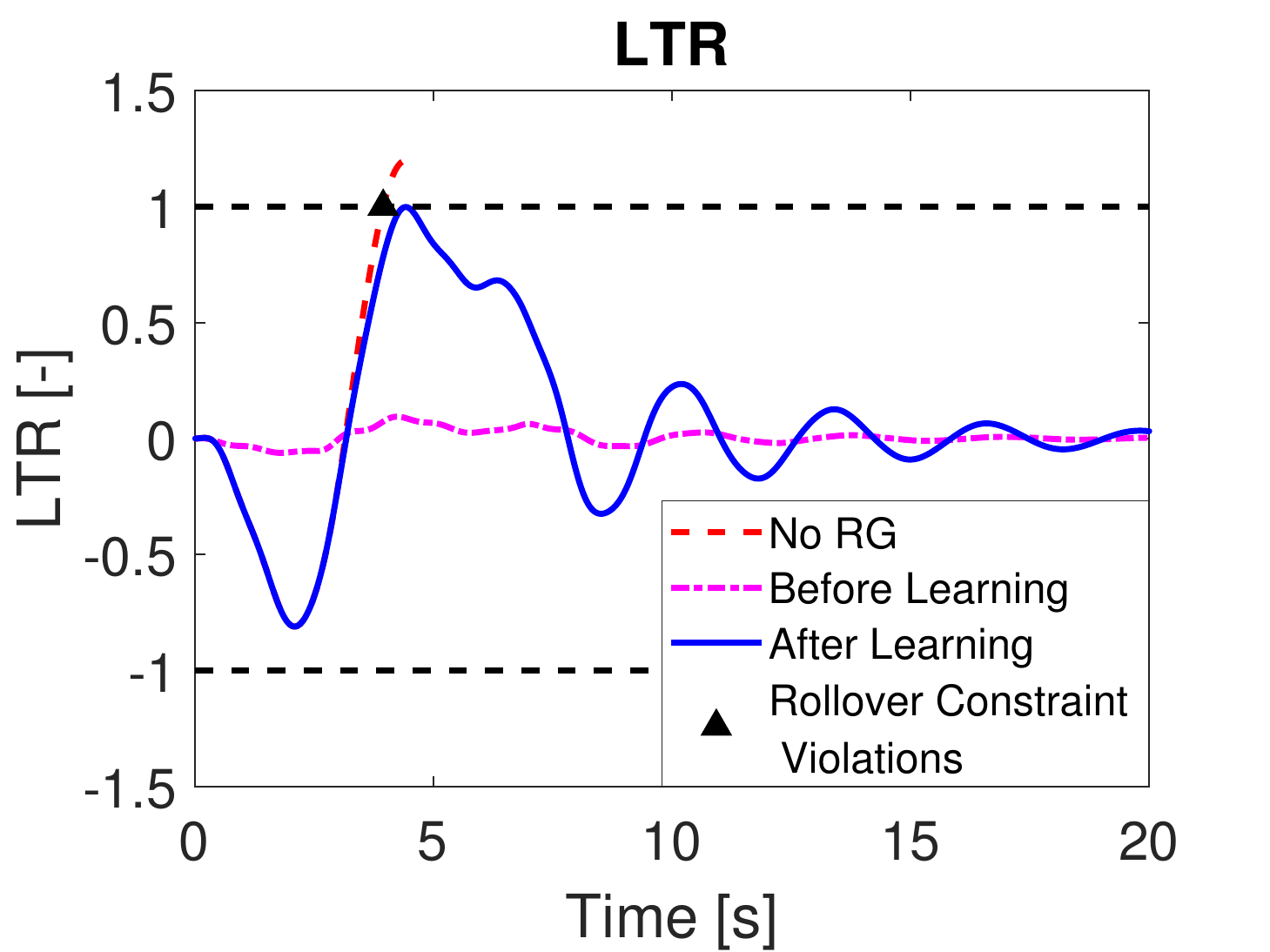,height=.195\textwidth}}  
\put(  125,  95){\epsfig{file=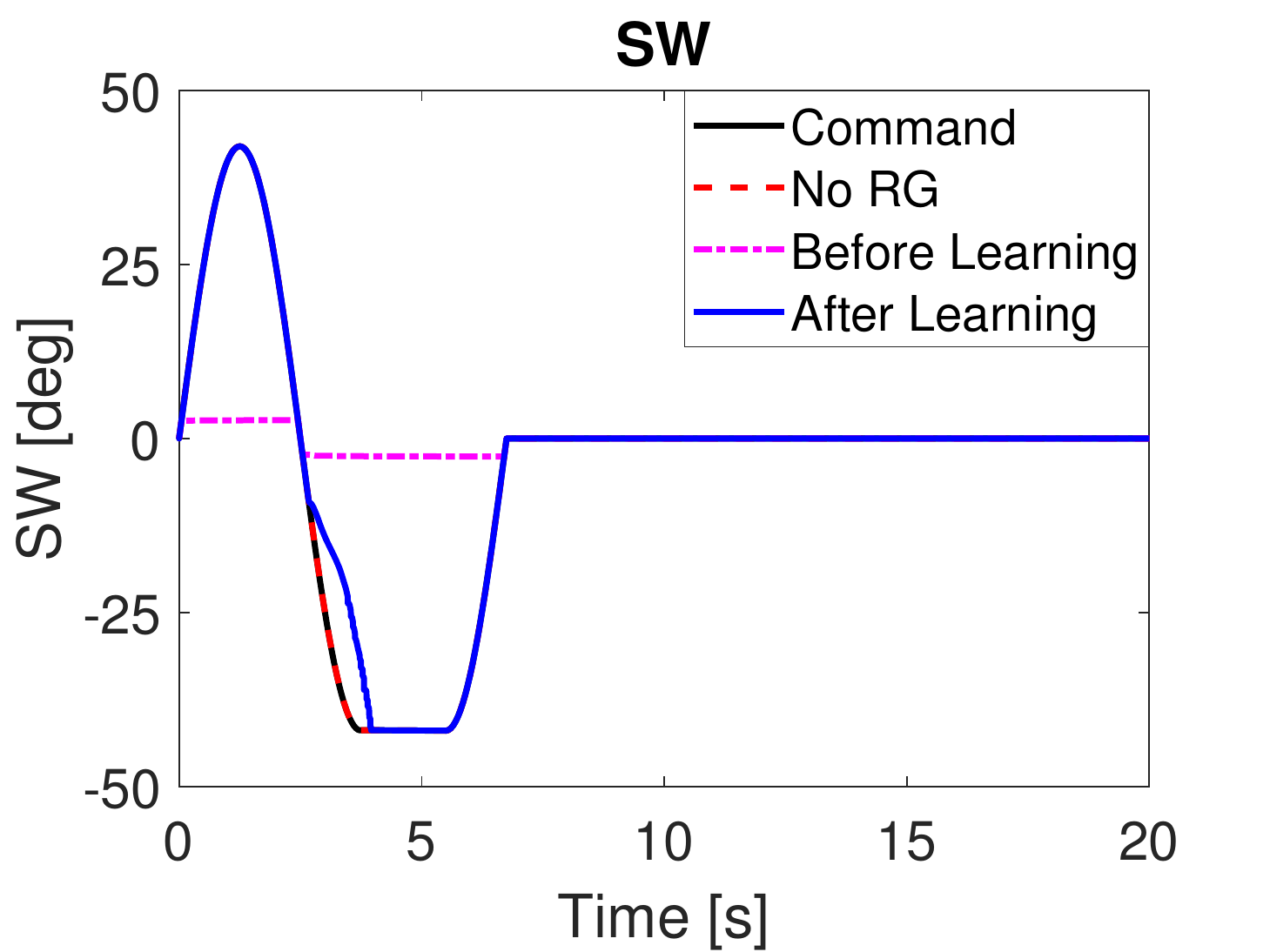,height=.195\textwidth}}  
\put(  125,  195) {\epsfig{file=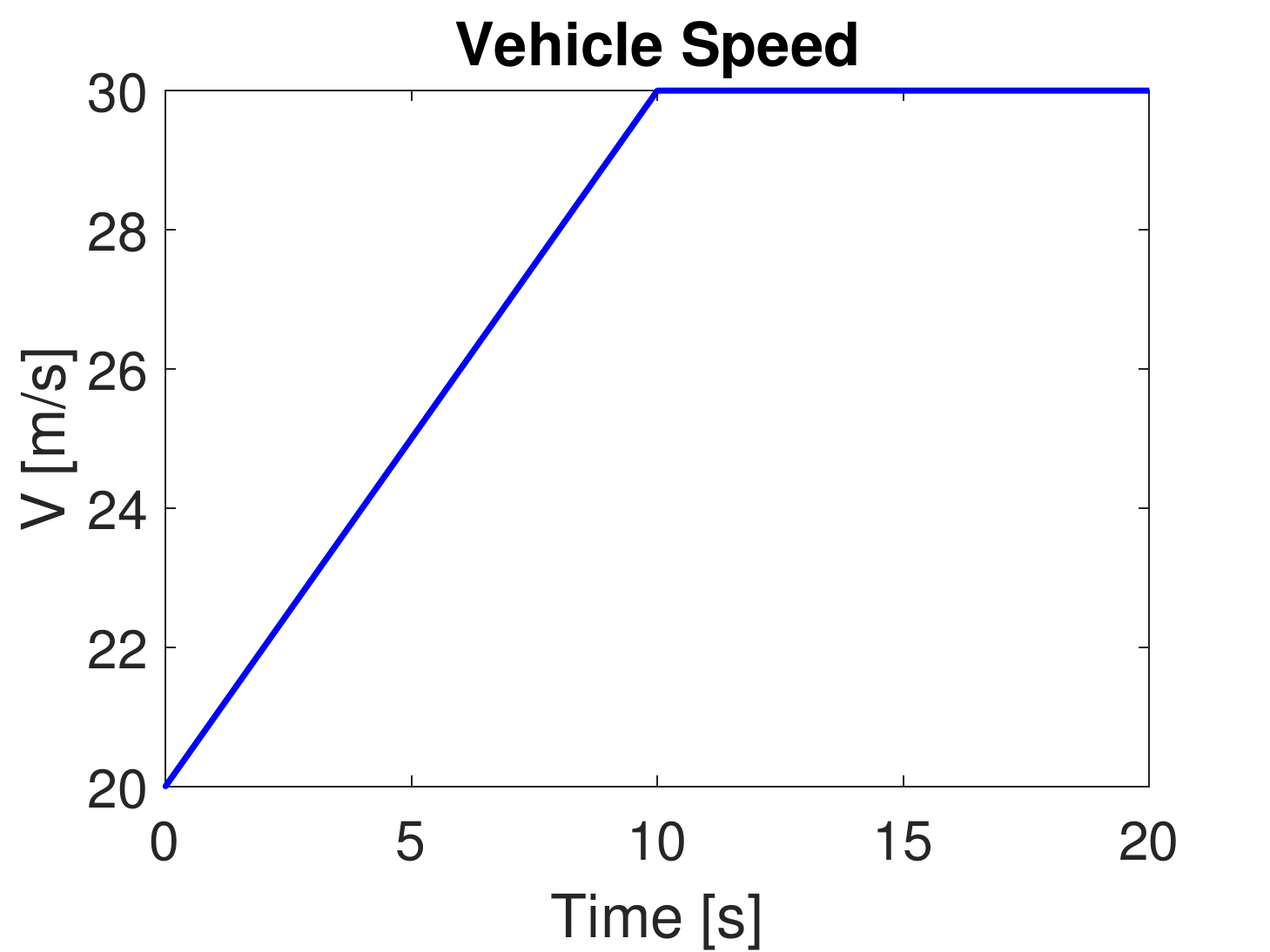,height=.195\textwidth}}  
\put( 14, 290){\small{(a)}}
\put( 144, 290){\small{(b)}}
\end{picture}
\end{center}
      \caption{Vehicle speed, steering angle, and LTR responses for the sine-and-dwell test when the truck is (a) decelerating from 30 (m/s) to 20 (m/s) at a rate of -3 (m/s$^2$) and (b) accelerating from 20 (m/s) to 30 (m/s) at a rate of 1 (m/s$^2$).}
      \label{fig:sin_and_dwell_spd_change}
     \vspace{0.2in}
\end{figure}

In Fig.~\ref{fig:sin_and_dwell_spd_change}(a), the sine-and-dwell test is performed while the truck is decelerating. Compared with Fig.~\ref{fig:sin_and_dwell}, where the truck is driving at a constant speed of 25 (m/s), rollover constraint violations occur when the truck is performing positive steering, which is attributed to the larger speed at the beginning. As a consequence, LRG modifies the positive steering command so that the constraints are enforced. Fig.~\ref{fig:sin_and_dwell_spd_change}(b) shows the sine-and-dwell test results when the truck is accelerating. LRG makes smaller modifications when responding to the negative steering command compared with Fig.~\ref{fig:sin_and_dwell} because of the lower speed before 5 (sec). In either Fig.~\ref{fig:sin_and_dwell_spd_change}(a) or (b), LRG is able to guard the fuel truck from violating rollover constraints in presence of speed changes.

%% file: sections/06_conclusion.tex
\section{Conclusion} \label{sec:conclusion}

In this paper, a learning reference governor (LRG) approach has been proposed to enforce state and control constraints through the modification of the reference command in systems for which an accurate model is unavailable. The LRG uses learning to improve the command tracking performance without causing constraint violations. Theoretical guarantees of safety, convergence of the learning algorithm and finite-time convergence of the modified reference command to the original constant reference command have been given. Finally, the application of the proposed approach to fuel truck rollover avoidance
under fuel sloshing effects has been considered. Simulation results have been presented that demonstrate LRG can protect the fuel truck from rollover and the conservatism can be reduced through learning, pushing this vehicle to its mobility limits. Furthermore, LRG can be configured to support vehicle operation at different tank fill ratios and varying vehicle speeds. A similar approach could be employed to handle variability due to road surface conditions, and bank and incline angles.

\section{Acknowledgements}

The authors would like to acknowledge the technical and financial support of the Automotive Research Center (ARC) in accordance with Cooperative Agreement W56HZV-19-2-0001 U.S. Army DEVCOM Ground Vehicle Systems Center (GVSC) Warren, MI and the Advanced Vehicle Power Technology Alliance (AVPTA) between the Department of Energy (DOE) and the Department of the Army (DA).

Unclassified: DISTRIBUTION A. Approved for public release; distribution unlimited. OPSEC\#: 4915.

Disclaimer: Reference herein to any specific commercial company, product, process, or service by trade name, trademark, manufacturer, or otherwise, does not necessarily constitute or imply its endorsement, recommendation, or favoring by the United States Government or the Department of the Army (DoA). The opinions of the authors expressed herein do not necessarily state or reflect those of the United States Government or the DoA, and shall not be used for advertising or product endorsement purposes.

%% file: sections/07_appendix.tex
\section*{Appendix}

Here we show that the H\"older continuity assumption, {\it (A4)}, for the function $D$ defined in \eqref{eq:def_D} holds true for all stable linear time-invariant systems. In particular, $D$ for stable linear time-invariant systems are actually Lipschitz continuous (H\"older continuous with $\beta = 1$).

{\bf Lemma 2:} For linear time-invariant systems in the following form: 
\begin{subequations}\label{eq:lti_sys}
\begin{align}
    \dot{\state}(t) &= A\state(t) + B\inputref(t), \\
    \outputsys(t) &= C\state(t) + F \inputref(t), 
\end{align} 
\end{subequations}
where $A$ is a stable matrix, i.e., every eigenvalue of $A$ has strictly negative real part. The function $D$ defined in \eqref{eq:def_D} for the system \eqref{eq:lti_sys} satisfies the H\"older continuity condition \eqref{eq:lipschitz} with $\beta = 1$ and some finite $L>0$.

{\bf Proof:} For any $(\inputref_1, \delta\inputref_1, \delta\state_1), (\inputref_2, \delta\inputref_2, \delta\state_2) \in \mathbb{R}^{n_\inputref} \times \mathbb{R}^{n_\inputref} \times \mathbb{R}^n$, we have
\small
\begin{align*}
&\, \big| D(\inputref_1, \delta\inputref_1, \delta\state_1) - D(\inputref_2, \delta\inputref_2, \delta\state_2) \big| \\[2pt]
= &\, \big| \sup_{t \in [0,\infty)} \big\|\phi(t, x_\inputref(\inputref_1)+\delta\state_1, \inputref_1+\delta\inputref_1 \big) - \outputsys_\inputref(\inputref_1) \big\|  \,- \\[-2pt]
&\sup_{t \in [0,\infty)} \big\|\phi(t, x_\inputref(\inputref_2)+\delta\state_2, \inputref_2+\delta\inputref_2) - \outputsys_\inputref(\inputref_2) \big\|\, \big| \\
\le &\, \sup_{t \in [0,\infty)} \big\| \phi(t, x_\inputref(\inputref_1)+\delta\state_1, \inputref_1+\delta\inputref_1) - \\[-2pt]
& \phi(t, x_\inputref(\inputref_2)+\delta\state_2, \inputref_2+\delta\inputref_2 ) - \big(\outputsys_\inputref(\inputref_1) - \outputsys_\inputref(\inputref_2)\big) \big \| \\[2pt]
= &\, \sup_{t \in [0,\infty)} \big\| C \psi\big(t, \state_\inputref(\inputref_1 - \inputref_2) + (\delta\state_1 - \delta\state_2), (\inputref_1 - \inputref_2) + \\[-2pt]
& (\delta\inputref_1 - \delta\inputref_2) \big) + F \big((\inputref_1 - \inputref_2) + (\delta\inputref_1 - \delta\inputref_2)\big) - \outputsys_\inputref(\inputref_1-\inputref_2) \big\| \\[2pt]
\le &\, \sup_{t \in [0,\infty)} \big\| C \psi\big(t, \state_\inputref(\inputref_1 - \inputref_2) + (\delta\state_1 - \delta\state_2), (\inputref_1 - \inputref_2) + \\[-2pt]
& (\delta\inputref_1 - \delta\inputref_2) \big) \big\| + \big\| F \big((\inputref_1 - \inputref_2) + (\delta\inputref_1 - \delta\inputref_2)\big) - \outputsys_\inputref(\inputref_1-\inputref_2) \big\|,
\end{align*}
\normalsize
where we have used the superposition property for linear systems and all suprema are finite as $A$ is stable.

Using $\state_\inputref(\inputref) = -A^{-1}B\inputref$, $\outputsys_\inputref(\inputref) = (-CA^{-1}B + F)\inputref$, $\psi(t, \state_0, \inputref) = e^{At} \state_0 + A^{-1} (e^{At}-I)B \inputref$, and the fact that $A$ is a stable matrix, it is easily seen from the above inequality that there exists some finite $L'>0$ that is independent of $(\inputref_1, \delta\inputref_1, \delta\state_1)$ and $(\inputref_2, \delta\inputref_2, \delta\state_2)$ such that
\begin{align*}
&\, \big| D(\inputref_1, \delta\inputref_1, \delta\state_1) - D(\inputref_2, \delta\inputref_2, \delta\state_2) \big| \\
\le &\, L' \big(\|\inputref_1-\inputref_2\| + \|\delta\inputref_1-\delta\inputref_2\| + \|\delta\state_1-\delta\state_2\|\big).
\end{align*}

Since $\|\inputref\| + \|\delta\inputref\| + \|\delta\state\|$ defines a norm on $\mathbb{R}^{n_\inputref} \times \mathbb{R}^{n_\inputref} \times \mathbb{R}^n$, using the equivalence of norms for finite-dimensional vector spaces, we obtain \eqref{eq:lipschitz} with some finite $L>0$ and exponent $\beta = 1$. Under Euclidean norm, $L'$ can be expressed as $L' = \max\Big(\eta \|C\|, (\eta+1)\|CA^{-1}\|\|B\| + \|F\|$\Big), where $\eta = \sup_{t \in [0,\infty)} \|e^{At}\|$. $\blacksquare$

{\color{black}
In the next lemma, we show an example of nonlinear systems, to which the corresponding function $D$ in \eqref{eq:def_D} satisfies our H\"older continuity assumption {\it (A4)} and is not necessarily Lipschitz continuous. This example also illustrates the fact that the H\"older continuity assumption made in this paper, {\it (A4)}, is a weaker and more general assumption than the Lipschitz continuity assumption relied upon in our previous work \cite{liu2020model} and \cite{liu2021tanktruck}.


{\bf Lemma 3:} For systems in the following form: 
\begin{subequations}\label{eq:lti_sys_holder}
\begin{align}
    \dot{\state}(t) &= A\state(t) + B\inputref(t)^\frac{1}{\beta}, \\
    \outputsys(t) &= C\state(t) + F \inputref(t)^\frac{1}{\beta}, 
\end{align} 
\end{subequations}
where $\beta \geq 1$, and $A$ is a stable matrix, and the state space $X \subset \mathbb{R}^n$ is bounded, and $\inputref(t) \in \mathbb{R}$ is a scalar. The function $D$ defined in \eqref{eq:def_D} for the system \eqref{eq:lti_sys} satisfies the H\"older continuity condition \eqref{eq:lipschitz} with exponent $1/\beta$ and some finite $L>0$.

{\bf Proof:} Let $u = \inputref^\frac{1}{\beta}$. Then we can regard $u$ as the input to the \eqref{eq:lti_sys_holder} and apply the results obtained in Lemma 2. For any $(\inputref_1, \delta\inputref_1, \delta\state_1), (\inputref_2, \delta\inputref_2, \delta\state_2) \in \mathbb{R} \times \mathbb{R} \times X$, we obtain
\begin{align*}
&\, \big| D(\inputref_1, \delta\inputref_1, \delta\state_1) - D(\inputref_2, \delta\inputref_2, \delta\state_2) \big| \\
\le &\, L' \big(\|u_1-u_2\| + \|\delta u_1-\delta u_2\| + \|\delta\state_1-\delta\state_2\|\big),
\end{align*}
where $u = \inputref^\frac{1}{\beta}$ and $\delta u = (\inputref + \delta\inputref)^\frac{1}{\beta} - \inputref^\frac{1}{\beta}$. Substitute $u$ and $\delta u$ with $\inputref$ and $\delta\inputref$, 
\begin{align*}
&\, \big| D(\inputref_1, \delta\inputref_1, \delta\state_1) - D(\inputref_2, \delta\inputref_2, \delta\state_2) \big| \\
\le &\, L' \big(\|\inputref_1^\frac{1}{\beta} - \inputref_2^\frac{1}{\beta}\| +  \|\delta\state_1-\delta\state_2\| + \\
&\, \| (\inputref_1 + \delta\inputref_1)^\frac{1}{\beta} - \inputref_1^\frac{1}{\beta} - (\inputref_2 + \delta\inputref_2)^\frac{1}{\beta} + \inputref_2^\frac{1}{\beta}\| \big) \\
\le &\, L' \big(2\|\inputref_1^\frac{1}{\beta} - \inputref_2^\frac{1}{\beta}\| +  \|\delta\state_1-\delta\state_2\| + \\
&\, \| (\inputref_1 + \delta\inputref_1)^\frac{1}{\beta} - (\inputref_2 + \delta\inputref_2)^\frac{1}{\beta}\| \big) \\
\le &\, L' \big(2\|\inputref_1 - \inputref_2\|^\frac{1}{\beta} +  \|\delta\state_1-\delta\state_2\| + \|\inputref_1 + \delta\inputref_1 - \inputref_2 - \delta\inputref_2\|^\frac{1}{\beta} \big) \\
\le &\, L' \big(3\|\inputref_1 - \inputref_2\|^\frac{1}{\beta} +  \|\delta\state_1-\delta\state_2\| + \|\delta\inputref_1 - \delta\inputref_2\|^\frac{1}{\beta} \big) \\
\le &\, L' \big(3\|\inputref_1 - \inputref_2\|^\frac{1}{\beta} +  K\|\delta\state_1-\delta\state_2\|^\frac{1}{\beta} + \|\delta\inputref_1 - \delta\inputref_2\|^\frac{1}{\beta} \big), 
\end{align*}
where the second and fourth inequalities are derived from the triangle inequality, and the third inequality is obtained from inverse triangle inequality with exponent $1/\beta$, and $K = \max_{\delta\state_{1,2}\in X} \|\delta\state_1 - \delta\state_2 \|^{1-\frac{1}{\beta}}$. 

Note that functions $f(x) = x^\frac{1}{\beta}$ with $\beta \geq 1$ is concave on $x \in [0, \infty)$. Then,
\begin{align*}
&\, \big| D(\inputref_1, \delta\inputref_1, \delta\state_1) - D(\inputref_2, \delta\inputref_2, \delta\state_2) \big| \\
\le &\, L' \big( a_1\|\inputref_1 - \inputref_2\| +  a_2\|\delta\state_1-\delta\state_2\| + a_3\|\delta\inputref_1 - \delta\inputref_2\| \big)^\frac{1}{\beta},
\end{align*}
where $a_1 = \frac{12^\beta}{4}$, $a_2 = \frac{4^\beta K^\beta}{4}$, and $a_3 = 2^{\beta-1}$. Since $a_1\|\inputref\| + a_2\|\delta\state\| + a_3\|\delta\inputref\|$ defines a norm on $\mathbb{R} \times \mathbb{R} \times X$, using the equivalence of norms for finite-dimensional vector spaces, we obtain \eqref{eq:lipschitz} with some finite $L>0$ and exponent $1/\beta$. $L'$ can be expressed as $L' = \max\Big(\eta \|C\|, (\eta+1)\|CA^{-1}\|\|B\| + \|F\|$\Big), where $\eta = \sup_{t \in [0,\infty)} \|e^{At}\|$. $\blacksquare$
}
